\documentclass[aps,superscriptaddress,twocolumn,twoside,floatfix,rmp,nofootinbib,a4paper]{revtex4-2}
\usepackage{times}
\usepackage{amsmath}
\usepackage{amssymb}
\usepackage{amsthm}
\usepackage{natbib}
\usepackage[T1]{fontenc}
\usepackage[utf8]{inputenc}
\usepackage{dsfont}
\usepackage[caption=false]{subfig}
\usepackage{kbordermatrix}
 \renewcommand{\kbldelim}{(}
 \renewcommand{\kbrdelim}{)}

\usepackage{pifont}
 \newcommand{\vmark}{\text{\ding{51}}}

\usepackage[colorlinks=true,linkcolor=blue,citecolor=magenta,urlcolor=blue]{hyperref}
\usepackage{overpic}

\theoremstyle{definition}
\newtheorem*{remark*}{Remark}

\DeclareMathOperator{\tr}{tr}
\newcommand{\id}{\mathds{1}}
\newcommand{\ket}[1]{|#1\rangle}
\newcommand{\bra}[1]{\langle#1|}
\newcommand{\braket}[2]{\langle#1|#2\rangle}
\newcommand{\bracket}[3]{\langle#1|#2|#3\rangle}
\newcommand{\ketbra}[2]{|#1\rangle\langle#2|}
\newcommand{\expect}[1]{\langle#1\rangle}
\DeclareMathOperator{\dint}{d\!}
\newcommand{\st}{\mathrm{s.t.}}
\begin{document}


\title{Semidefinite programming relaxations for quantum correlations}

\author{Armin Tavakoli}
\affiliation{\mbox{Physics Department and NanoLund, Lund University, Box 118, 22100 Lund}, \mbox{Sweden}}

\author{Alejandro Pozas-Kerstjens}
\affiliation{\mbox{Group of Applied Physics, University of Geneva, 1211 Geneva 4, Switzerland}}
\affiliation{\mbox{Instituto de Ciencias Matem\'aticas (CSIC-UAM-UC3M-UCM), 28049 Madrid}, \mbox{Spain}}

\author{Peter Brown}
\affiliation{\mbox{T\'el\'ecom Paris~-~LTCI, Inria, Institut Polytechnique de Paris, 91120 Palaiseau}, \mbox{France}}

\author{Mateus Araújo}
\affiliation{\mbox{Departamento de Física Teórica, Atómica y Óptica, Universidad de Valladolid}, \mbox{47011 Valladolid, Spain}}

\begin{abstract}
Semidefinite programs are convex optimisation problems involving a linear objective function and a domain of positive-semidefinite matrices.
Over the past two decades, they have become an indispensable tool in quantum information science.
Many otherwise intractable fundamental and applied problems can be successfully approached by means of relaxation to a semidefinite program.
Here, we review such methodology in the context of quantum correlations.
We discuss how the core idea of semidefinite relaxations can be adapted for a variety of research topics in quantum correlations, including nonlocality, quantum communication, quantum networks, entanglement, and quantum cryptography.
\end{abstract}


\maketitle

\tableofcontents

\section{Introduction}\label{sec:intro}
Understanding and explaining the correlations observed in nature is a central task for any scientific theory.
For quantum mechanics, the study of correlations has a crucial role in both its concepts and its applications.
It broadly concerns the foundations of quantum theory, quantum information science and nowadays also the emerging quantum technologies.
Although quantum correlations is an umbrella term, under which many different physical scenarios are accommodated, it establishes a common focus on the investigation of probability distributions describing physical events.
Naturally, the various expansive topics focused on quantum correlations have warranted review articles of their own and we refer to them for specific in-depth discussions; see e.g.~\citet{Genovese2005, BrunnerCavalcantiReview2014, TavakoliPozas2022} for nonlocality, \citet{GuhneReview2009, HorodeckiReview2009, Friis2019} for entanglement, \citet{Budroni2022review} for contextuality, \citet{Brassard2003, Gisin2007, Buhrman2010} for quantum communication and \citet{Gisin2002, Scarani2009, Xu2020, Portmann2022, Pirandola20020} for quantum cryptography.

Studies of quantum correlations take place in a given scenario, or experiment, where events can influence each other according to some causal structure and the influences are potentially subject to various physical limitations.
For example, this can be a Bell experiment where two parties act outside each other's light cones but are nevertheless connected through a pre-shared entangled state.
Another example is a communication scenario where the channel connecting the sender to the receiver only supports a given number of bits per use.
The fundamental challenge is to characterise the set of correlations predicted by quantum theory.
This applies directly to a variety of basic questions, e.g.~determining the largest violation of a Bell inequality or the largest quantum-over-classical advantage in a communication task.
It also applies indirectly to several problems that rely on bounding such correlations, e.g.~benchmarking a desirable property of quantum devices or computing the secret key rate in a quantum cryptographic scheme.
Unfortunately, the characterisation of quantum correlations is typically difficult and can only be solved exactly, by analytical means, in a handful of convenient special cases.
Therefore, it is of pivotal interest to find other, more practically viable, methods for characterising quantum correlations.

Over the past two decades, semidefinite programs (SDPs) have emerged as an efficient and broadly useful tool for investigating quantum theory in general, and quantum correlations in particular.
An SDP is an optimisation task in which a linear objective function is maximised over a cone of positive-semidefinite (PSD) matrices subject to linear constraints.
They rose to prominence in convex optimisation theory in the early 1990s through the development of efficient interior-point evaluation methods; see \citet{Vandenberghe1996} for a review.
Today they are an indispensable tool for the field of quantum correlations.
However, it is frequently the case that quantum correlation problems cannot directly be cast as SDPs, thus impeding a straightforward solution.
Sometimes the reason is that the problem is simply not convex, e.g.~bilinear optimisation.
Sometimes the reason is that although the problem is convex, it cannot be represented exactly as an SDP, e.g.~optimisation of von Neumann entropy.

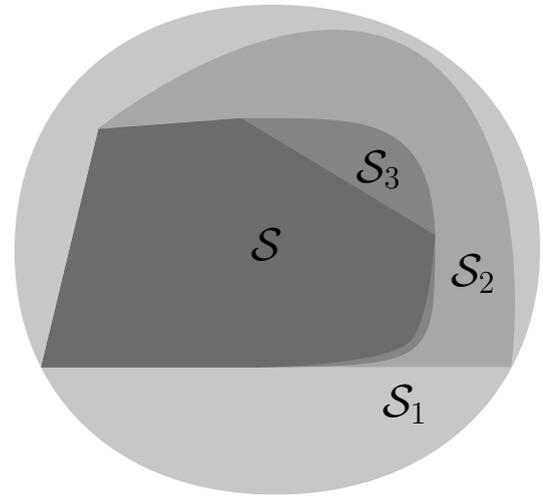
\begin{figure}
	\centering
	\begin{overpic}[width=0.8\columnwidth]{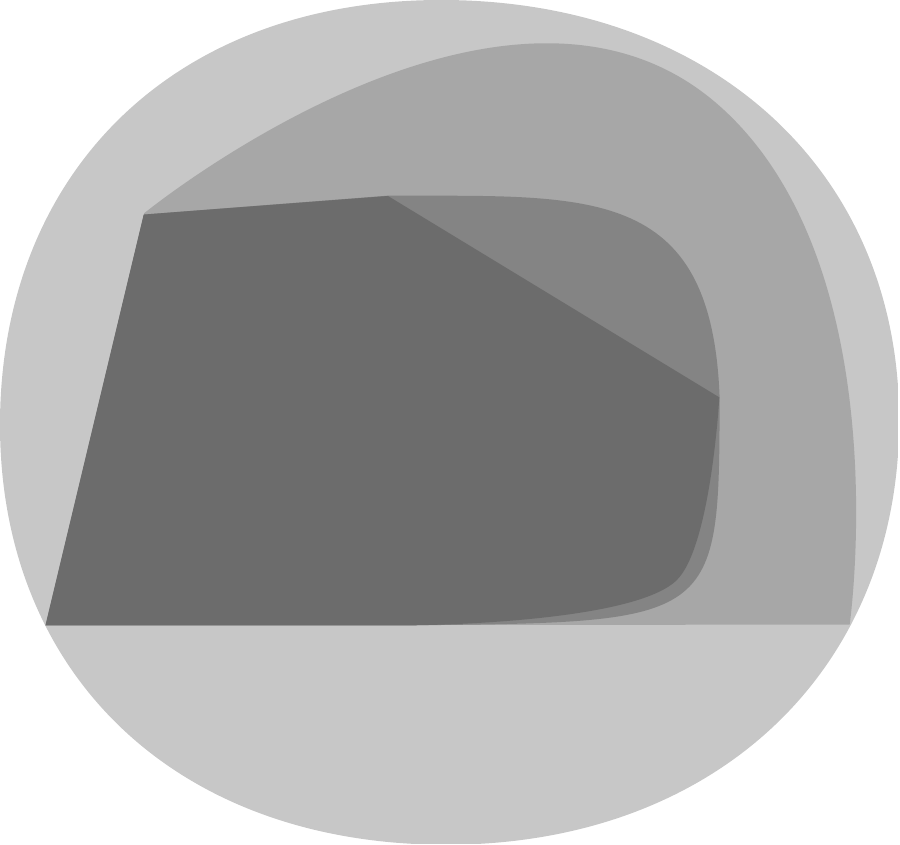}
		\put(45,45){\LARGE$\mathcal{S}$}
		\put(70,15){\LARGE$\mathcal{S}_1$}
		\put(83,40){\LARGE$\mathcal{S}_2$}
		\put(65,60){\LARGE$\mathcal{S}_3$}
	\end{overpic}
	\caption{Schematic representation of SDP relaxation methods.
Instead of directly characterising a complicated set, $\mathcal{S}$, one approximates it with supersets $\mathcal{S}_i$, each of which can be characterised via SDP.
Often there exists sequences $\{\mathcal{S}_i\}_i$ such that the next set is contained in the previous, thus giving a more precise approximation of $\mathcal{S}$.}
	\label{fig:relaxation}
\end{figure}

Nevertheless, SDPs offer a powerful path out of such difficulties because they can be employed to approximate solutions that would otherwise remain obscure.
Specifically, more sophisticated quantum correlation problems, that are not immediately solvable by an SDP, can still be approached through sequences of increasingly precise relaxations, each of which is itself an SDP (see Fig.~\ref{fig:relaxation}).
In this way, one can obtain approximations that are accurate enough for practical purposes and sometimes even exact.
From the methodology perspective, these SDP relaxation methods have attained a prominent role in quantum information science.
Their success derives in part from the fact that today there exist powerful, practical and easily accessible algorithms for their evaluation, and partly from the fact that they offer a single methodology that pertains to most forms of quantum correlation, even though the physics underpinning experiments can be vastly different.

The purpose of this article is to review SDP relaxation methods for quantum correlations.
We discuss how this methodology can be adapted for a variety of conceptual and applied problems.
In Section~\ref{sec:background}, we introduce the basics of semidefinite programming and some of the main correlation scenarios.
In Section~\ref{sec:sdp}, we present a general framework for semidefinite relaxation hierarchies that can be applied to many of the later, physically motivated, considerations.
Sections~\ref{sec:separability} and~\ref{sec:nonlocality} discuss SDP relaxation methods in the context of entanglement theory and nonlocality, respectively, including device-independent applications.
Section~\ref{sec:communication} focuses on correlations from quantum communication and their applications.
Section~\ref{sec:qkd} concerns SDP methods for evaluating the performance of protocols in random number generation and quantum key distribution.
Section~\ref{sec:networks} focuses on networks comprised of independent sources of entanglement and discusses SDP methods for assessing their nonlocality.
Section~\ref{sec:further} gives an overview of some related topics where SDP relaxations are prominent.
Finally, Section~\ref{sec:conclusions} provides a concluding outlook.
A guide to free and publicly available SDP solvers and relevant quantum information software packages is found in Appendix~\ref{sec:implementations}.

\section{Background}\label{sec:background}
We begin with a basic introduction to semidefinite programming, referring the reader to relevant books and review articles, e.g.~\citet{Wolkowicz2000, Klerk2002, Boyd2004}, for in-depth discussions.
In particular, for their use in quantum correlations, see the recent book \citet{SDPbook}, which offers a didactic approach to SDP using basic quantum information tasks, and \citet{mironowicz2023}, which focuses on the mathematical foundations of SDP.

\subsection{Primals and duals}
A semidefinite program is an optimisation problem in which a linear objective function is optimised over a convex domain consisting of the intersection of a cone of PSD matrices with hyperplanes and half-spaces.
In general this can be written as
\begin{equation}\label{primal}
	\begin{aligned}
		\max_X \quad & \langle C, X\rangle \\
		\st \quad & \langle A_i, X \rangle = b_i\quad\forall\,i, \\
		& X \succeq 0,
	\end{aligned}
\end{equation}
where $C$, $X$, and the $A_i$ are Hermitian matrices, $b$ is a real vector with components $b_i$, $\langle \cdot,\cdot\rangle$ denotes the inner product, and $X\succeq 0$ means that $X$ is PSD.
In addition to the above linear equality constraints, SDPs can also include linear inequality constraints.
These can always be converted into linear equality constraints, as appearing in Eq.~\eqref{primal}, by introducing additional parameters known as slack variables.
Whilst we have chosen to define SDPs as optimizations of the form presented in Eq.~\eqref{primal}, many alternative definitions exist in the literature, e.g., see~\cite{Watrous2018} for a definition based on Hermitian-preserving maps\footnote{A Hermitian-preserving map has a Hermitian Choi state representation.}.
Importantly, however, all such definitions are equivalent, and in particular any SDP presented can be rewritten in the form of Eq.~\eqref{primal}.

SDPs are generalisations of the more elementary linear programs (LPs) for which the PSD constraint is replaced by an element-wise positivity constraint.
This is achieved by restricting the matrix $X$ to be diagonal.
It is well known that LPs can be efficiently evaluated using interior-point methods \cite{Karmarkar1984} and such methods also generalise to the case of SDPs \cite{Alizadeh1992, Kamath1992, Kamath1993, NesterovBook}.

To every SDP of the form of Eq.~\eqref{primal}, one can associate another SDP of the form
\begin{equation}\label{dual}
	\begin{aligned}
		\min_{y} \quad & \langle b, y \rangle \\
		\st \quad & \sum_i A_i y_i \succeq C,
	\end{aligned}
\end{equation}
where the optimisation is now over the real vector $y$ with components $y_i$.
This is known as the dual SDP corresponding to the primal SDP in Eq.~\eqref{primal}.
Every feasible point of the dual SDP gives a value $\langle b, y \rangle$ that is an upper bound on the optimal value of the primal SDP.
Thus, also every feasible point of the primal SDP gives a value $\langle C, X \rangle$ that is a lower bound on the optimal value of the dual SDP.
This relation is known as weak duality.

A fundamental question is whether the bounds provided by weak duality can be turned into equality, i.e.,~when does the optimal value of the primal in Eq.~\eqref{primal} coincide with the optimal value of the dual in Eq.~\eqref{dual}? When they are equal we say that strong duality holds.
Strong duality always holds for LPs but in general not for SDPs.
However, a sufficient condition for strong duality is that the primal or the dual SDP is strictly feasible \cite{Slater1950}: in the primal formulation \eqref{primal} this means that there exists an $X^* \succ 0$ such that $\langle A_i, X^* \rangle = b_i$ for all $i$, and in the dual formulation \eqref{dual} that there exists a $y^*$ such that $\sum_i A_i y^*_i \succ C$.

If one wants to solve an SDP numerically one needs, in addition, that the optimal values of the primal and dual problems are attained, i.e., that there exist finite $X$ and $y$ that produce the optimal values.
A sufficient condition for their existence is that the SDP is strictly feasible and its feasible region is bounded, or that both the primal and dual problems are strictly feasible \cite{NesterovBook, Drusvyatskiy2017}.

\subsection{Correlation scenarios and quantum theory}
This section provides a brief introduction to some often studied quantum correlation scenarios.
We first discuss scenarios based on entanglement and then address scenarios featuring communication.
The presentation is geared towards highlighting the relevance of LPs and SDPs.

\subsubsection{Entanglement-based scenarios}
\label{sec:introduction:entanglement}
\begin{figure}
	\centering
	\includegraphics[width=0.95\columnwidth]{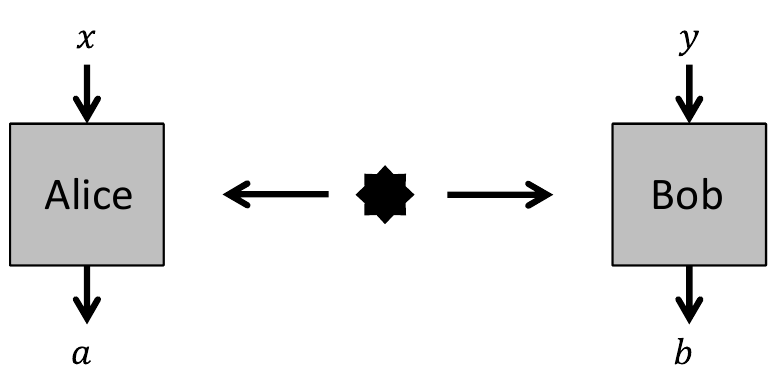}
	\caption{The standard bipartite entanglement-based scenario.
A source emits a pair of particles shared between the parties Alice and Bob.
They each privately select inputs $x$ and $y$ and perform associated measurements that produce outcomes $a$ and $b$, respectively.}\label{FigBellScenario}
\end{figure}

The standard scenario for investigating quantum correlations harvested from the shares of a bipartite state is illustrated in Fig.~\ref{FigBellScenario}.
A source emits a pair of particles in some state $\rho_{AB}$ that is shared between two parties, Alice and Bob.
Formally, a state is a PSD operator $\rho_{AB}\succeq 0$ of unit trace $\tr(\rho_{AB})=1$.
Alice and Bob can independently select classical inputs, $x$ and $y$, respectively from finite sets $X$ and $Y$, and perform corresponding quantum measurements on their systems $A$ and $B$.

In general, a quantum measurement with $N$ possible outcomes is represented by a positive operator-valued measure (POVM), i.e.,~a set of PSD operators $\{E_i\}_{i=1}^N$ that sums to identity: $E_i\succeq 0$ and $\sum_{i=1}^N E_i=\id$.
These conditions ensure the positivity and normalisation of probabilities, respectively.
We write the measurements of Alice and Bob as POVMs $\{A_{a|x}\}$ and $\{B_{b|y}\}$ respectively, where $a$ and $b$ denote their respective outcomes.
The probability distribution of their outcomes, for a specific choice of inputs, is given by Born's rule,
\begin{equation}\label{correlations}
	p(a,b|x,y)=\tr\left((A_{a|x}\otimes B_{b|y})\rho_{AB}\right).
\end{equation}
This conditional probability distribution is interchangeably referred to as the distribution or the correlations.
Such entanglement-based correlations are often studied in three different scenarios, namely those of entanglement, steering and nonlocality:

\textbf{Entanglement.---} A bipartite quantum state is called separable if it can be written as a probabilistic mixture of states individually prepared by Alice and Bob, namely \cite{nielsen2010}
\begin{equation}\label{separable}
	\rho_{AB}=\sum_{\lambda} p(\lambda) \phi_\lambda\otimes \varphi_\lambda,
\end{equation}
where $p(\lambda)$ is a probability distribution and $\phi_\lambda$ and $\varphi_\lambda$ are arbitrary quantum states of Alice and Bob, respectively.
Importantly, some bipartite states cannot be decomposed in this way, and are called entangled.
For an in-depth discussion of entanglement, we refer to \citet{HorodeckiReview2009}.

Assume that Alice and Bob, as in Fig.~\ref{FigBellScenario}, perform known quantum measurements on some unknown shared state $\rho_{AB}$.
Can we determine if the state is separable or entangled? This is done by inspecting the correlations in Eq.~\eqref{correlations}.
One approach is that both Alice and Bob perform a set of tomographically complete local measurements; most famously exemplified by a complete set of mutually unbiased bases \cite{Ivanovic1981, Wootters1989} or a symmetric, informationally complete POVM \cite{Renes2004}.
Then they can reconstruct the density matrix $\rho_{AB}$ and try to decide its separability through some analytical criterion.
Unfortunately, the separability problem is known to be NP-hard\footnote{More precisely, it is NP-hard to decide whether a quantum state is $\epsilon$-close to the set of separable states when $\epsilon$ is an inverse polynomial of the dimension.} \cite{Gurvits2003, Gharibian2010} and a necessary and sufficient criterion is only known for qubit-qubit or qubit-qutrit systems.
This is the well-known positive partial transpose (PPT) criterion \cite{Peres1996, Horodecki1996}, which more generally is a necessary condition for separability in all dimensions: a bipartite system is entangled if $\rho_{AB}^{T_A}\nsucceq 0$, where $T_A$ denotes transposition on system $A$.\footnote{If $\rho_{AB}=\sum_{ijkl}\rho_{ijkl}\ket{ij}\bra{kl}$, then $\rho_{AB}^{T_A}=\sum_{ijkl}\rho_{ijkl}\ket{kj}\bra{il}$.} However, many entangled states go undetected by this criterion \cite{Horodecki1998}.

The PPT criterion can be used to quantify the amount of entanglement in a quantum state, for example by computing how much of the maximally mixed state needs to be mixed with $\rho_{AB}$ to make the resulting state PPT.
This is known as the random robustness of entanglement with respect to the PPT criterion, and can be computed via a simple SDP\footnote{The random robustness is then given by $-t d^2$.} \cite{Vidal1999}:
\begin{equation}\label{entanglementprimal}
	\begin{aligned}
		\max_t \quad & t \\
		\st \quad & \rho_{AB}^{T_A} \succeq t \id.
	\end{aligned}
\end{equation}
By the above discussion, if the optimal value of the SDP is negative then it must be the case that the state $\rho_{AB}$ is entangled.
Note that since having non-positive partial transposition is not a necessary condition for a state to be entangled, this SDP is a \emph{relaxation} of the separability problem, and the random robustness it computes is only a lower bound on the actual amount of entanglement in $\rho_{AB}$.
This is a simple example of the fundamental idea behind the methods explored in this review: in order to tackle an intractable problem, one finds partial conditions for its solution that are tractable to compute and provide bounds on the quantity of interest.
Ideally, one should find a sequence of tighter and tighter partial conditions that in the infinite limit correspond exactly to the problem one is trying to solve.
In Section~\ref{sec:DPS} we will see how this can be done for the separability problem.

It is also useful to consider the dual of the SDP in Eq.~\eqref{entanglementprimal}, namely\footnote{Note that a direct application of the definition of dual would give $W^{T_A}$ as the optimization variable instead of $W$, so we changed it for convenience. Throughout the review we will do such trivial simplifications without comment.}
\begin{equation}\label{entanglementwitness}
	\begin{aligned}
		\min_W \quad & \tr\left(W\rho_{AB}\right) \\
		\st \quad & \tr(W) = 1, \\
		& W^{T_A} \succeq 0\,.
	\end{aligned}
\end{equation}
Consider any feasible point $W$ of the dual SDP, it then follows from weak duality of SDPs that for any state $\rho_{AB}$, we have that $\tr(W \rho_{AB})$ is an upper bound on the random robustness of entanglement.
In particular, as $W$ is Hermitian and hence an observable, if we measure $W$ and find that $\tr(W \rho_{AB}) < 0$, this implies that the state $\rho_{AB}$ is entangled.
Thus the dual SDP provides us with an operational procedure to detect and quantify entanglement~\cite{Brandao2005}.
The operator $W$ is known as an entanglement witness \cite{Terhal2000, Lewenstein2000}.
In particular, one does not need to perform full tomography of the quantum state, which is often impractical as the number of required measurements grows rapidly with the dimension of the state.
In order to measure the entanglement witness, one would need to decompose it in the form $W=\sum_{a,b,x,y}c_{abxy} A_{a|x}\otimes B_{b|y}$ for some real coefficients $c_{abxy}$ and some POVMs for Alice and Bob.
Such a decomposition requires in general many fewer measurements than tomography, so a witness allows to detect entanglement from partial knowledge of the quantum state.

It is important to emphasise that entanglement witnesses do not come only from the partial transposition criterion.
In principle, for any entangled state $\rho_{AB}$ one can construct a witness $W$ such that $\tr(W\rho_{AB}) < 0$, but that for any separable state $\sigma_{AB}$ it holds that $\tr(W \sigma_{AB}) \ge 0$.
The construction of the witness operator is, however, not straightforward.
Witness methods can sometimes detect entangled states even using just two local measurement bases, see e.g.~\citet{Toth2005, Bavaresco2018}.
A common approach is to construct entanglement witnesses through the estimation of the fidelity between the state prepared in the laboratory and a pure target state \cite{Bourennane2004}.
While this method is practical for particular types of entanglement, see e.g.~\citet{Leibfried2005, Haffner2005, Lu2007, Wang2016b}, it fails to detect the entanglement of most states \cite{Weilenmann2020}.
Independently of using the density matrix or partial knowledge of it, determining whether a state is separable or entangled is difficult.

\textbf{Steering.---} By performing measurements on her share of a suitable entangled state and keeping track of the outcome $a$, Alice can remotely prepare any ensemble of states for Bob \cite{Hughston1993, Gisin1984}.
The discussion of how entanglement allows one system to influence (or steer) another system traces back to Schr\"odinger's remarks \cite{Schrodinger1935} on the historical debate about ``spooky action at a distance'' \cite{Einstein1935}.
Consider again the situation in Fig.~\ref{FigBellScenario} but this time we ask whether Bob can know that Alice is quantumly steering his system.
The set of states remotely prepared by Alice for Bob, when her outcome is made publicly known, along with the probabilities of her outcomes, is described by a set of subnormalised states of Bob, $\varrho_{a|x}=\tr_A\left((A_{a|x}\otimes \id) \rho\right)$.
This set is known as an assemblage \cite{Pusey2013}.
The assemblage can be modelled without a quantum influence from Alice to Bob if there exists a local hidden state decomposition \cite{Wiseman2007}, namely
\begin{equation}
	\varrho_{a|x}=\sum_\lambda p(\lambda) p(a|x,\lambda) \sigma_\lambda,
	\label{eq:assemblage}
\end{equation}
for some probabilities $p(\lambda)$ and $p(a|x,\lambda)$, and quantum states $\sigma_\lambda$.
One can interpret this as a source probabilistically generating the pair $(\lambda,\sigma_\lambda)$, sending the former to Alice, who then classically decides her output, and delivering the latter to Bob.
If no model of the form of Eq.~\eqref{eq:assemblage} is possible, then we say that the assemblage demonstrates steering and that consequently $\rho_{AB}$ is steerable.
For in-depth reviews on steering, we refer to \citet{UolaReview2020, CavalcantiReview2016}.

Deciding the steerability of an assemblage is an SDP.
To see this, note that for a given number of inputs and outputs for Alice, there are finitely many functions $r$ mapping $x$ to $a$.
Indexing them by $\lambda$, we can define deterministic distributions $D(a|x,\lambda)=\delta_{r_\lambda(x),a}$.
A strictly feasible formulation of the SDP is
\begin{equation}\label{lhsprimal}
	\begin{aligned}
		\max_{\{\tilde{\sigma}_\lambda\},t} \quad & t\\
		\st \quad & \varrho_{a|x}=\sum_\lambda \tilde{\sigma}_\lambda D(a|x,\lambda) \qquad \forall\,a,\,x,\\
		& \tilde{\sigma}_\lambda \succeq t\id \hspace{2.94cm} \forall\,\lambda.
	\end{aligned}
\end{equation}
A local hidden state model is possible if and only if the optimal value of Eq.~\eqref{lhsprimal} is nonnegative.
Notice that normalisation of the assemblage is implicitly imposed by the equality constraint.

Moreover, it is interesting to consider the SDP dual to Eq.~\eqref{lhsprimal},
\begin{equation}\label{lhsdual}
	\begin{aligned}
		\min_{\{W_{a,x}\}} \quad & \sum_{a,x}\tr\left(W_{a,x}\varrho_{a|x}\right)\\
		\st \quad & \sum_{a,x,\lambda}\tr\left(W_{a,x}\right)D(a|x,\lambda) =1, \\
		& \sum_{a,x}W_{a,x}D(a|x,\lambda)\succeq 0 \qquad \forall\,\lambda.
	\end{aligned}
\end{equation}
The first constraint is a normalisation for the dual variables $\{W_{a,x}\}$ and the second constraint ensures that all local hidden state models return nonnegative values.
Thus, if the assemblage demonstrates steering, the dual gives us an inequality,
\begin{equation}\label{steeringwitness}
	\sum_{a,x}\tr\left(W_{a,x}\varrho_{a|x}\right)\geq 0,
\end{equation}
which is satisfied by all local hidden state models and violated in particular by the assemblage $\{\varrho_{a|x}\}$ but also by some other assemblages.
Indeed, the inequality \eqref{steeringwitness} can be viewed as the steering equivalent of an entanglement witness (recall Eq.~\eqref{entanglementwitness}), i.e.,~a steering witness.
However, it is important to note that steering is a stronger notion than entanglement because it is established only from inspecting the assemblage, i.e.,~Bob's measurements are assumed to be characterised whereas Alice's measurements need not even follow Born's rule.
Note that as in the case of entanglement witnesses one does not need to perform full tomography of Bob's states to test steering witnesses.

\textbf{Nonlocality.---} Bell's theorem \cite{Bell1964} proclaims that there exist quantum correlations \eqref{correlations} that cannot be modelled in any theory respecting local causality\footnote{A discussion of the historical debate on the interpretation of Bell's theorem can be found in \cite{Laudisa2022}.} \cite{Bell1975}.
Such a theory, known as a local (hidden variable, LHV) model, assigns the outcomes of Alice and Bob based on their respective inputs and some shared classical information $\lambda$.
A local model for their correlation takes the form
\begin{equation}\label{LHV}
	p(a,b|x,y)=\sum_\lambda p(\lambda) p(a|x,\lambda)p(b|y,\lambda).
\end{equation}
Correlations admitting such a decomposition are called local whereas those that do not are called nonlocal.
For an in-depth discussion of nonlocality, we refer to \citet{BrunnerCavalcantiReview2014, ScaraniBook}.

The response functions $p(a|x,\lambda)$ and $p(b|y,\lambda)$ can be written as probabilistic combinations of deterministic distributions but, in analogy with the case of the local hidden state model, any randomness can be absorbed into $p(\lambda)$.
Thus, without loss of generality, we can focus on deterministic response functions and their convex combinations enabled by the shared common cause.
Geometrically, the set of local correlations forms a convex polytope \cite{Fine1982}.
Deciding whether a given distribution $p(a,b|x,y)$ is local can therefore be cast as an LP,
\begin{equation}\label{LHVLP}
	\begin{aligned}
		\max_{\{p(\lambda)\},t} \quad & t\\
		\st \quad & \sum_{\lambda} p(\lambda) D(a|x,\lambda)D(b|y,\lambda)=p(a,b|x,y),\\
		& p(\lambda)\geq t \qquad \forall\,\lambda,
	\end{aligned}
\end{equation}
where the cardinality of $\lambda$ is the total number of deterministic distributions.
The correlations are local if and only if the optimal value of Eq.~\eqref{LHVLP} is nonnegative.
As before, it is also interesting to consider the dual LP,
\begin{equation}\label{LHVdual}
	\begin{aligned}
		\min_{\{c_{abxy}\}} \quad & \sum_{a,b,x,y} c_{abxy}p(a,b|x,y)\\
		\st \quad & \sum_{\lambda,a,b,x,y} c_{abxy}D(a|x,\lambda)D(b|y,\lambda)=1, \\
		& \sum_{a,b,x,y}c_{abxy} D(a|x,\lambda)D(b|y,\lambda)\geq 0 \qquad \forall\, \lambda.
	\end{aligned}
\end{equation}
This is clearly reminiscent of the steering dual in Eq.~\eqref{lhsdual}.
The first constraint normalises the coefficients $\{c_{abxy}\}$ and the second constraint ensures that if $p(a,b|x,y)$ is local then the value of the dual is nonnegative.
Hence it implies the inequality
\begin{equation}\label{BellInequality}
	\sum_{a,b,x,y}c_{abxy}p(a,b|x,y)\geq 0,
\end{equation}
which is satisfied by all local distributions and violated by some nonlocal distributions, in particular the target distribution $p(a,b|x,y)$ whenever it is nonlocal.
This can be seen as the nonlocality equivalent of an entanglement and steering witness, but inequalities like Eq.~\eqref{BellInequality} are more well known under the name Bell inequalities.
The violation of a Bell inequality in quantum theory is the strongest sense of entanglement certification, as it requires no assumptions on the measurements of Alice or Bob.

The most famous and widely used Bell inequality is the Clauser-Horne-Shimony-Holt (CHSH) inequality \cite{Clauser1969}.
It applies to the simplest scenario in which nonlocality is possible, namely when Alice and Bob have two inputs each ($x,y\in\{0,1\}$) and two possible outcomes each ($a,b\in\{0,1\}$).
The CHSH inequality reads\footnote{Note, in contrast to Eq.~\eqref{BellInequality}, that the right-hand side is nonzero and the inequality sign is flipped. This is only a matter of convention. The notation of Eq.~\eqref{CHSHInequality} can be obtained from Eq.~\eqref{BellInequality} by setting $c_{abxy}=\frac12-(-1)^{a+b+xy}$.}
\begin{equation}
	\label{CHSHInequality}
	S_\text{CHSH}\equiv\sum_{a,b,x,y} (-1)^{a+b+xy}p(a,b|x,y)\leq 2.
\end{equation}
This is equivalent to the historical formulation of this inequality, in terms of expectation values of observables, $\expect{A_0B_0}+\expect{A_0B_1}+\expect{A_1B_0}-\expect{A_1B_1}\leq 2$.
For this reformulation, one simply writes $p(a,b|x,y)=(1+(-1)^a\expect{A_x}+(-1)^b\expect{B_y}+(-1)^{a+b}\expect{A_xB_y})/4$ by inverting the definition of expectation value.
A quantum model based on a singlet state and particular pairs of anticommuting qubit measurements can achieve the violation $S_\text{CHSH}=2\sqrt{2}$.
This is the maximum violation achievable with quantum systems \cite{Tsirelson1980}.

More generally, one can employ a simple optimisation heuristic known as seesaw \cite{Werner2001, Liang2007, Pal2010} to search for the largest quantum violation of any given Bell inequality.
Formally, this is obtained by replacing $p(a,b|x,y)$ by Born's rule in the left-hand side of Eq.~\eqref{BellInequality} and optimizing over the state and measurements.
The main observation is that for a fixed state and fixed measurements on (say) Bob's side, the optimal value of the resulting object is a linear function of Alice's measurements and thus can be evaluated as the SDP\footnote{When the outcomes are binary, this is just an eigenvalue problem and hence does not require an SDP formulation.} \cite{Audenaert2002}\begin{equation}\label{seesawbell1}
	\begin{aligned}
		\max_{\{A_{a|x}\}} \quad &\sum_{a,b,x,y}c_{abxy}\tr\left(A_{a|x}\otimes B_{b|y} \rho_{AB}\right)\\
		\st \quad & \sum_{a} A_{a|x}=\id \qquad \forall\,x,\\
		&A_{a|x}\succeq 0 \hspace{1.35cm} \forall\,a,\,x.
	\end{aligned}
\end{equation}
Given the optimised POVMs of Alice, an analogous SDP then evaluates the optimal value of the Bell parameter over Bob's POVMs.
Then, given the optimised POVMs of Alice and Bob, the optimal state can be obtained as an eigenvector with maximal eigenvalue of the operator\footnote{As any maximal eigenvalue problem, this can also be formulated as an SDP: computing the maximum of $\tr(\rho\mathcal S)$ such that $\rho \succeq 0$ and $\tr(\rho) = 1$.}
\begin{equation}\label{seesawbell2}
	\mathcal S = \sum_{a,b,x,y}c_{abxy} A_{a|x}\otimes B_{b|y}.
\end{equation}
One then starts with a random choice for the state and measurements, and iterates the three optimizations until the value of the Bell parameter converges.
From any starting point it will converge monotonically to a local optimum, but one cannot guarantee it will reach the global optimum, i.e., the largest quantum value.
Nevertheless, this heuristic is very useful in practice, and when repeated with several different starting points it often does find the optimal quantum model.

Notably, the routine can be reduced to only two optimisations per iteration.
This is achieved by considering the measurements of just one party and the ensemble of sub-normalised states remotely prepared by the other party (i.e., the assemblage).
Optimisation over the latter can also be cast as the SDP 
\begin{equation}\label{seesawbell3}
	\begin{aligned}
		\max_{\{\rho_{a|x}\}} \quad &\sum_{a,b,x,y}c_{abxy}\tr\left(\rho_{a|x}B_{b|y}\right)\\
		\st \quad & \sum_{a} \rho_{a|x}=\sum_{a} \rho_{a|x'} \qquad \forall\,x,\,x',\\
		& \sum_a \tr\left(\rho_{a|x}\right)=1 \qquad\quad\,\, \forall\,x,\\
		& \rho_{a|x}\succeq 0 \qquad \qquad\qquad\quad \forall\,a,\,x.
	\end{aligned}
\end{equation}
This result seems to be part of the folklore; the earliest mention we are aware of is in Appendix C of \citet{Quintino2014}.
The crucial point is that every assemblage has a quantum realization.
This is a well-known consequence of the Schrödinger-GHJW theorem \cite{Schrodinger1935, Hughston1993, Gisin1984}, as discussed for example in \cite{Sainz2015}.

One is often interested in the maximal value of a Bell inequality when the correlations are not constrained to come from LHVs or quantum theory, but only to obey the principle of no-signaling.
No-signaling is the assumption that the one party's outcome cannot depend on the input of the other, which can be physically justified e.g.~through space-like separation of the parties.
This is formalised as
\begin{equation}\label{eq:ns-conditions}
	\begin{aligned}
		&\sum_{b} p(a,b|x,y)= \sum_{b'} p(a,b'|x,y') \quad \forall\,a,\,x,\,y,\,y',\\
		&\sum_{a} p(a,b|x,y)= \sum_{a'} p(a',b|x',y) \quad \forall\,b,\,x,\,x',\,y.
	\end{aligned}
\end{equation}
Since these conditions are linear, the set of all distributions satisfying them (known as no-signaling correlations) can be characterised by LP.
The maximal value one obtains is in general larger than with LHVs or quantum theory.
For instance, no-signaling correlations can achieve the higher-than-quantum CHSH violation of $S_\text{CHSH}=4$ \cite{Khalfin1985, Popescu1994}.

\subsubsection{Communication-based scenarios}
\label{seccomscenario}
\begin{figure}[t!]
	\centering
	\includegraphics[width=0.95\columnwidth]{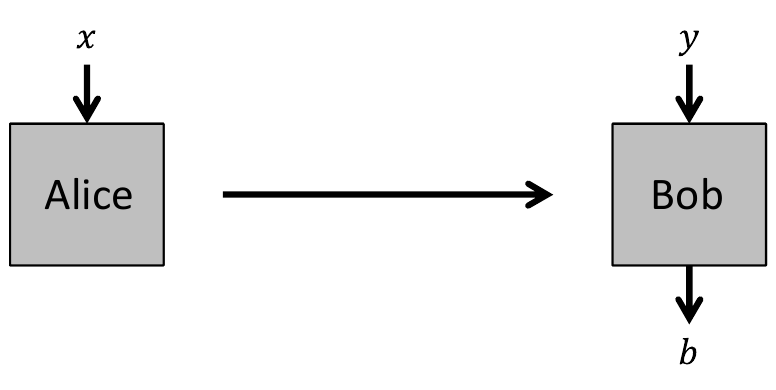}
	\caption{The standard communication scenario.
A sender (Alice) and a receiver (Bob) hold inputs $x$ and $y$, respectively.
Alice chooses a message that is sent to and measured by Bob, producing an outcome $b$.}\label{FigCommunicationScenario}
\end{figure}

An important family of scenarios is those in which physical systems are not shared, but communicated from some parties to others.
The simplest communication scenario is known as the prepare-and-measure scenario and it is illustrated in Fig.~\ref{FigCommunicationScenario}.
A sender, Alice, privately selects an input $x$ and encodes it into a message that is sent over a communication channel to a receiver, Bob, who privately selects an input $y$ and performs an associated decoding to receive an outcome $b$.
In a quantum model, the message is described by a quantum state, i.e.,~$\rho_x$, and the measurements by POVMs $\{M_{b|y}\}$.
Hence, the scenario amounts to preparing a number of different quantum states, labeled by $x$, and then measuring them with a number of different measurement settings, labeled $y$.
The quantum correlations established are given by Born's rule,
\begin{equation}\label{BornCom}
	p(b|x,y)=\tr\left(\rho_x M_{b|y}\right).
\end{equation}

In contrast, a classical model describes the messages as distinguishable, and can without loss of generality be assigned integer values, but potentially also mixed via classical randomness.
Adopting the notations of quantum models, such classical messages are written $\rho_x=\sum_{m} p(m|x)\ketbra{m}{m}$ for some conditional message distribution $p(m|x)$.
Since classical models admit no superpositions, all classical measurements are restricted to the same basis, namely $M_{b|y}=\sum_{m}p(b|y,m) \ketbra{m}{m}$.
Moreover, it is common to also consider a shared classical cause, $\lambda$, between Alice and Bob.
Following Born's rule, classical correlations then take the form
\begin{equation}\label{classicalPM}
	p(b|x,y)=\sum_{m,\lambda} p(\lambda)p(m|x,\lambda)p(b|y,m,\lambda).
\end{equation}
Any correlation that does not admit such a model is called nonclassical.
In order for the correlations to be interesting, a restricting assumption must be introduced.
Otherwise Alice can always send $x$ to Bob, who can then output $b$ according to any desired $p(b|x,y)$.
Typically, the restriction is put on the channel connecting the parties.
For this purpose, various approaches have been proposed, all closely linked to SDP techniques.
We discuss them in Section~\ref{sec:communication-distinguish}.

Here, we exemplify the most well-studied case, namely when the Hilbert space dimension of the message is assumed, or equivalently for classical models, when the cardinality of the message alphabet is known.
For a classical model with a message alphabet of size $d$, the set of correlations in Eq.~\eqref{classicalPM} can be described by an LP \cite{Gallego2010}.
In analogy with discussions in the previous section, any randomness in the encoding function $p(m|x,\lambda)$ and the decoding function $p(b|y,m,\lambda)$ can be absorbed into $p(\lambda)$.
Since $d$ is fixed, there are only finitely many different encoding and decoding functions and they can be enumerated by $\lambda$.
The LP for deciding whether a given $p(b|x,y)$ admits a classical model based on a $d$-dimensional message is
\begin{equation}\label{PMLP}
	\begin{aligned}
		\max_{\{p(\lambda)\},t} \quad & t\\
		\st \quad & \sum_\lambda p(\lambda)\sum_{m=1}^dD(m|x,\lambda)D(b|y,m,\lambda)=p(b|x,y),\\
		& p(\lambda)\geq t \qquad \forall\,\lambda,
	\end{aligned}
\end{equation}
where the normalisation of $p(\lambda)$ is implicit in the equality constraint.
For reasons analogous to the discussion of local models, the dual of this LP, when $p(b|x,y)$ is nonclassical, provides a hyperplane in the space of correlations which separates it from the classical polytope, i.e.,~an inequality of the form
\begin{equation}\label{PMineq}
	S\equiv \sum_{b,x,y}c_{bxy}p(b|x,y)\geq  0,
\end{equation}
for some coefficients $\{c_{bxy}\}$, satisfied by all classical models based on $d$-dimensional messages but violated by the target nonclassical distribution.

Interestingly, it is known that correlations obtained from $d$-dimensional quantum systems can violate the limitations of $d$-dimensional classical messages.
The earliest example, based on comparing a bit message against a qubit message, appeared in \citet{Wiesner1983} and was later re-discovered in \citet{Ambainis2002}.
This is known as a quantum random access code.
To see that it is possible, consider that Alice holds two bits, $x=x_0x_1\in\{0,1\}^2$ and Bob holds one bit, $y\in\{0,1\}$, and that Bob is asked to output the value of Alice's $y$th bit.
However, Alice can send only one (qu)bit to Bob.
Classically, one can convince oneself that, on average, the success probability can be no larger than $p_\text{suc}\equiv \frac{1}{8}\sum_{x_0,x_1,y}p(b=x_y|x_0,x_1,y)\leq 3/4$.
This is achieved by Alice sending $x_0$ and Bob outputting $b=x_0$ irrespective of $y$.
In contrast, a quantum model can achieve $p_\text{suc}=\frac{1}{2}(1+\frac{1}{\sqrt{2}})$ by having Bob measure the Pauli observables $\sigma_X$ and $\sigma_Z$ while Alice communicates the qubit states with Bloch vectors\footnote{The Bloch vector of a qubit state $\rho$ uniquely determines the state, and is given by $\left(\tr(\rho \sigma_X),\tr(\rho \sigma_Y),\tr(\rho \sigma_Z)\right)$.} $((-1)^{x_0},0,(-1)^{x_1})/\sqrt{2}$.
Such quantum communication advantages are also known to exist for any value of $d$ \cite{Brunner2013q, Tavakoli2015}.

If we are given an inequality of the form of Eq.~\eqref{PMineq} and asked to violate it in quantum theory, we can use a seesaw heuristic to numerically search for the optimal value of $S$, in analogy with the case of Bell inequalities \cite{Tavakoli2017}.
In the prepare-and-measure scenario, the optimisation of $S$ becomes an SDP when the states $\rho_x$ are fixed, and a simple set of eigenvalue problems when the measurements $M_{b|y}$ are fixed.
Specifically, for fixed states the problem becomes\footnote{When the outcomes are binary, this too is an eigenvalue problem and hence does not require an SDP formulation.}
\begin{equation}\label{seesaw1}
	\begin{aligned}
		\max_{\{M_{b|y}\}} \quad &\sum_{b,x,y}c_{bxy}\tr\left(\rho_xM_{b|y}\right)\\
		\st \quad & \sum_{b} M_{b|y}=\id \qquad \forall\,y,\\
		& M_{b|y}\succeq 0,
	\end{aligned}
\end{equation}
and for fixed measurements it reduces to computing the eigenvectors with maximal eigenvalue of the operators
\begin{equation}\label{seesaw2}
	\mathcal S_x = \sum_{b,y}c_{bxy} M_{b|y}
\end{equation}
for each $x$.
Thus, by starting from a randomised initial set of states, one can run the SDP in Eq.~\eqref{seesaw1} and use the returned measurements to compute the optimal states from Eq.~\eqref{seesaw2}.
The process is iterated until the value converges.

\begin{table*}[t]
	\begin{tabular}{c||c|c|c|c}
		Reference          & Scenario            & Convergence   & Selected application   & Section \\ \hline\hline
		\citet{Navascues2008}    & Bell nonlocality        & \phantom{$^\ast$}\vmark$^\ast$  & Bell correlations                          & \ref{sec:NPA}    \\\hline
		\citet{Moroder2013}     & Bell nonlocality        & \vmark      & Entanglement quantification                     & \ref{sec:nonlocality:di:moroder}    \\\hline
		\citet{Doherty2004}     & Entanglement          & \vmark      & Entanglement detection                       & \ref{sec:DPS}    \\\hline
		\citet{Navascues2015b}   & Bell nonlocality        & \vmark      & Dimension restrictions   & \ref{sec:entdim}    \\\hline
		\citet{Navascues2015b}   & Prepare-and-measure      & \textbf{?}    & Dimension restrictions   & \ref{sec:NV}    \\\hline
		\citet{Tavakoli2022a}    & Prepare-and-measure      & \textbf{?}    & Information restrictions                      & \ref{sec:communication-distinguish}   \\\hline
		\citet{Tavakoli2021c}    & \begin{tabular}[c]{@{}c@{}}Entanglement-assisted\\ prepare-and-measure\end{tabular} & \vmark    & Dimension restrictions      & \ref{sec:EAPM}    \\\hline
		\citet{Brown2021b}     & Bell nonlocality        & \textbf{?}  & Quantum key distribution                      & \ref{sec:qkd_di_vn}   \\\hline
		\citet{Pozas2019}      & Network nonlocality      & \phantom{$^\ast$}\vmark$^*$    & Network Bell tests                         & \ref{sec:networks:swapping}    \\\hline
		\citet{Wolfe2019}\footnote{Note that this is a hierarchy of LPs, that characterises classical correlations.}
		& Network nonlocality      & \vmark      & \begin{tabular}[c]{@{}c@{}} Causal inference \\ \& network Bell tests \end{tabular}   & \ref{sec:networks:inflation}    \\\hline
		\citet{Wolfe2021}      & Network nonlocality      & \phantom{$^\ast$}\vmark$^*$    & \begin{tabular}[c]{@{}c@{}} Causal inference \\ \& network Bell tests \end{tabular}   & \ref{sec:networks:inflation}    \\\hline
		\citet{Ligthart2021}    & Network nonlocality      & \vmark      & \begin{tabular}[c]{@{}c@{}} Causal inference \\ \& network Bell tests \end{tabular}   & \ref{sec:networks:inflation}    \\\hline
		\citet{Tavakoli2021a}    & Prepare-and-measure      & \textbf{?}    & Operational contextuality                      & \ref{sec:contextuality}   \\ \hline
		\citet{Chaturvedi2021}   & Prepare-and-measure      & \textbf{?}    & Operational contextuality                      & \ref{sec:contextuality}   \\ \hline
	\end{tabular}
	\caption{Overview of some of the SDP relaxation hierarchies for quantum correlations.
	The table presents the main reference, the considered correlation scenario, whether convergence to the quantum set is known to hold (\vmark) or it is unknown whether it holds (\textbf{?}), the main domain of application and the relevant section of this article.
	The symbol \vmark$^\ast$ means that convergence is achieved to the set of quantum correlations under the additional relaxation that the tensor product structure of the Hilbert space is replaced with a commutation structure.
	}
	\label{TableOverview}
\end{table*}

\subsection{Overview of semidefinite relaxation hierarchies}
In the previous section we have seen how some classical correlation sets can be characterised via LPs and how SDPs facilitate some quantum correlation problems.
However, the characterisation of the set of quantum correlations in most scenarios cannot be achieved with a single SDP, but rather requires SDP relaxation hierarchies.
These relaxation hierarchies and their applications are a major focus of the upcoming sections.
Here, we provide in Table~\ref{TableOverview} an overview of SDP relaxation hierarchies encountered in the study of quantum correlations, the scenario to which they apply, their convergence properties, their main domain of application and the section in this article where they are further discussed.
The overview is not comprehensive, as there are also other correlation scenarios where such techniques apply and some of them are discussed in Section~\ref{sec:further}.
Furthermore, the hierarchies are not unique; there can be several different SDP hierarchies addressing the same problem, as is the case for instance in the two final rows of the table.

Whether an SDP hierarchy converges to the targeted set of correlations is an interesting question, but it can come with noteworthy subtleties.
For instance, as we will see later, in Bell nonlocality the tensor product structure of the Hilbert space is relaxed to a single-system commutation condition.
Several hierarchies converge to this latter characterisation, which is known to be a strict relaxation of the bipartite tensor-product structure when considering infinite-dimensional systems \cite{Yuen2020}.
Importantly, even if a hierarchy converges to the quantum set, but also when it does not, what is often of practical interest is how fast useful correlation bounds can be obtained, since it is commonly the case that one cannot evaluate more than a few levels of relaxation.


\section{Semidefinite relaxations for polynomial optimisation}
\label{sec:sdp}
In this section we review the mathematical preliminaries for some of the SDP relaxation methods used in the subsequent sections of this review.
A crucial fact about SDPs is that they can be used to approximate solutions to optimisation problems that themselves are not SDPs.
That is, some optimisation problems can be relaxed into a sequence, or hierarchy, of increasingly complex SDPs, each providing a more accurate bound on the solution than the previous.

One particular example of this is polynomial optimisation, which can be relaxed to a sequence of SDPs via the so-called \emph{moment approach}, or its dual, known as \emph{sum-of-squares programming}.
Considering the various semidefinite programming relaxations discussed in this review, many of them fall into this framework of semidefinite relaxations for polynomial optimisation.
In such cases the original problem can be either viewed as (or closely approximated by) some polynomial optimisation problem which can then be transformed into an SDP hierarchy by the aforementioned methods.
In fact, polynomial optimisation is at the core of many of the results discussed in all of the remaining sections.
In light of this we will now dedicate some time to give an overview of the SDP relaxations of such optimisation problems.

\subsection{Commutative polynomial optimisation}
\label{sec:lasserre}
Consider the following optimisation problem
\begin{equation}\label{eq:pop}
	\begin{aligned}
		\max_{\{x_j\}} \quad & f(x_1, \dots, x_n) \\
		\st \quad & g_i(x_1, \dots, x_n) \geq 0 \qquad \forall\,i,
	\end{aligned}
\end{equation}
where $f$ and $g_i$ are all polynomials in the variables $x_1,\dots, x_n \in \mathbb{R}$.
This type of problem is known as a (commutative) \emph{polynomial optimisation} problem.
Apart from the applications discussed in this review, this family of optimisation problems has found applications in control theory~\cite{Henrion2005}, probability theory~\cite{Bertsimas2005} and machine learning~\cite{Hopkins2016}.
However, polynomial optimisation is known to be NP hard~\cite{Nesterov2000}.
The moment and sum of square hierarchies, first proposed in \citet{Lasserre2001} and \citet{Parrilo2000}, offer a recipe to formulate a sequence of SDPs that, under mild conditions, will converge to the optimal value of Eq.~\eqref{eq:pop}.
We will now describe both hierarchies at a high level and refer interested readers to the survey article of \citet{Laurent2009} for a more precise treatment.

\subsubsection{Moment matrix approach}
The moment matrix approach, commonly known as the Lasserre hierarchy, relaxes Eq.~\eqref{eq:pop} into a sequence of SDPs.
In the following we will describe how these relaxations can be constructed and towards this goal we must first introduce some notation.
A \emph{monomial} is any product of the variables $\{x_j\}_j$ and the \emph{length} of a monomial denotes the number of terms in the product, e.g., $x_1x_3^2$ has length $3$ and $x_4x_5x_6^3$ has length $5$.
We define the constant $1$ to have length $0$.
For $k \in \mathbb{N}$, let $\mathcal{S}_k$ denote the set of monomials with length no larger than $k$.
For a feasible point $x = (x_1, \dots, x_n)$ of the problem \eqref{eq:pop} let us define its moment matrix of level $k$, $G^k$, to be a matrix indexed by monomials in $\mathcal{S}_k$ whose element in position $(u,v)$ is given by
\begin{equation}
	G^k(u,v) = u(x) v(x),
\end{equation}
where $u,v \in \mathcal{S}_k$\,.
One crucial feature of moment matrices is that they are necessarily positive semidefinite, as 
\begin{equation}\label{PSDcon}
	G^k = \left(\sum_u u(x) \ket{u}\right)\left(\sum_{u'} u'(x) \ket{u'}\right)^\dagger.
\end{equation}
Furthermore, the value of any polynomial of degree no larger than $2 k$ can be evaluated at the point $x$ by an appropriate linear combination of the elements of $G^k$.
Thus, by taking $k$ large enough, the value of $f(x_1,\dots,x_n)$ can be reconstructed from the moment matrix.

In addition to $G^k$, for each $g_i$ appearing in the constraints of Eq.~\eqref{eq:pop} we introduce a \emph{localising moment matrix} of level \mbox{$k_i\in \mathbb{N}$}, denoted $G^{k_i}_{g_{i}}$, which will act as a relaxation of the constraint $g_{i}(x) \geq 0$.
This new matrix is indexed by elements of $\mathcal{S}_{k_i}$ and the element at index $(u,v)$ is given by
\begin{equation}
	G^{k_i}_{g_i}(u,v) = u(x) v(x) g_i(x)\,.
\end{equation}
One natural choice of $k_i$ is $\lfloor k - \deg(g_i)/2\rfloor$, since this ensures that the polynomial $u(x) v(x) g_i(x)$ is of a degree small enough to be expressed as a linear combination of the elements of the original moment matrix $G^k$.
We will assume in the remainder of this section that $k_i$ is chosen this way but we refer the reader to the remark in Section~\ref{rem:monomial-sets} for further discussion on choices of indexing sets.
Finally, for any feasible point $(x_1,\dots,x_n)$ of Eq.~\eqref{eq:pop} one again necessarily has that $G^{k_i}_{g_i} \succeq 0$.

The core idea of the Lasserre hierarchy is that, instead of directly optimising Eq.~\eqref{eq:pop}, for each level $k$ one can optimise over all PSD matrices that satisfy the same constraints as the level-$k$ moment matrices of a feasible point of Eq.~\eqref{eq:pop}.
When taking $k$ large enough so that all the polynomials in the problem can be expressed as linear combinations of the moment matrix elements, e.g., $f(x) = \sum_{u,v \in \mathcal{S}_{k}} c_{uv} u(x)v(x)$ for some coefficients $c_{uv} \in \mathbb{R}$, then one arrives at the semidefinite program
\begin{equation}\label{eq:lasserre_primal}
	\begin{aligned}
		\max \quad & \sum_{u,v \in \mathcal{S}_k} c_{uv} G^k(u, v) \\
		\st \quad & G^k \succeq 0, \\
		& G^{k_i}_{g_i} \succeq 0 \qquad \forall\,i,
	\end{aligned}
\end{equation}
where there are many implicit equality constraints relating the elements of $G_{g_i}^{k_i}$ matrices to linear combinations of the elements of $G^k$.
Additionally there are other constraints based on the construction of the moment matrices, e.g., $G^k(uw,v) = G^k(u, wv)$ for all monomials $u$, $v$, $w$ such that $uw,wv\in S_{k}$ as well as the normalisation constraint $G^k(1,1) = 1$.
Note that, as every feasible point of Eq.~\eqref{eq:pop} defines a feasible point of Eq.~\eqref{eq:lasserre_primal}, this new optimisation problem is a relaxation and its optimal value constitutes an upper bound on the optimal value of Eq.~\eqref{eq:pop}.
Furthermore, \citet{Lasserre2001} proved that under certain conditions the sequence of optimal values of Eq.~\eqref{eq:lasserre_primal} indexed by the relaxation level $k$ will converge to the optimal value of Eq.~\eqref{eq:pop}.
Note however that the size of the SDPs grows rapidly\footnote{The moment matrix of level $k$ is of size $|\mathcal{S}_k|\times |\mathcal{S}_k|$ with $|\mathcal{S}_k| = \frac{(k+n-1)!}{(n-1)!\, k!}$.} with $k$.
Nevertheless, in many practical problems of interest it has been observed that small relaxation levels can give accurate, and sometimes tight, bounds.

To better illustrate this method let us demonstrate its use on the following problem
\begin{equation}\label{eq:lasserre_ex}
	\begin{aligned}
		\max \quad & x_2^2 - x_1\, x_2 - x_2\\
		\st \quad & x_1 - x_1^2 \geq 0, \\
		& x_2 - x_2^2 \geq 0.
	\end{aligned}
\end{equation}
The monomial set for level $k=1$ is $\mathcal{S}_1 = \{1, x_1, x_2\}$.
The corresponding relaxation of Eq.~\eqref{eq:lasserre_ex} at this level is the SDP\footnote{Throughout this review, when representing Hermitian matrices we omit the elements below the diagonal since they are determined by the elements above the diagonal.}
\begin{equation}\label{eq:pop_ex_l1}
	\begin{aligned}
		\max \quad & y_{22} - y_{12} - y_{02} \\
		\mathrm{s.t.} \quad & \begin{pmatrix}
			1 & y_{01} & y_{02} \\
			& y_{11} & y_{12} \\
			& 		& y_{22}
		\end{pmatrix} \succeq 0, \\
		& y_{01} - y_{11} \geq 0, \\
		& y_{02} - y_{22} \geq 0,
	\end{aligned}
\end{equation}
which has an optimal value of $1/8$.
The monomial set for the second level, $k=2$, is $\mathcal{S}_2 = \{1, x_1,x_2,x_1^2, x_1x_2, x_2^2\}$, and the corresponding relaxation is
\begin{equation}\label{eq:pop_ex_l2}
	\begin{aligned}
		\max \quad & y_{05} - y_{04} - y_{02} \\
		\mathrm{s.t.} \quad & \begin{pmatrix}
			1 & y_{01} & y_{02} & y_{03} & y_{04} & y_{05} \\
			& y_{03} & y_{04} & y_{13} & y_{14} & y_{15} \\
			& & y_{05} & y_{14} & y_{15} & y_{25} \\
			& & & y_{33} & y_{34} & y_{35} \\
			& & & & y_{35} & y_{45} \\
			& & & & & y_{55}
		\end{pmatrix} \succeq 0, \\
		& \begin{pmatrix}
			y_{01} - y_{03} & y_{03} - y_{13} & y_{04} - y_{14} \\
			& y_{13} - y_{33}& y_{14} - y_{34} \\
			& & y_{15} - y_{35}
		\end{pmatrix} \succeq 0, \\
		& \begin{pmatrix}
			y_{02} - y_{05} & y_{04} - y_{15} & y_{05} - y_{25} \\
			& y_{14} - y_{35} & y_{15} - y_{45} \\
			& & y_{25} - y_{55}
		\end{pmatrix} \succeq 0\,.
	\end{aligned}
\end{equation}
At this level one can now see how the entries of the localising moment matrices are linear combinations of the elements of the original moment matrix.
If we solve the above example numerically we find that it gives an objective value of $0.000021$.
In particular, as we increase the relaxation level the objective values converge towards the optimal value of the original problem, which is $0$ and is achieved when $x_1 = 0$ and $x_2=1$.

\subsubsection{Sum of squares approach}

The dual problems to the moment matrix relaxations also have an interesting interpretation in terms of optimising over sum-of-squares (SOS) polynomials~\cite{Parrilo2000} (see also the survey of \citet{Laurent2009} for a discussion on the duality of the two approaches).
A polynomial $p(x)$ is an SOS polynomial if it can be written as $p(x) = \sum_i r_i(x)^2$ for some polynomials $r_i(x)$.
Note that an SOS polynomial is necessarily nonnegative, i.e., $p(x) \geq 0\,\forall\,x$.
We can therefore upper bound our original problem, given in Eq.~\eqref{eq:pop}, by the SOS problem
\begin{equation}\label{eq:pop_dual}
	\begin{aligned}
		\min \quad & \lambda \\
		\mathrm{s.t.} \quad & \lambda - f(x) = s_0(x) + \sum_{i=1}^{n-1} s_i(x) g_i(x), \\
		& s_j \in \mathrm{SOS} \qquad \forall\,j=0,\dots,n-1,\\
		& \lambda \in \mathbb{R},
	\end{aligned}
\end{equation}
where the optimisation is over $\lambda$ and SOS polynomials $s_j$.
Notice that whenever we have an $x$ such that $g_i(x) \geq 0$ for every $i$ (i.e., $x$ is a feasible point of Eq.~\eqref{eq:pop}) we know that the right-hand side of the equality constraint must be nonnegative and hence $f(x) \leq \lambda$.
Therefore this dual problem gives an upper bound on the maximum of $f(x)$.
Like the original problem~\eqref{eq:pop}, this is not necessarily an easy problem to solve.
Nevertheless one can again relax it to a hierarchy of SDPs.

The key idea is to notice that for any SOS polynomial $p(x)$ one can always write it in the form $p(x) = w^T M w$, where $M$ is a PSD matrix and $w$ is a vector of monomials.
Thus, one can obtain a hierarchy of relaxations by bounding the length of the monomials in the vector $w$.
Let $\mathrm{SOS}_k$ be the set of all the SOS polynomials generated when $w$ is the vector of all the monomials in $\mathcal{S}_k$.
We then have the following hierarchy of relaxations for $k \in \mathbb{N}$.
\begin{equation}\label{eq:lasserre_dual}
	\begin{aligned}
		\min \quad & \lambda \\
		\mathrm{s.t.} \quad & \lambda - f(x) = s_0(x) + \sum_{i=1}^{n-1} s_i(x) g_i(x), \\
		& s_j \in \mathrm{SOS}_{2 k - \deg(g_j)} \qquad \forall\,j=0,\dots,n-1,\\
		& \lambda \in \mathbb{R},
	\end{aligned}
\end{equation}
where $\deg(g_0) = 0$.
This gives a sequence of SDP relaxations for Eq.~\eqref{eq:pop_dual}.
Moreover, the SDPs in Eq.~\eqref{eq:lasserre_dual} are precisely the dual SDPs of the moment matrix relaxations of Eq.~\eqref{eq:lasserre_primal} \cite{Pironio2010}.

By solving these SDPs it is possible to extract an SOS decomposition of $\lambda - f(x)$, which gives a \emph{certificate} that $f(x) \leq \lambda$ whenever $g_i(x) \geq 0\,\forall\,i$.
For instance, solving the level-$1$ relaxation of our previous example, given in Eq.~\eqref{eq:lasserre_ex}, we find that for $\lambda = \frac18$ we can write $\frac18 - x_2^2 + x_1 x_2 + x_2$ as
\begin{equation}
	\frac12 \left(\frac12 - x_1 -x_2\right)^2 + \frac12(x_1 - x_1^2) + \frac{3}{2}(x_2 - x_2^2)\,.
\end{equation}
Whenever the constraints $x_1 \geq x_1^2$ and $x_2 \geq x_2^2$ are satisfied the above polynomial is nonnegative and hence, as it is equal to $\frac18 - x_2^2 + x_1 x_2 + x_2$ it must be that $x_2^2 - x_1 x_2 - x_2 \leq \frac18$.
This is an analytical proof of the upper bound which can be extracted from the numerics.

\subsection{Noncommutative polynomial optimisation}
\label{sec:NoncommutativePoly}
The polynomial optimisation problems of the previous section can also be extended to the setting wherein the variables do not commute.
Historically it was discovered through the study of quantum nonlocality: based on the work of Tsirelson \cite{Tsirelson1980}, Wehner showed that the correlations of two-outcome Bell inequalities without marginals can be characterized via SDP \cite{Wehner2006}.
The general case, which requires an SDP hierarchy, was discovered soon afterwards by Navascués et al. \cite{Navascues2007}.
Only later the connection to the commutative case and the extension to arbitrary polynomials was realized \cite{Navascues2008,Doherty2008b,Pironio2010}.

Given some Hilbert space $\mathcal{H}$ we can now consider polynomials of bounded operators $X_1, \dots, X_n$ on $\mathcal{H}$.
In particular, consider the following optimisation problem
\begin{equation}\label{eq:ncpop_primal}
	\begin{aligned}
		\max \quad & \tr\left(\rho\, f(X_1, \dots, X_n)\right) \\
		\mathrm{s.t.} \quad & \tr \left(\rho\, h_i(X_1, \dots, X_n)\right) \geq 0 \qquad \forall\,i,\\
		& \,g_j(X_1,\dots,X_n) \succeq 0 \qquad \qquad \,\,\, \forall\,j, \\
		& \tr(\rho) = 1, \\
		& \rho \succeq 0,
	\end{aligned}
\end{equation}
where the optimisation is over all Hilbert spaces $\mathcal{H}$, all states $\rho$ on $\mathcal{H}$ and all bounded operators $X_1,\dots, X_n$ on $\mathcal{H}$, and the polynomials $f$, $h_i$ and $g_j$ are all Hermitian\footnote{A polynomial $f(X_1, \dots, X_n)$ is called Hermitian if $f = f^\dagger$.} -- although the variables $X_1, \dots X_n$ need not necessarily be Hermitian.
This noncommutative generalisation of Eq.~\eqref{eq:pop} rather naturally captures many problems in quantum theory and, as we shall see in later sections, it forms the basis for characterising nonlocal correlations (see Section~\ref{sec:nonlocality}), communication correlations (see Sections~\ref{sec:communication-dimension} and \ref{sec:communication-distinguish}), computing key rates in cryptography (see Section~\ref{sec:qkd}) and characterising network correlations (see Section \ref{sec:networks}).

As in the commutative case, this problem is in general very difficult to solve.
Indeed, the noncommutative setting is a generalisation of the former and hence inherits its complexity.
Nevertheless, \citet{Pironio2010} showed that relatively natural extensions of the moment and sum-of-squares hierarchies can be derived that lead to a hierarchy of SDPs that (under mild conditions\footnote{A sufficient condition for convergence is that the constraints of the problem imply a bound on the operator norm of feasible points $(X_1,\dots,X_n)$.
Following the formulation in \citet{Pironio2010}, one should be able to determine some constant $C$ such that $C^2 - \sum_{i=1}^n X_{i}^{\dagger} X_i \succeq 0$ for all feasible points $(X_1,\dots,X_n)$.
For example if $X_i$ are all projectors then we can take $C=\sqrt{n}$.}) will converge to the optimal value of Eq.~\eqref{eq:ncpop_primal}.

\subsubsection{Moment matrix approach}
Following the previous section closely, a \emph{monomial} is any product of the operators $X_1, \dots, X_n$ and its length is the number of elements in the product.
We define the length of the identity operator to be $0$.
For $k \in \mathbb{N}$ let $\mathcal{S}_k$ denote the set of monomials of length no larger than $k$, noting that if a variable $X_i$ is not Hermitian then we also include its adjoint $X_i^{\dagger}$ in the set of variables generating the monomials in $\mathcal{S}_k$.

For any feasible point $(\mathcal{H}, \rho, X_1,\dots, X_n)$ of the problem it is possible to define a moment matrix, $\Gamma^k$, of level $k$ which is a matrix indexed by elements of $\mathcal{S}_k$ and whose $(M,N)$ entry for $M, N \in \mathcal{S}_k$ is given by
\begin{equation}
	\Gamma^k(M,N) = \tr\left(\rho \, M^\dagger N\right).
\end{equation}
As in the commutative case, this moment matrix is necessarily PSD as for any vector $\ket{w}$ we have
\begin{equation}
	\bra{w} \Gamma^k \ket{w} = \tr\left(\rho R^\dagger R\right) \geq 0,
\end{equation}
where $R = \sum_{N \in \mathcal{S}_k} \braket{N}{w} N$.
Note that for any polynomial $p(X)$ of degree no larger than $2k$ in the variables $X_1,\dots, X_n$ we have that $\tr\left(\rho\, p(X)\right)$ is a linear combination of the elements of $\Gamma^k$.
For each polynomial $g_i$ appearing in the constraints of Eq.~\eqref{eq:ncpop_primal} we also introduce a localising moment matrix of level $k_i$, denoted $\Gamma^{k_i}_{g_i}$, whose $(M,N)$ entry is
\begin{equation}
	\Gamma_{g_i}^{k_i}(M, N) = \tr \left(\rho \, M^{\dagger} g_i(X) N\right).
\end{equation}
As in the commutative case a natural choice of $k_i$ is $\lfloor k - \deg(g_i)/2\rfloor$ to ensure that all elements of $\Gamma_{g_i}^{k_i}$ can be expressed as linear combinations of elements of $\Gamma^k$.
Note that if $g_i(X)$ is PSD then its corresponding moment matrix is also PSD.

As in the case of the Lasserre hierarchy, it is possible to relax the problem~\eqref{eq:ncpop_primal} to a hierarchy of SDPs by optimising over semidefinite matrices that resemble moment matrices and localising moment matrices of level $k$.
In particular if $\deg(f),\, \deg(h_i) \leq 2k$, we can write $f(X) = \sum_{M,N \in \mathcal{S}_k} f_{MN} M^{\dagger} N$ and $h_i(X) = \sum_{M,N \in \mathcal{S}_k} h^i_{MN} M^{\dagger} N$ where $f_{MN}, h^i_{MN} \in \mathbb{C}$.
Then for $k\in \mathbb{N}$ such that $\deg(f), \deg(h_i) \leq 2k$ we define the level-$k$ relaxation of Eq.~\eqref{eq:ncpop_primal} to be the SDP
\begin{equation}
	\begin{aligned}
		\max \quad & \sum_{M,N \in \mathcal{S}_k} f_{MN} \Gamma^k(M,N) \\
		\mathrm{s.t.} \quad & \sum_{M,N} h^i_{MN} \Gamma^k(M,N) \geq 0 \qquad \forall\,i, \\
		& \Gamma^{k_j}_{g_j} \succeq 0 \qquad\qquad\qquad\qquad\quad \forall\,j, \\
		& \Gamma^k \succeq 0\,.
	\end{aligned}
\end{equation}
As in the case of the Lasserre hierarchy, there are many implicit equality constraints in the above SDP, e.g., $\Gamma^k(A, BC) = \Gamma(B^{\dagger}A, C)$, and the normalisation condition $\Gamma^k(\id,\id) = 1$.
Moreover, if $f_{MN}$ and $h^i_{MN}$ are all real, we can without loss of generality restrict $\Gamma^k$ to be real valued, as explained in Section \ref{sec:symmetries}.

Let us take a look at a noncommutative extension of the example we introduced in the previous subsection (see problem~\eqref{eq:lasserre_ex}).
Suppose that $X_1$ and $X_2$ are now Hermitian operators, and that we are interested in solving the following problem
\begin{equation}\label{eq:ncpop_ex}
	\begin{aligned}
		\max \quad & \tr \left[\rho \, \left(X_2^2 - \frac12 X_1 X_2 - \frac12 X_2 X_1 - X_2\right)\right] \\
		\mathrm{s.t.} \quad & X_1 - X_1^2 \succeq 0, \\
		& X_2 - X_2^2 \succeq 0, \\
		& \tr(\rho) =1, \\
		& \rho \succeq 0\,.
	\end{aligned}
\end{equation}
The main difference between Eqs.~\eqref{eq:lasserre_ex} and \eqref{eq:ncpop_ex} is that the monomial $x_1x_2$ is replaced by a noncommutative generalization, $(X_1X_2+X_2X_1)/2$.
Note that if we were to add the condition $[X_1,X_2]=0$, then the optimal value of the problem would coincide with that of Eq.~\eqref{eq:lasserre_ex}.
Considering the indexing set $\{\id, X_1, X_2\}$ the level-1 relaxation of~\eqref{eq:ncpop_ex} corresponds to the SDP
\begin{equation}\label{eq:ncpop_ex_l1}
	\begin{aligned}
		\max \quad & y_{22} - \frac12 y_{12} - \frac12 y_{21} - y_{02} \\
		\mathrm{s.t.} \quad & \begin{pmatrix}
			1 & y_{01} & y_{02} \\
			& y_{11} & y_{12} \\
			& 		& y_{22}
		\end{pmatrix} \succeq 0, \\
		& y_{01} - y_{11} \geq 0, \\
		& y_{02} - y_{22} \geq 0.
	\end{aligned}
\end{equation}
As the moment matrix is Hermitian and can be taken to be real (and thus $y_{12} = y_{21}$), we see that the SDP in~\eqref{eq:ncpop_ex_l1} is equivalent to the level-1 relaxation for the corresponding commutative problem (see~\eqref{eq:pop_ex_l1}).
Thus, like in the commutative setting, we find a value of $1/8$ at level 1.
Interestingly, we see a difference between the commutative and noncommutative problems emerge at level 2.
The level-2 relaxation is based on the indexing set $\{\id,X_1,X_2,X_1^2, X_1 X_2, X_2 X_1, X_2^2\}$, which is larger than the corresponding commutative indexing set, and results in the SDP relaxation
\begin{equation}\label{eq:ncpop_ex_l2}
	\begin{aligned}
		\max \quad & y_{06} - \frac12 y_{04} - \frac12 y_{05} - y_{02} \\
		\mathrm{s.t.} \quad & \begin{pmatrix}
			1 & y_{01} & y_{02} & y_{03} & y_{04} & y_{05} & y_{06} \\
			& y_{03} & y_{04} & y_{13} & y_{14} & y_{15} & y_{16} \\
			& & y_{06} & y_{14} & y_{24} & y_{16} & y_{26} \\
			& & & y_{33} & y_{34} & y_{35} & y_{36} \\
			& & & & y_{44} & y_{45} & y_{46} \\
			& & & & & y_{55} & y_{56} \\
			& & & & & & y_{66}
		\end{pmatrix} \succeq 0, \\
		& \begin{pmatrix}
			y_{01} - y_{03} & y_{03} - y_{13} & y_{04} - y_{14} \\
			& y_{13} - y_{33}& y_{14} - y_{34} \\
			& & y_{24} - y_{44}
		\end{pmatrix} \succeq 0, \\
		& \begin{pmatrix}
			y_{02} - y_{06} & y_{05} - y_{16} & y_{06} - y_{26} \\
			& y_{15} - y_{55} & y_{16} - y_{56} \\
			& & y_{26} - y_{66}
		\end{pmatrix} \succeq 0\,,
	\end{aligned}
\end{equation}
where by taking the moment matrices to be real we find that $y_{04}=y_{05}$.
Running this SDP we again find that the optimal value is $1/8$ and it is possible to show that the hierarchy had already converged at level 1.
This value is achieved by the qubit state $\rho = \ketbra{0}{0}$ together with the projectors 
\begin{equation}
	X_1 = \frac14 \begin{pmatrix}
		1 & \sqrt{3} \\
		\sqrt{3} & 3
	\end{pmatrix}, \quad X_2 = \frac14 \begin{pmatrix}
		1 & -\sqrt{3} \\
		-\sqrt{3} & 3
	\end{pmatrix}\,.
\end{equation}
This implies that the optimal value of $\frac18$ for the noncommutative problem~\eqref{eq:ncpop_ex} is different from the optimal value of the commutative problem~\eqref{eq:lasserre_ex}, which was $0$.

\subsubsection{Sum of squares approach}
In the same spirit as Section \ref{sec:lasserre}, the dual problem to the moment matrix approach can be seen as an optimisation over SOS polynomials, in this case with noncommuting variables.
Given a polynomial of operators $p(X_1, \dots, X_n)$ we say that $p$ is a sum of squares if it can be written in the form
\begin{equation}
	p(X_1,\dots, X_n) = \sum_i r_i^{\dagger}(X_1,\dots,X_n) r_i(X_1,\dots, X_n)
\end{equation}
for some polynomials $r_i$.
It is evident that SOS polynomials are necessarily PSD.
It is thus possible to find an upper bound on the problem~\eqref{eq:ncpop_primal} by instead solving the problem
\begin{equation}\label{eq:ncpop_dual}
	\begin{alignedat}{3}		\min & \quad \lambda \\
		\mathrm{s.t.}& \quad \lambda - f(X) = \,\,&&s_0(X) + \sum_i \nu_i h_i(X) \\
		&\quad &&+ \sum_{i,j} r^\dagger_{ij}(X) g_j(X) r_{ij}(X), \\
		& \quad \nu_i \geq 0 &&\forall\,i, \\
		& \quad s_0 \in \mathrm{SOS},  \\
		& \quad \lambda \in \mathbb{R},
	\end{alignedat}
\end{equation}
where the optimisation is over $\lambda$, the nonnegative real numbers $\nu_i$, the sum-of-squares polynomial $s_0$, and arbitrary polynomials $r_{ij}(X)$.
Given a feasible point of the problem~\eqref{eq:ncpop_dual} and any quantum state $\rho$, it is clear that if $g_j(X) \succeq 0$ and if $\tr(\rho h_i(X)) \geq 0$ then we must have $\lambda \geq \tr(\rho f(X))$.
Therefore any feasible point of Eq.~\eqref{eq:ncpop_dual} provides an upper bound on the optimal value of Eq.~\eqref{eq:ncpop_primal}.

This SOS optimisation can furthermore be relaxed to a hierarchy of SDPs.
To see this note first that a polynomial $p(X)$ is a sum of squares if and only if there exists a PSD matrix $M$ such that $p(X) = w^{\dagger} M w$, where $w$ is a vector of monomials.
Thus, by considering vectors $w$ whose entries are monomials up to degree $k$ (i.e., elements of $\mathcal{S}_k$), one optimises over SOS polynomials up to degree $2k$ and the constraint $s_0 \in \mathrm{SOS}$ is relaxed to the SDP constraint $s_0 \in \mathrm{SOS}_{2k}$.
The real variables $\lambda$ and $\nu_i$ all appear linearly in the problem and are therefore valid variables for an SDP problem.
Finally we have the terms of the form
\begin{equation}
	\sum_{i} r_i^\dagger(X) g(X) r_i(X).
\end{equation}
This is similar to an SOS polynomial, $\sum_{i} r_i^\dagger(X) r_i(X)$, except that it is centered around a polynomial $g(X)$.
Like in the case of an SOS polynomial, for a bounded degree of $r_i$ this quantity can be rewritten as a PSD matrix $M$ with its entries multiplied by $g(X)$, creating a new matrix $M_g$ that satisfies $w^\dagger M_g w = \sum_{i} r_i^\dagger(X) g(X) r_i(X)$.
Thus this term can also be reinterpreted as a PSD condition.
Following the notation for SOS polynomials we denote the set of $g$-centered SOS polynomials of degree up to $d$ by $\mathrm{SOS}^{g}_{d}$.

For each $k \in \mathbb{N}$ large enough, one arrives at the following hierarchy of semidefinite programming relaxations for Eq.~\eqref{eq:ncpop_dual}
\begin{equation}\label{eq:ncpop_dual_relaxed}
	\begin{aligned}
		\min \quad & \lambda \\
		\mathrm{s.t.} \quad & \lambda - f(X) = s_0(X) + \sum_i \nu_i h_i(X) + \sum_j s_j(X) \\
		& \nu_i \geq 0 \qquad\qquad\qquad\quad\,\, \forall\,i, \\
		& s_0 \in \mathrm{SOS}_{2k}, \\
		& s_j \in \mathrm{SOS}^{g_j}_{ 2k - \deg(g_j)} \qquad \forall\,j, \\
		& \lambda \in \mathbb{R}\,.
	\end{aligned}
\end{equation}

By relaxing the noncommutative problem~\eqref{eq:ncpop_ex} to level 1 of the SOS hierarchy we find that the polynomial $\frac18 - X_2^2 + \tfrac12 (X_1X_2 + X_2X_1) + X_2^2$ can be written as
\begin{equation}
	\begin{aligned}
		\frac12\left(\frac12 - X_1 - X_2\right)^\dagger \left(\frac12 - X_1 - X_2\right) 
		&+ \frac12(X_1 - X_1^2) \\
		&+ \frac{3}{2}(X_2 - X_2^2),
	\end{aligned}
\end{equation}
which provides an analytical proof that for any Hermitian operators $(X_1,X_2)$ that satisfy $X_1-X_1^2 \succeq 0$ and $X_2-X_2^2 \succeq 0$ we must have that $\tr \left(\rho \, \left(X_2^2 - \frac12 (X_1 X_2 + X_2 X_1) - X_2\right)\right)\leq\frac18$.

\begin{remark*}\label{rem:monomial-sets}
	Throughout this section we have repeatedly used a monomial indexing which was chosen up to some degree $k$.
	In both the moment and SOS approach this $k$ defines the index of the SDP hierarchy.
	It is important to note however that it is not necessary to construct a hierarchy with these sets and in general indexing by any set of monomials $\mathcal{S}$ will lead to a valid semidefinite relaxation of the problem.
	Such constructions can lead to more accurate bounds with fewer computational resources or to interesting physical constraints \cite{Moroder2013}.
	Note that this also applies to the indexing sets of the localising moment matrices.
\end{remark*}


\section{Entanglement}\label{sec:separability}
Entangled states are fundamental in quantum information science.
In this section we discuss the use of SDP relaxation methods for detecting and quantifying entanglement.

\subsection{Doherty-Parrilo-Spedalieri hierarchy}\label{sec:DPS}
Recall that a bipartite state is separable when it can be written in the form given by Eq.~\eqref{separable}.
Otherwise, it is said to be entangled.
This leads to an elementary question: is a given bipartite density matrix separable or entangled? While a general solution is very challenging~\cite{Gurvits2003,Gharibian2010}, the problem can be solved through a converging hierarchy of semidefinite relaxations of the set of separable states known as the Doherty-Parrilo-Spedalieri (DPS) hierarchy \cite{Doherty2002,Doherty2004}.
As we shall see in Section~\ref{sec:entanglement:multipartite:detection}, the DPS hierarchy can also be adapted to entangled states of many parties.

Consider a quantum state $\rho_{AB} \in \mathcal{H}_A \otimes \mathcal{H}_B$.
If the state is separable, then for any positive integer $n$ we can construct a \textit{symmetric extension} of this quantum state, that is, a quantum state $\rho_n \in \mathcal{H}_A \otimes \mathcal{H}_B^{\otimes n}$ that is invariant under permutation of the subsystems in $\mathcal{H}_B^{\otimes n}$, and such that taking the partial trace over the additional subsystems recovers $\rho_{AB}$.
For a separable state written as in Eq.~\eqref{separable}, such an extension is given by
\begin{equation}
	\rho_n^\text{sep}=\sum_{\lambda} p(\lambda) \phi_\lambda\otimes \varphi^{\otimes n}_\lambda.
	\label{sepSDP:extension}
\end{equation}
The main idea behind the DPS hierarchy is that this does not hold for entangled states: for any fixed entangled $\rho_{AB}$ there is a threshold $n_0$ such that symmetric extensions with $n>n_0$ do not exist.

Testing whether such a symmetric extension exists can be cast as an SDP, and therefore this gives a complete SDP hierarchy for testing entanglement.
However, this test can quickly become computationally demanding.
Two more ideas can be used to make it more tractable.
The first is combining the test for symmetric extensions with the PPT criterion, described in Section \ref{sec:introduction:entanglement}: we add the requirement that the extension $\rho_n$ must have positive partial transposition across all possible bipartitions\footnote{Note that because of the symmetry of $\rho_n$ only $n$ partial transpositions need to be considered, instead of the $2^n$ possible ones.}.
This is satisfied by $\rho_n^\text{sep}$.
The second idea is to make use of the symmetry of $\rho_n^\text{sep}$ in order to reduce the size of the problem\footnote{Symmetrisation techniques are useful for a wide class of SDPs and are further discussed in Section \ref{sec:symmetries}.}.
The key observation is that $\rho_n^\text{sep}$ is not only invariant under the permutation of $\mathcal{H}_B^{\otimes n}$, but satisfies a stronger condition\footnote{To see that this is stronger, consider the state $\ket{\psi^-} = \frac1{\sqrt2}(\ket{01}-\ket{10})$. It is not symmetric, as $\text{SWAP}\ket{\psi^-} = -\ket{\psi^-}$, but it is permutation invariant, as $\text{SWAP} \ket{\psi^-}\bra{\psi^-} \text{SWAP} = \ket{\psi^-}\bra{\psi^-}$.} known as Bose symmetry, that is,
\begin{equation}
	\rho_n^\text{sep} = \left(\id_A \otimes P_B \right)\rho_n^\text{sep} = \rho_n^\text{sep} \left(\id_A \otimes P_B\right)
	\label{sepSDP:bose}
\end{equation}
for any permutation $P$.
This implies that we can additionally require that $\rho_n^\text{sep}$ belongs to the symmetric subspace over the copies of $B$ \cite{Doherty2004}, which has dimension $s_n = \binom{d_B + n -1}{n}$, as opposed to the dimension $d_B^n$ of the whole space.
Let then $V_B$ be an isometry from $\mathbb{C}^{s_n}$ to the symmetric subspace of $\mathcal{H}_B^{\otimes n}$.
With all the pieces in place, we can state the DPS SDP:
\begin{equation}\label{eq:dps}
	\begin{aligned}
		\max_{\sigma,t} \quad &t \\
		\st \quad & \sigma \succeq t\id, \\
		& \rho_n = \big(\id_A \otimes V_B\big) \sigma \big(\id_A \otimes V_B^\dagger \big), \\
		& \rho_{AB} = \tr_{B_2,\dots,B_n}(\rho_n), \\
		& \rho_n^{T_{B_1}\ldots T_{B_k}} \succeq t\id \quad \forall\,k,
	\end{aligned}
\end{equation}
where $\sigma$ is an auxiliary operator used to characterise the symmetric subspace.

The variable $t$ has been introduced to make the problem strictly feasible as in Section \ref{sec:intro}.
The dimension of $\sigma$ is $d_A s_n$, which for fixed $d_B$ increases exponentially fast with $n$, reflecting the fact that determining separability is an NP-hard problem.
It is possible to compute convergence bounds on the DPS hierarchy, i.e., until which $n$ does one need to go in order to test whether a given quantum state is $\epsilon$-close to the set of separable states \cite{Navascues2009b}.

From the dual of the DPS hierarchy one can in principle obtain an entanglement witness for any entangled state.
Moreover, the dual of the DPS hierarchy can also be interpreted as a commutative sum-of-squares hierarchy~\cite{Fang2021}.

The hierarchy collapses at the first level if $d_Ad_B \le 6$, as in this case the PPT criterion is necessary and sufficient for determining whether a state is entangled~\cite{Horodecki1996}.
A natural question is then whether it also collapses at a finite level for larger dimensions.
Surprisingly, the answer is negative, and moreover one can show that no single SDP can characterise separability in these dimensions \cite{Fawzi2019}.
It is possible, however, to solve a weaker problem with a single, albeit very large, SDP: optimizing linear functionals over the set of separable states \cite{Harrow2017}.

While the DPS hierarchy gives converging outer SDP relaxations of the separable set, it is also possible to construct converging inner relaxations of the same set \cite{Navascues2009}.
These relaxations closely follow the ideas of the DPS relaxations and are based on the observation that small linear perturbations can destroy the entanglement of states with $n$-fold Bose symmetric extension.
However, they differ in the fact that the resulting set of SDPs is not a hierarchy, as the next criterion is not always strictly stronger than the previous.

\subsection{Bipartite entanglement}
In this subsection we review the application of SDP methods to the simplest entanglement scenario, namely that of entanglement between two systems.

\subsubsection{Quantifying entanglement}
Once a quantum state is known to be entangled, a natural question is to quantify its entanglement; see e.g.~the review \citet{Plenio2005}.
In the standard paradigm, where the parties can perform local operations assisted by classical communication (LOCC) and have access to asymptotically many copies of the state, it is natural to consider conversion rates between a given state and the maximally entangled state as quantifiers of entanglement.
Two important quantifiers are the distillable entanglement, $E_D$, and the entanglement cost, $E_C$.

The distillable entanglement, $E_D$, addresses the largest rate, $R$, at which one can convert, by means of LOCC, a given bipartite state $\rho_{AB}$ into a $d$-dimensional maximally entangled state $\phi^+_d$ \cite{Bennett1996b}.
This is equivalent to asking how many copies of a maximally entangled qubit pair can be extracted asymptotically from $\rho_{AB}$.
While this definition may appear to be somewhat arbitrary, in the asymptotic setting many alternative definitions turn out to be equivalent to it \cite{Rains1999}.
Therefore,
\begin{equation}
	\begin{aligned}
		E_D(\rho_{AB})= \sup\quad & R\\
		\st\quad &\lim_{n\rightarrow \infty} \inf_\mathcal{L} \|\mathcal{L}\left(\rho_{AB}^{\otimes n}\right)-\phi^+_{2^{\lfloor nR\rfloor }}\|_1 =0,
	\end{aligned}
\end{equation}
where $\|O\|_1=\tr \sqrt{O^\dagger O}$ is the trace norm and $\mathcal{L}$ is the set of LOCC operations.

The entanglement cost is the smallest rate required to convert maximally entangled states into a given state by means of LOCC,
\begin{equation}\label{entcost}
	\begin{aligned}
		E_C(\rho_{AB})= \inf\quad & R\\
		\st \quad&\lim_{n\rightarrow \infty} \inf_\mathcal{L} \|\rho_{AB}^{\otimes n}-\mathcal{L}\left(\phi^+_{2^{\lfloor nR\rfloor}}\right)\|_1 =0.
	\end{aligned}
\end{equation}
This definition remains unaltered by changing the distance measure \cite{Hayden2001}.
In general $E_D\leq E_C$, with equality holding for pure states \cite{Vidal2001, Horodecki2003}.
In fact, a large class of entanglement measures can be shown to be bounded from above by $E_C$ and from below by $E_D$ \cite{Matthew2002, Horodecki2000b}.
Thus, computing these quantities is of particular interest.
Unfortunately, due the difficulty of characterising $\mathcal{L}$ \cite{Chitambar2014}, such computations are very hard \cite{Huang2014}, but they can be efficiently bounded using SDP methods.

A frequently used entanglement measure is the logarithmic negativity of entanglement \cite{Vidal2002}.
It is defined as $E_{\mathcal{N}}=\log \|\rho^{T_B}\|_1$ and it bounds the distillable entanglement as $E_D(\rho)\leq E_\mathcal{N}(\rho)$.
It can be computed as the following SDP,
\begin{equation}
	\begin{aligned}
		\|\rho^{T_B}\|_1=\min_{\sigma_\pm} \quad & \tr\left(\sigma_+\right) + \tr\left(\sigma_-\right)\\
		\st \quad & \rho^{T_B} = \sigma_+-\sigma_-,\\
		& \sigma_\pm \succeq 0.
	\end{aligned}
	\label{eq:negativity}
\end{equation}
To see the connection between the trace norm and SDP, note that every Hermitian operator, $O$, can be written as $O=\sigma_+-\sigma_-$ for some PSD operators $\sigma_\pm$.
A related SDP-computable quantity is the tempered negativity, which is defined for a given $\rho$ as $\sup\{\tr\left(\rho X\right): \|X^{T_A}\|_{\infty} \leq 1, \|X\|_{\infty}=\tr\left(\rho X\right) \}$.
This is a lower bound on both the negativity and the entanglement cost $E_C$.
It was introduced to show the irreversibility of entanglement theory when the free operations are not restricted to LOCC but instead can be arbitrary non-entangling operations \cite{Lami2023}.

Consider that we are given one copy of a non-maximally entangled state and we want to distill a state with as large a fidelity with the maximally entangled state as possible.
By relaxing the LOCC paradigm to the (technically more convenient) superset of global operations that preserve PPT, the fidelity can be bounded by an SDP \cite{Rains2001}.
However, this bound is not additive \cite{Wang2017}.
Therefore, once we move into the many-copy regime, the size of the SDP grows with the number of copies $n$, making it unwieldy for the asymptotic limit $n\rightarrow \infty$.
Notably, in this LOCC-to-PPT relaxed setting, the irreversibility of entanglement (i.e.,~$E_D\neq E_C$) still persists as shown through SDP in \citet{Wang2017b}; see also \citet{Ishizaka2005}.
In \citet{Wang2020}, an SDP-computable measure is introduced for quantifying non-PPT entanglement under global operations that are completely PPT preserving.
This can be used to bound from above and below the one-shot exact entanglement cost under such free operations\footnote{The exact entanglement cost is a more restrictive variation of Eq.~\eqref{entcost} where $\rho_{AB}$ is generated exactly instead of with asymptotically vanishing error \cite{Audenaertb2003}.}.

An alternative upper bound on $E_D$ is reported in \citet{Wang2016} which is fully additive under tensor products, thus resolving the limit issue, and computable by SDP.
It is given by $E_W=\log W(\rho_{AB})$ where
\begin{equation}\label{WangSDP}
	\begin{aligned}
		W(\rho_{AB})= \min \quad & \|\sigma^{T_B}_{AB}\|_1 \\
		\st \quad & \sigma_{AB}\succeq \rho_{AB}.
	\end{aligned}
\end{equation}
This is bounded from below by the bound in \cite{Rains2001} and from above by the logarithmic negativity.

While the asymptotic setting is conceptually interesting, a more applied approach often considers imperfect conversions between states using finitely many copies.
In this so-called one-shot setting, SDP methods have been used for bounding the rate of entanglement distillation for a given degree of error \cite{Fang2019}.
This has been considered using many different relaxations of LOCC which admit either LP or SDP formulations \cite{Regula2019}.
In \citet{Rozp2018} SDPs are used for entanglement distillation under realistic limitations on the number of copies, error and exchange of messages in LOCC, including also the setting in which success is only probabilistic.

Another interesting entanglement measure is the squashed entanglement \cite{Christandl2004}.
It has several desirable properties: it is fully additive under tensor products, it obeys a simple entanglement monogamy relation \cite{Koashi2004} and it is faithful, i.e.,~it is non-zero if and only if the state is entangled \cite{Brandao2011}.
The definition draws inspiration from quantum key distribution by considering the smallest possible quantum mutual information between Alice and Bob upon conditioning on a third, ``eavesdropper'', system $E$ with which the state may be correlated.
The squashed entanglement is defined as
\begin{equation}
	\begin{aligned}
		E_{sq}(\rho_{AB})= \min_{\rho_{ABE}} \quad & \frac{1}{2}I(A:B|E)\\
		\st \quad & \rho_{AB}=\tr_E\left(\rho_{ABE}\right),
	\end{aligned}
\end{equation}
where the quantum conditional mutual information can be given in terms of the conditional von Neumann entropy as $I(A:B|E)=H(A|E)-H(A|BE)$.
While it is NP-hard to compute~\cite{Huang2014}, it can be bounded from below by means of a hierarchy of SDPs \cite{Fawzi2022}.
By defining a new system $D$ such that $\rho_{ABED}$ is pure, it follows from entropic duality relations that $I(A:B|E) = H(A|E) + H(A|D)$.
This transforms the objective function into a nonnegative sum of von Neumann entropies which can then be lower bounded using techniques similar to those discussed in Section~\ref{sec:qkd}.
It is unknown whether the hierarchy converges to $E_{sq}$ but non-trivial lower bounds can already be obtained at the first level for particular states.

A complementary class of entanglement measures are based on convex roof constructions.
This means that one considers every possible decomposition of a mixed state, $\rho=\sum_i p_i \ketbra{\psi_i}{\psi_i}$, and evaluates the minimal entanglement as averaged over the entanglement of the pure states in the decomposition, i.e.,~$E_\text{roof}(\rho)=\min \sum_i p_i E(\ket{\psi_i})$, for some pure-state entanglement measure $E$.
Examples of this are the entanglement of formation \cite{Bennett1996, Wootters1998} and geometric measures of entanglement \cite{Wei2003}.
In \citet{Toth2015} it is shown that convex roofs of polynomial entanglement measures can be viewed as separability problems.
An illustrative example is the linear entropy, $E(\ket{\psi_{AB}})=1-\tr\left(\rho^2_A\right)$.
By observing that $E(\ket{\psi_{AB}})=\tr\left(\mathbb{P}^\text{asym}_{AA'} \ketbra{\psi}{\psi}_{AB} \otimes \ketbra{\psi}{\psi}_{A'B'} \right)$ where $\mathbb{P}^\text{asym}$ is the projector onto the antisymmetric subspace, the convex roof of the linear entropy can be written as $E_\text{roof}(\rho)=\tr\left(\mathbb{P}^\text{asym} \sigma\right)$ where $\sigma=\sum_i p_i \ketbra{\psi_i}{\psi_i}_{AB} \otimes \ketbra{\psi_i}{\psi_i}_{A'B'}$.
The state $\sigma$ is separable with respect to $AB|A'B'$, symmetric under swapping these systems, and its marginal is $\tr_{A'B'}(\sigma)=\rho$.
Thus, by relaxing separability to e.g.~PPT, $E_\text{roof}$ can be bounded through an SDP.

Finally, we mention that the SDP-based discussion of entanglement quantification and conversion can be extended to many other quantum resource theories, e.g.~fidelity distillation of basis-coherence (instead of entanglement as the resource) under incoherent operations (instead of LOCC as the free operation) \cite{Lami2019}.
Similarly, conversion rates can be addressed by SDPs for resource theories of Gaussian states under Gaussian operations \cite{Lami2018}, basis-coherent states \cite{Napoli2016, Bischof2019}, entanglement in complex versus real Hilbert spaces \cite{Kondra2022} and asymmetry of states under group actions \cite{Piani2016}.

\subsubsection{Detecting the entanglement dimension}
Suppose that a bipartite state with local dimension $d$ is certified to be entangled.
Does the preparation of the state truly require one to generate entanglement between $d$ degrees of freedom? For pure states, this idea of an entanglement dimension is formalised in the Schmidt rank of a state $\ket{\psi}$.
Every pure bipartite state, up to local unitaries, admits a Schmidt decomposition, $\ket{\psi}=\sum_{i=1}^s \lambda_i\ket{i,i}$, for some real and normalised, nonnegative coefficients $\{\lambda_i\}$.
The Schmidt rank is the number of non-zero terms ($1\leq s\leq d$) in the Schmidt decomposition.
For mixed states this concept is extended to the Schmidt number.
Let $\rho=\sum_{i}p_i \ketbra{\psi_i}{\psi_i}$ be a decomposition.
The Schmidt number is the largest Schmidt rank of the pure states $\{\ket{\psi_i}\}$ minimised over all possible decompositions of $\rho$ \cite{Terhal2000b}.

One way to witness the Schmidt number is based on the range of $\rho$.
If the range of $\rho$ is not spanned by pure states that have a Schmidt rank of at most $s$ then $\rho$ must have a Schmidt number of at least $s+1$.
However, verifying that the range cannot be spanned by such states is not easy in general.
In \citet{Johnston2022} it is shown that the more general question of whether a given subspace of pure quantum states contains any product states (or states with Schmidt rank $s$) can be addressed by means of a hierarchy of LPs.
This method exploits elementary properties of local antisymmetric projections applied to tensor products of the basis elements of the considered space.
Every entangled subspace is detected at some finite level of this hierarchy and it is more efficient to compute in comparison to SDP-based approaches.

In analogy with entanglement witnesses, a relevant endeavour is to witness the Schmidt number from partial information about the state $\rho_{AB}$, i.e.,~to find an observable $O$ such that $\tr\left(O\rho\right)\leq \alpha$ holds for all states with Schmidt number at most $s$ but is violated for at least one state with a larger Schmidt number.
Determining the value of $\alpha$ for any given $O$ can be related to a separability problem in a larger Hilbert space \cite{Hulpke2004}.
Specifically
\begin{equation}
	\alpha = \max_{\sigma} \,\,\, s^2\tr\left(O_{AB}\otimes \ketbra{\phi^+_s}{\phi^+_s}_{A'B'} \sigma_{AA'BB'}\right),
\end{equation}
where the four-partite state $\sigma$ is separable with respect to the bipartition $AA'|BB'$.
This is a useful connection because, among other things, it allows us to use known SDP-compatible relaxations of separability to compute bounds on $\alpha$.
However, it has the drawback that the dimension of global Hilbert space scales as $(ds)^2$ and thus evaluating a bound for a larger Schmidt number becomes more demanding.
This type of approach, based on auxiliary spaces $A'B'$, can also be used to address the Schmidt number of $\rho_{AB}$ directly, without a witness observable, via a hierarchy of SDPs that naturally generalises the DPS construction \cite{Weilenmann2020, Guhne2021b}.
One treats the density matrix $\sigma_{AA'BB'}$ as a variable and imposes that $\rho_{AB}=s\Pi_{A'B'}^\dagger \sigma \Pi_{A'B'}$ and that $\sigma$ has a $k$-symmetric extension that is PPT in the sense of DPS.
Notably, constructions of this sort, which connect Schmidt number witnessing to separability problems, can also be leveraged to certify higher-dimensional entanglement in the steering scenario \cite{Gois2022}.
Furthermore, one can systematically search for adaptive Schmidt number witness protocols, that use one-way LOCC from Alice to Bob.
This has been proposed in a hypothesis testing framework that aims to minimise the total probability of false positives and false negatives for the Schmidt number detection scheme \cite{Hu2021}.
To achieve this, one can employ the SDP methods of \citet{Weilenmann2021}, and in particular the dual of the DPS-type approach to Schmidt numbers, to relax the set of possible witnesses.

An alternative approach to Schmidt number detection is to do away with the computational difficulty associated to the auxiliary spaces $A'B'$ by trading it for other relaxations.
For example, a Schmidt number no larger than $s$ implies that $\bracket{\phi^+_d}{\rho}{\phi^+_d}\leq \frac{s}{d}$ for every maximally entangled state $\phi^+_d$ \cite{Terhal2000b}.
Knowing that $\rho$ is close to a particular maximally entangled state thus yields a potentially useful semidefinite relaxation of states with Schmidt number $s$.
Another option is to use the positive but not completely positive generalised reduction map $R(\sigma)=\tr\left(\sigma\right)\id-\frac{1}{s}\sigma$.
Applied to one share of a state $\rho_{AB}$ with Schmidt number $s$ it still returns a valid quantum state \cite{Tomiyama1985}, which constitutes a semidefinite constraint on $\rho_{AB}$.
Either of these conditions can be incorporated into an SDP, now of size only $d^2$, for computing an upper bound on an arbitrary linear witness.
An iterative SDP-based algorithm that constructs Schmidt number witnesses by leveraging this type of ideas appears in \citet{Wyderka2022}.

Further, we note that SDP relaxation methods are used in many other contexts of entanglement detection.
This includes, for example, the construction of entanglement witnesses from random measurements in both discrete \cite{Szangolies2015} and continuous variables \cite{Mihaescu2020}, the evaluation of perturbations to known entanglement witnesses due to small systematic measurement errors \cite{Morelli2022}, unifying semidefinite criteria for entanglement detection via covariance matrices \cite{Gittsovich2008} and the problem of determining the smallest number of product states required to decompose a separable state \cite{Gribling2021a}.

\subsection{Multipartite entanglement}
When considering states of more than two subsystems, one must deal both with an exponentially growing Hilbert space dimension and with an increasing number of qualitatively different entanglement configurations.
SDP methods can be useful in both these regards.

\subsubsection{Entanglement detection}\label{sec:entanglement:multipartite:detection}
Multipartite systems are said to be entangled if they are not fully separable, i.e.,~ when they cannot be expressed as convex combinations of individual states held by each of the parties.
Fully separable states of $N$ subsystems take the form
\begin{equation}
	\rho=\sum_\lambda p(\lambda) \psi_\lambda^{(1)}\otimes \ldots\otimes \psi_\lambda^{(N)}.
	\label{eq:fullyseparable}
\end{equation}

It is possible to extend the original, bipartite, DPS hierarchy discussed in Section~\ref{sec:DPS} to the multipartite case \cite{Doherty2005} and thereby to decide the multipartite separability problem via SDP in the limit of large levels in the hierarchy.
Essentially, one considers symmetric extensions of the form of Eqs.~\eqref{sepSDP:extension} and \eqref{sepSDP:bose} for all but one of the parties.
Considering also dual problem leads to witnesses of multipartite entanglement \cite{Brandao2004, Brandao2004b}.
Due to the aforementioned increase in computational cost, a naive use of this approach is limited in practice to the study of small multipartite systems, both in the number of constituents and in their dimension.
A way to circumvent this problem is by limiting the state space by considering representations of multipartite states in the form of tensor networks \cite{Navascues2020tn}.
This approach is used for detecting both, entanglement and nonlocality, in systems composed of hundreds of particles.
Another approach is to use all of the symmetries that arise when considering the existence of symmetric extensions of the global state.
\citet{Navascues2021b} finds hierarchies that are efficient in time and space requirements, that allow to detect entanglement from two-body marginals in systems of hundreds of particles, and that can be used even for infinite systems with appropriate symmetries.
The approach followed in \citet{Navascues2021b}, namely formulating entanglement detection as asking whether given marginals are consistent with a joint separable state, is an instance of the \textit{quantum marginal problem}, that we will review in the following section.

It is also possible to construct SDP relaxations of the set of separable states from the interior.
\citet{Ohst2023} develops a seesaw-like method in which single-system state spaces are approximated with a polytope.
By considering larger polytopes, one obtains better inner relaxations of separability at the price of computing more demanding SDPs.
This was for instance used to compute bounds on visibilities and robustness measures against full separability for systems up to five qubits or three qutrits.

SDP hierarchies have also been proposed for deciding the full separability of specific states.
One example is multipartite Werner states.
These states have the defining property that they are invariant under the action of any $n$-fold unitary $U^{\otimes n}$.
For such states it is possible, using representation theory \cite{Huber2021}, to provide a characterisation that does not depend on the dimension, and which can be tested via Lasserre's hierarchy or via SDP hierarchies for trace polynomials \cite{Klep2022}.
These hierarchies, therefore, give entanglement witnesses that are valid independently of the local dimensions.
Another class of examples is pure product states.
These can be characterised in terms of suitable degree-3 polynomials in commuting variables \cite{Eisert2004}.
Thus, optimisations under the set of multipartite pure product states can be solved via the Lasserre SDP hierarchy.
This approach has been followed for computing entanglement measures for three- and four-qubit states.

The multipartite separability problem has also been formulated as an instance of the \textit{truncated moment} problem \cite{Bohnet2017,Frerot2022} (see also \citet{Milazzo2020} for an application to the separability of quantum channels).
This problem consists in obtaining a probability measure that reproduces some finite number of observed moments, and can be solved via SDP \cite{Laurent2009}.
In the context of entanglement, this translates into determining whether there exists a separable quantum state that reproduces some observed expectation values.
\citet{Frerot2022}, building on the results of \citet{Bohnet2017}, addresses this problem by developing an NPA-like hierarchy of matrices that are all PSD if the observed expectation values can be reproduced by a separable state.
This gives a tool for detecting many-body entanglement, that recovers the covariance matrix criterion of \citet{Gittsovich2010} and the spin-squeezing inequalities of \citet{Toth2009} at concrete finite levels of the hierarchy.
However, this tool fails to address finer notions of entanglement (i.e.,~failure of $k$-separability for $k>2$).
Multipartite entanglement detection has also been formulated in terms of adaptive strategies \cite{Weilenmann2021}, which can be formulated in terms of Lasserre-like SDP hierarchies \cite{Weilenmann2023}.

According to Eq.~\eqref{eq:fullyseparable}, a state is already entangled if two particles are entangled, even if all the rest remain in a separable state.
A stronger requirement is called genuine multipartite entanglement (GME).
A state has GME if it cannot be generated by classically mixing quantum states that are separable with respect to some bipartition of the subsystems.
We can let $\tau$ denote a bipartition of all the particle labels $\{1,\ldots,N\}$ and associate a state $\sigma_\tau$ which is bi-separable across $\tau$.
If no model of the form $\rho=\sum_{\tau} p(\tau) \sigma_\tau$ exists, where $\tau$ ranges over all the possible bipartitions, then $\rho$ has GME.

While some simple witnesses of GME can be systematically constructed without SDPs (see e.g.~\citet{Bourennane2004, Zhang2023}), SDP methods offer a powerful approach for reasonably small particle numbers.
A sufficient condition for GME is obtained from replacing the separable states $\sigma_\tau$ with quantum states $\omega_\tau$ that are PPT with respect to the bipartition~$\tau$.
Then, by defining the subnormalised operators $\tilde{\omega}_\tau=p(\tau)\omega_\tau$ and adding the normalisation condition $\sum_{\tau} \tr\left(\tilde{\omega}_\tau\right)=1$, one obtains an SDP relaxation of GME \cite{Jungnitsch2011}.
If no such decomposition of $\rho$ is found, SDP duality allows the construction of an inequality that witnesses GME.
This method has been found to be practical for detecting GME in systems up to around seven qubits.
Sometimes, this SDP can even be reduced to an LP, enabling easier computations \cite{Jungnitsch2011b}.

The procedure above can be generalised to any positive map that acts on only one element of a bipartition \cite{Lancien2015}: namely, one relaxes each state $\sigma_\tau=\sum_i p_i \sigma^{(i)}_{\tau_1}\otimes\sigma^{(i)}_{\tau_2}$ by another state, $\sigma_{\Lambda_\tau}$, that satisfies $\Lambda_{\tau_1}\otimes\id_{\tau_2}[\sigma_{\Lambda_\tau}]\succeq 0$ for a given positive map $\Lambda_\tau$.
This has the apparent disadvantage that one would potentially need to run over all possible $\Lambda_\tau$ in order to prove that a state admits such a decomposition, but it turns out that, in practice, simple maps such as the aforementioned transposition map, the Choi map, or the Hall-Breuer map \cite{Clivaz2017}, allow to identify large families of GME states.
This connection between separable states and positive maps is also exploited in order to build witnesses of GME from witnesses of bipartite entanglement \cite{Huber2014}.
Moreover, this connection has also been used in linear algebra, where the Lasserre hierarchy is used for checking whether linear maps and matrices are positive and separable, respectively \cite{Nie2016}.

\subsubsection{Quantum marginal problems}\label{sec:qmp}
The quantum marginal problem (QMP) asks whether there exists a global entangled state that is compatible with a given collection of few-body states.
That is, given a collection of quantum states $\{\rho_i\}_{i=1}^I$, each supported in a set of quantum systems $K_i\subset[1,\dots,N]$ (with, in general, $K_i\cap K_j\not=\emptyset$), the QMP asks whether a joint state $\rho\in\mathcal{H}_1\otimes\cdots\otimes\mathcal{H}_N$ exists that satisfies $\rho_i=\tr_{\setminus K_i}(\rho)$ for all $i$, where $\tr_{\setminus K_i}$ denotes the partial trace over all systems except $K_i$.
The QMP is very naturally cast as an SDP \cite{Hall2007}, since it involves only positive operators (the marginal quantum states) and linear constraints between them.
However, solving these SDPs is in general expensive due to the exponential growth of its size with $N$ and the dimension of the subsystems.
In fact, in terms of computational complexity, the QMP is a QMA-complete problem \cite{liu2007qma}.
Roughly speaking, QMA is one proposed quantum computing counterpart of the complexity class NP (see \citet[Ch. 14]{KitaevBook}, \citet{Gharibian2023}).
Nevertheless, particular cases are tractable or admit tractable relaxations.

One such particular case is that in which the global state is pure, i.e.,~$\rho=\ket{\psi}\bra{\psi}$.
In this case, the QMP can be connected to a separability problem.
This restriction makes the problem no longer an SDP, since the requirement of having a global pure state introduces a nonlinear constraint $\rho^2=\rho$.
\citet{Yu2021} overcomes this issue by considering a symmetric extension of the complete global state, where pure states can be characterised by the restriction $\tr(S_{AB}\rho\otimes\rho)=1$, with the operator $S_{AB}$ denoting the swap operator $S_{AB}=\sum_{i,j}\ket{ij}\bra{ji}$ (note that $\tr(S_{AB}\rho\otimes\sigma)=\tr(\rho\sigma)$).
Then, the separability of $\rho\otimes\rho$ can be relaxed via the DPS hierarchy to an SDP.
This procedure is generalised in \citet{Huber2022}, which formulates the compatibility problem in terms of spectra.
Namely, rather than asking for a joint state that reproduces some given reduced density matrices, \citet{Huber2022} asks whether there exists a joint state such that the spectra of marginals coincides with some given set.
Working with spectra instead of density matrices allows to exploit symmetries in order to reduce the computational load of the problem.
This approach, moreover, produces witnesses of incompatibility for arbitrary local dimensions.

A case where the characterisation can be done exactly in terms of an efficient SDP is that of states that are invariant under permutation of parties \cite{Aloy2021}.
In such a case, first, the symmetries reduce greatly the number of marginals: namely, there is only one possible marginal for each number of subsystems.
Thus it suffices to consider only the problem of the compatibility with a single marginal.
Moreover, the number of parameters required for the description of the joint state is also very small.
This allows \citet{Aloy2021} to give necessary and sufficient conditions for the QMP as a single, tractable SDP for systems composed of up to 128 particles.

The compatibility problem has also been formulated in terms of quantum channels.
On one hand, \citet{Haapasalo2021} considers the problem of whether a global broadcasting channel $\Phi_{A\rightarrow B_1\dots B_N}$ exists that has a given set of channels $\Phi_{A\rightarrow B_i}$ as marginals.
This problem can be connected to a state compatibility problem via the Choi-Jamio\l{}kowski isomorphism \cite{Choi1975,Jamiolkowski1972}, allowing to use the methods outlined above.
On the other, \citet{Hsieh2022} considers the more general problem of whether a global evolution is compatible with a set of local dynamics, giving a measure of robustness that can be computed exactly via SDP.

A problem related to the QMP is determining, from a set of marginals of a joint state, properties of other marginals.
For instance, one can ask, given some entangled states that are marginals of an unknown joint state, whether the remaining marginals must be entangled as well.
This problem can be addressed via entanglement witnesses whose optimisation can be cast as SDPs \cite{Tabia2022}.
The QMP has also been tackled via tools from the study of nonlocality \cite{Bermejo2023}, that are the subject of the next section.


\section{Quantum nonlocality}\label{sec:nonlocality}
In this section we discuss SDP relaxation methods for quantum nonlocality and their applications to quantum information.

\subsection{The Navascués-Pironio-Acín hierarchy}\label{sec:NPA}
\label{sec:nonlocality:npa}
A fundamental question in Bell nonlocality is to characterise the set of distributions that are predicted by quantum theory in a Bell scenario (recall Fig.~\ref{FigBellScenario}) with a given number of inputs ($X$ and $Y$) and outputs ($N$ and $M$).
This corresponds to deciding whether for any given distribution $p(a,b|x,y)$ there exists a bipartite quantum state of any dimension, $\ket{\psi}$, and local measurements for Alice and Bob, $\{A_{a|x}\}_{a,x}$ and $\{B_{b|y}\}_{b,y}$, respectively, such that the distribution can be written in the form\footnote{Note that since there are no restrictions on the dimension, we can without loss of generality assume the quantum state to be pure and the measurements to be projective.}
\begin{equation}\label{eq:npaborn}
	p(a,b|x,y)= \bra{\psi} A_{a|x}\otimes B_{b|y} \ket{\psi}.
\end{equation}
In contrast to the set of correlations associated with local models (see Section \ref{sec:introduction:entanglement}), the set of quantum correlations for arbitrary input/output scenarios, denoted by $\mathcal{Q}$, admits in general no simple and useful characterisation: checking membership is known to be undecidable \cite{Yuen2020}.
Importantly, however, $\mathcal{Q}$ can be approximated by a sequence of supersets $\{\mathcal{Q}_k\}_{k=1}^\infty$ in such a way that
\begin{equation}\label{sequence}
	\mathcal{Q}_1 \supseteq \mathcal{Q}_2 \supseteq \ldots \supseteq \mathcal{Q}_\infty \supseteq \mathcal{Q},
\end{equation}
where the membership of $p$ in $\mathcal{Q}_k$ can be decided by an SDP.
Thus, if there exists some $k$ for which a hypothesised distribution $p$ fails to be a member of $\mathcal{Q}_k$, it follows that no quantum model for $p$ exists.
This hierarchy of outer SDP relaxations to the quantum set of Bell nonlocal correlations is known as the Navascués-Pironio-Acín (NPA) hierarchy \cite{Navascues2007}.
In the limit of the sequence, $k\rightarrow \infty$, the NPA hierarchy converges to a set of distributions $\mathcal{Q}_\infty$.
This set corresponds to quantum models in which the tensor product, demarcating the separation of parties, is relaxed to the commutation condition $[A_{a|x},B_{b|y}]=0$ \cite{Navascues2008, Doherty2008b}.
This corresponds to changing Eq.~\eqref{eq:npaborn} to 
\begin{equation}
	p(a,b|x,y)= \bra{\psi} A_{a|x}B_{b|y} \ket{\psi}.
\end{equation}
In finite dimensions, the commutation condition is equivalent to the tensor product, and thus $\mathcal{Q}_\infty=\mathcal{Q}$.
In infinite dimensions, the conditions are inequivalent \cite{Scholz2008} and, moreover, a strict separation exists \cite{Yuen2020}.
Therefore, the NPA hierarchy in general does not converge to $\mathcal{Q}$.
However, for a specific distribution, in a specific input/output scenario, it sometimes is the case that there exists some finite $k$ for which the membership of $p$ in $\mathcal{Q}_k$ is necessary and sufficient for a quantum model.

The NPA hierarchy is a noncommutative polynomial relaxation hierarchy of the type discussed in section \ref{sec:NoncommutativePoly}.
Define a list of all linearly independent measurement operators in the Bell scenario, $L=\{\id, \mathbf{A},\mathbf{B}\}$ where $\mathbf{A}=\left(A_{1|1},\ldots,A_{N-1|X}\right)$ and $\mathbf{B}=\left(B_{1|1},\ldots,B_{M-1|Y}\right)$.
Then let $\mathcal{S}_k$ be the set of all products of length at most $k$ of the operators appearing in $L$.
Note that one is free to also include some but not all products of a given length; in such cases one speaks of intermediate hierarchy levels, which are often useful in practice (see the remark in Section~\ref{rem:monomial-sets}).
We associate to level $k$ an $|\mathcal{S}_k|\times |\mathcal{S}_k|$ moment matrix whose rows and columns are indexed by elements $u,v\in \mathcal{S}_k$,
\begin{equation}\label{momentmatrixNPA}
	\Gamma(u,v)=\bracket{\psi}{u^\dagger v}{\psi},
\end{equation}
for some unknown state $\ket{\psi}$.
This moment matrix encodes constraints from quantum theory, namely that $\Gamma(\id,\id) = 1$, $\Gamma(u,v) = \Gamma(s,t)$ whenever $u^\dagger v = s^\dagger t$, and $\Gamma(u,v) = 0$ whenever $u^\dagger v = 0$.
This implies for example that $\Gamma(B_{b|y},B_{b'|y}A_{a|x}) = \delta_{b,b'} \Gamma(A_{a|x},B_{b|y})$, as we are assuming without loss of generality that the measurements are projective.
Additionally, we want to constrain the moment matrix to reproduce the probability distribution in question, which requires $\Gamma({A_{a|x},B_{b|y}})=p(a,b|x,y)$, $\Gamma({A_{a|x},\id})=p_A(a|x)$ and $\Gamma(\id,B_{b|y}) = p_B(b|y)$.
Following Section~\ref{sec:sdp}, the key observation is that a necessary condition for the existence of a quantum model for $p$ is that the remaining free variables comprising the moment matrix, i.e.,~all variables not fixed by $p$ and the additional equality constraints, can be chosen such that $\Gamma \succeq 0$.
Finally, since these constraints are all real, one can without loss of generality restrict $\Gamma$ to be real valued, as explained in Section \ref{sec:symmetries}.

For example, the first level of the hierarchy ($k=1$) corresponds to the moment matrix\footnote{The symbols outside the matrix in Eq.~\eqref{level1}, and equivalently in Eq.~\eqref{eq:scalarextension}, denote the element of $\mathcal{S}_k$ that indexes the corresponding row or column.\label{fn:bordermatrix}}
\begin{equation}
	\Gamma = \kbordermatrix{
		& \id & \mathbf{A} & \mathbf{B} \\[0.3ex]
		\id & 1 & P_A & P_B \\[1ex]
		\mathbf{A}^\dagger & P_A^T & S_A & P_{AB} \\[1ex]
		\mathbf{B}^\dagger & P_B^T & P_{AB}^T & S_B
	}.
	\label{level1}
\end{equation}
This matrix has size \mbox{$|\mathcal{S}_1|=1+(N-1)X+(M-1)Y$}, where $P_A = \left[p_A(1|1), p_A(2|1),\ldots,p_A(N-1|X)\right]$ and $P_B = \left[p_B(1|1), p_B(2|1),\ldots,p_B(M-1|Y)\right]$ are Alice's and Bob's marginal probabilities, ${P_{AB}}^{(ax)(by)}=\bra{\psi}A_{a|x}B_{b|y}\ket{\psi}=p(a,b|x,y)$ is the table of joint probabilities, and ${S_A}^{(ax)(a'x')}=\bra{\psi}A_{a|x}A_{a'|x'}\ket{\psi}$ and ${S_B}^{(by)(b'y')}=\bra{\psi}B_{b|y}B_{b'|y'} \ket{\psi}$ are the matrices of second moments of Alice and Bob.
Thus, the sub-matrices $P_A$, $P_B$ and $P_{AB}$ are completely fixed by $p$ whereas the matrices $S_A$ and $S_B$ are entirely comprised of unknown variables except on their diagonals.
If they can be completed such that $\Gamma \succeq 0$, then $p\in \mathcal{Q}_1$.

This SDP is, however, in general \emph{not} strictly feasible, which sometimes causes numerical difficulties when checking for membership in $\mathcal Q_k$ \cite{Araujo2023}.
A straightforward variation is always strictly feasible (see Appendix \ref{sec:strict}): instead of constraining $\Gamma$ to reproduce a particular probability distribution, one leaves those terms as free variables, and optimises a Bell functional over them.
This allows one to compute bounds on the optimal quantum violation of a Bell inequality.
The generic Bell functional in Eq.~\eqref{BellInequality} can be expressed in terms of the moment matrix as $\sum_{a,b,x,y}c_{abxy}\Gamma({A_{a|x},B_{b|y}})$.
In our example for $\mathcal{Q}_1$, this would correspond to $P_A, P_B, P_{AB}$ not being fixed matrices but instead matrices composed of free variables, a linear combination of which is the objective function of the SDP.

It is also interesting to consider the dual of this SDP.
As the primal SDP can be considered a particular case of noncommutative polynomial optimisation, the dual can be considered a particular case of optimisation over SOS polynomials.
Let then $y$ be a vector of all the free variables in the moment matrix, and write $\Gamma = \Gamma_0 + \sum_i y_i \Gamma_i$.
The primal SDP is thus given by
\begin{equation}
	\begin{aligned}
		\max_{y} \quad & \langle b, y \rangle \\
		\st \quad & \Gamma_0 + \sum_i y_i \Gamma_i \succeq 0,
	\end{aligned}
\end{equation}
where $b$ encodes the Bell functional in question.
The dual SDP is then given by
\begin{equation}
	\begin{aligned}
		\min_X \quad & \langle \Gamma_0, X\rangle \\
		\st \quad & \langle \Gamma_i, X\rangle = -b_i, \\
		& X \succeq 0.
	\end{aligned}
\end{equation}
Any feasible solution to the dual SDP then gives an SOS proof that the optimal quantum violation is bounded above by $\langle \Gamma_0, X\rangle$.
To see this let $w$ be the vector of all monomials in $\mathcal{S}_k$ (ordered using the same ordering as the primal problem), then $S = w^{\dagger} X w = w^{\dagger} Q^{\dagger} Q w$ is an SOS polynomial where we have written $X=Q^\dagger Q$ as $X$ is PSD.
In this notation $Qw$ is a vector of polynomials $P_i$ such that $S = \sum_i P_i^\dagger P_i$.
Finally, if $\Gamma$ is the level $k$ moment matrix for any feasible quantum model then we have that 
\begin{equation}
	\langle \Gamma_0, X \rangle - \langle b, y \rangle = \langle \Gamma, X \rangle = \tr(\rho S) \geq 0\,.
\end{equation} Besides proving bounds on the optimal violation, this is also useful for self-testing, as we shall see in Section \ref{sec:selftesting}.
Lastly, note that the NPA hierarchy applies equally well to scenarios that feature more than two parties.

\subsubsection{Macroscopic Locality \& Almost-Quantum Correlations}
There is a large body of research aiming to identify the physical principles that constrain quantum correlations.
Some of these correspond to certain low levels of the NPA hierarchy.
One such principle is Macroscopic Locality \cite{Navascues2010}, which stipulates that classicality must re-emerge in the macroscopic limit of a Bell experiment.
Consider that the source in the Bell scenario does not emit one pair of particles but $N$ independent and identical pairs.
When Alice and Bob perform their measurements, they will respectively direct the incoming beam of particles onto their detectors, causing them all to fire, but with different detection rates.
Assuming that intensity fluctuations in the beam can be detected to the order $\sqrt{N}$, one can define the intensity fluctuation around the mean as $I_u=\frac{1}{\sqrt{N}}\sum_{i=1}^N \left(d^u_i-p(u)\right)$ where $u$ indicates the input/output pair for either Alice or Bob and $d^u_i$ is the indicator function for whether particle number $i$ impinged on the corresponding detector.
When $N\rightarrow \infty$, the central limit theorem implies that Alice and Bob will observe a Gaussian intensity fluctuation with vanishing mean and covariance matrix $\Gamma_{uv}=\expect{I_uI_{v}}=\frac{1}{N}\sum_{i,j=1}^N \expect{d^u_i d^{v}_j}$.
Recalling that the particle pairs are identical and independent one has that $\Gamma_{uv}=\expect{d^u_1 d^{v}_1}$.
Using the fact that the mean and the covariance matrix completely characterise the Gaussian, and the latter is always PSD, one can show that Macroscopic Locality is characterised by $\mathcal{Q}_1$, i.e.,~the existence of a matrix of the form of Eq.~\eqref{level1} that is PSD \cite{Navascues2010}.
However, macroscopically local correlations are insufficiently restrictive to capture the limitations of quantum theory.
From a physical point of view, this follows, for instance, from the fact that such correlations violate \cite{Cavalcanti2010} the principle of Information Causality which quantum theory is known to obey \cite{Pawlowski2009}.
Alternatively, this same follows immediately from the fact that in general $\mathcal{Q}_1 \neq \mathcal{Q}$.

A more precise constraint on the quantum set is known as almost-quantum correlations \cite{Navascues2015}.
These correlations satisfy several established elementary principles\footnote{However, almost-quantum correlations distinguish between measurements that are mathematically well-defined and measurements that are physically allowed, i.e.,~they violate the so-called no-restriction hypothesis \cite{Sainz2018}.} in addition to Macroscopic Locality, namely no trivial communication complexity \cite{,vanDam1999,Brassard2006}, no advantage for nonlocal computation \cite{Linden2007} and local orthogonality \cite{Fritz2013}.
Almost-quantum correlations are those that can be written in the form $p(a,b|x,y)=\bracket{\psi}{\tilde{A}_{a|x}\tilde{B}_{b|y}}{\psi}$ where $\sum_a \tilde{A}_{a|x}=\sum_b \tilde{B}_{b|y}=\id$ (normalisation) and $A_{a|x}B_{b|y}\ket{\psi}=B_{b|y}A_{a|x}\ket{\psi}$ for all $a$, $x$, $b$, and $y$.
Interestingly, this natural relaxation of quantum theory admits a simple SDP characterisation that is equivalent to choosing a monomial indexing set $\mathcal{S}_{1+AB}=\{\id,\mathbf{A},\mathbf{B},\mathbf{A}\times \mathbf{B}\}$ in the NPA hierarchy \cite{Navascues2015}.
This is a level that is intermediate between
$k=1$ and $k=2$ and it is often referred to as level ``$1+AB$'' (see the remark in Section~\ref{rem:monomial-sets}).

For practical purposes, an important problem is the speed of convergence of the NPA hierarchy.
In bipartite Bell scenarios with a small number of inputs and outputs, it has been systematically investigated by empirical means in \citet{Lin2022naturallyrestricted}.
Typically, almost-quantum correlations are significantly more constraining than macroscopically local correlations, and the relative difference often increases with the number of outputs.

\subsubsection{Tsirelson bounds}
A natural application of the NPA hierarchy is to compute bounds on the optimal quantum violation of Bell inequalities.
This value is known as the Tsirelson bound, named after Boris Tsirelson's analytical derivation \cite{Tsirelson1980, Tsirelson1993} of the maximal quantum violation of the CHSH inequality \cite{Clauser1969}.
However, with the exception of particularly convenient families of Bell inequalities (some examples are found for instance in \citet{Epping2013, Salavrakos2017, Tavakoli2021MUB, Augusiak2019}) the Tsirelson bound is too difficult to determine analytically.
Therefore, the NPA hierarchy provides a practically viable approach to bounding it.
Many concrete instances of Bell inequalities have been analysed by means of the NPA hierarchy, for example in the context of noise tolerance of nonlocality \cite{Vertesi2008, Tavakoli2020platonicsolids, Lin2022naturallyrestricted}, many outcomes in Bell tests \cite{Liang2009, TavakoliMarques2016, Tavakoli2017}, multipartite Bell tests \cite{Vertesi2011b, Grandjean2012, Lopez2016, Vallins2017}, Bell tests with additional constraints \cite{Bernards2020}, nonlocal games with conflicting party interests \cite{Pappa2015} and the detection loophole of Bell experiments \cite{Branciard2011, VertesiPironio2010, Szangolies2017, Cope2021}.

Interestingly, for a simple but large class of Bell inequalities, the Tsirelson bound is guaranteed \cite{Navascues2010} to coincide with the bound associated to the first level of the NPA hierarchy (i.e.,~Macroscopic Locality, $\mathcal{Q}_1$).
These Bell tests have binary outputs for Alice and Bob, and correspond to Bell functionals of the form $\sum_{x,y} c_{xy}\expect{A_xB_y}$ where $c_{xy}$ are arbitrary real coefficients and $A_x\equiv A_{1|x}-A_{2|x}$ and $B_y\equiv B_{1|y}-B_{2|y}$ are observables for Alice and Bob.
The quantum model corresponding to the value associated to $\mathcal{Q}_1$ is known as the Tsirelson construction.
This construction stipulates that for any set of dichotomic quantum observables one can find unit vectors $\vec{u}_x,\vec{v}_y\in\mathbb{R}^{X+Y}$ such that $\expect{A_xB_y}=\vec{u}^T_x \vec{v}_y$, and conversely for any unit vectors $\vec{u}_x,\vec{v}_y\in\mathbb{R}^{n}$ one can find observables $A_x$ and $B_y$ such that $\expect{A_xB_y}_\psi=\vec{u}^T_x \vec{v}_y$ where $\psi$ is a maximally entangled state of dimension $2^{\lceil \frac{n}{2}\rceil}$ \cite{Tsirelson1980, Tsirelson1987}.
This construction also provides a connection between Tsirelson bounds and the Grothendieck constant \cite{Grothendieck1953}.

The link between Tsirelson's construction and SDPs has been used to derive the Tsirelson bound for the Braunstein-Caves Bell inequalities \cite{Wehner2006} and analyses based on $\mathcal{Q}_1$ has yielded quantum Bell inequalities (i.e.,~inequalities satisfied by all quantum nonlocal correlations) for dichotomic observables \cite{Yang2011,Ishizaka2020,Mikos2023}.
Notably, for the simplest scenario with binary inputs and outputs, a quantum Bell inequality in the spirit of CHSH, which gives a complete characterisation of the two-point correlators, was known well before the advent of SDP relaxation methods \cite{Landau1988, Tsirelson1987, Masanes2003}.
However, an approach based already on $\mathcal{Q}_1$ leads to more accurate characterisations which now also take the marginal probabilities into account.
An example of such a $\mathcal{Q}_1$-based quantum Bell inequality in the spirit of CHSH is
\begin{equation}\label{qbell}
	\arcsin D_{11}+\arcsin D_{12}+\arcsin D_{21}-\arcsin D_{22}\leq \pi,
\end{equation}
where \resizebox{0.88\hsize}{!}{$D_{xy}=\left(\expect{A_xB_y}-\expect{A_x}\expect{B_y}\right)/\sqrt{(1-\expect{A_x}^2)(1-\expect{B_y}^2)}$}.
However, these inequalities are still not tight as there exist quantum correlations in the binary input/output scenario that satisfy Eq.~\eqref{qbell} but are not members of $\mathcal{Q}_2$ \cite{Navascues2007}.
Moreover, it is interesting to note that some more general classes of Bell inequalities, including some with many outputs, can be efficiently approximated by SDP methods without the need for hierarchies \cite{Masanes2005, Kempe2010}.

An illuminating application of the NPA hierarchy is to the Bell inequality known as $I_{3322}$ \cite{Froissart1981, Sliwa2003, Collins2004}.
It pertains to the second simplest Bell scenario: it is the only non-trivial facet of the local polytope for the bipartite Bell scenario with three inputs and two outputs (other than a lifting\footnote{Every Bell inequality that is a facet for a given number of inputs and outputs can be lifted to a facet when the number of parties, inputs or outputs is increased \cite{Pironio2005}.} of CHSH).
It can be written as
\begin{equation}
	\begin{aligned}
		I_{3322} = -p_{11} - p_{22} - p_{12} - p_{21} - p_{13} - p_{31}& \\ + p_{23} + p_{32} + p^A_1 + p^B_1 &\leq 1,
	\end{aligned}
\end{equation}
where we write $p_{xy} = p(1,1|x,y)$.
While the Tsirelson bound of the simplest Bell inequality, namely CHSH, is straightforward to obtain analytically, the opposite is the case for $I_{3322}$.
If one restricts to qubit systems, the maximal violation is $I_{3322}=\frac{5}{4}$ but the seesaw heuristic (discussed at the end of Section \ref{sec:introduction:entanglement}) has revealed that larger violations are possible by employing higher-dimensional systems.
In fact, it is conjectured that the Tsirelson bound is saturated only by an infinite-dimensional quantum state \cite{Pal2010}.
Upper bounds to the Tsirelson bound of $I_{3322}$ have been computed via the NPA hierarchy.
These are illustrated in Table~\ref{TableI3322}.
Relaxations up to $\mathcal{Q}_3$ were evaluated in \citet{Navascues2008}.
\citet{Pal2009, Pal2010} then computed $\mathcal{Q}_4$, and \citet{Rosset2018} evaluated $\mathcal{Q}_5$.
The computational requirements scale rapidly in the relaxation level, as is typically the case with SDP relaxation hierarchies, whereas the bounds rapidly converge.
The results for $\mathcal{Q}_4$ and $\mathcal{Q}_5$ are identical up to at least 17 digits, and match the best known quantum violation of $I_{3322}$ to within $10^{-16}$ \cite{Pal2010}.

\begin{table}
	\begin{tabular}{|c|c|c|}
		\hline
		Relaxation level & Value of SDP & Size of SDP matrix \\ \hline
		1        & 1.375 000 00 & 7         \\ \hline
		1+AB       & 1.251 470 90 & 16         \\ \hline
		2        & 1.250 939 72 & 28         \\ \hline
		3        & 1.250 875 56 & 88         \\ \hline
		4        & 1.250 875 38 & 244        \\ \hline
		5        & 1.250 875 38 & 628        \\ \hline
	\end{tabular}
	\caption{Upper bounds on the Tsirelson bound of the $I_{3322}$ Bell inequality for different levels of the NPA hierarchy, along with the size of the corresponding SDP matrix.}\label{TableI3322}
\end{table}

\subsection{Device-independent certification}\label{sec:selftesting}
Device-independent quantum information is the study of quantum information protocols executed under minimal assumptions \cite{Pironio2016}.
This typically amounts to assuming only the validity of quantum mechanics in otherwise uncharacterised experiments, or sometimes even just the no-signaling principle.
Here, we consider the former assumption and discuss how SDP relaxation methods for quantum nonlocality can be employed to device-independently certify properties of the underlying quantum systems.

\subsubsection{Self-testing}\label{sec:selftest}
Self-testing is a sophisticated form of quantum certification where, ideally, Alice and Bob are able to pinpoint their shared quantum state and measurements only from examining their nonlocal correlations \cite{Mayers2004}.
Naturally, these cannot be precisely deduced because quantum correlations in Bell scenarios are invariant under collective changes of reference frame.
Hence, at best they can be determined up to local transformations that leave the correlations invariant.
Such transformations are those that preserve inner products, and are called isometries\footnote{An isometry can be seen as a linear map that consists of a possible appending of additional degrees of freedom to the system followed by a unitary transformation. They are defined as linear operators $V$ satisfying $V^\dagger V = \id$.}.
A simple example is that from any correlations achieving the Tsirelson bound of the CHSH inequality one can deduce that the state is a singlet up to local isometries \cite{Summers1987, Braunstein1992, Popescu1992, Tsirelson1993}.
For a review of self-testing, we refer the reader to \citet{Supic2020selftestingof}.

SDP techniques offer a powerful approach to self-testing.
To showcase this, let us first define an operator 
\begin{equation}\label{constructselftest}
	\mathcal{B}=\beta_Q\id - \sum_{a,b,x,y}c_{abxy}A_{a|x}\otimes B_{b|y}.
\end{equation}
The second term is the Bell operator associated to a quantum model for a generic Bell functional.
$\beta_Q$ denotes the Tsirelson bound of that Bell inequality, i.e.,~the maximal quantum value of the Bell parameter.
Hence, one can think of Eq.~\eqref{constructselftest} as a shifted Bell operator tailored such that $\bracket{\psi}{\mathcal{B}}{\psi}\geq 0$ for all quantum states, i.e.,~the operator is PSD.
Assume now that we are able to find a decomposition of $\mathcal{B}$ as a sum-of-squares of some operators $\{P_l\}$,
\begin{equation}\label{SOSdecomp}
	\mathcal{B}=\sum_l P_l^\dagger P_l,
\end{equation}
where $P_l$ are some polynomials of the local measurement operators $\{A_{a|x}\} $ and $\{B_{b|y}\}$.
Witnessing the Tsirelson bound, namely $\bracket{\psi}{\mathcal{B}}{\psi}=0$, therefore implies that $P_l\ket{\psi}=0$ for all $l$.
These relations can be very useful for deducing properties of the local measurements \cite{Yang2013,Bamps2015}.
Take for example the CHSH inequality, for which finding an SOS decomposition for Eq.~\eqref{constructselftest} is particularly simple: working with the observables instead of the POVM elements, one can choose $P_1=\frac{A_1+A_2}{\sqrt{2}}-B_1$ and $P_2=\frac{A_1-A_2}{\sqrt{2}}-B_2$.
Following the given procedure, one can deduce that $\{A_1,A_2\}\ket{\psi}=\{B_1,B_2\}\ket{\psi}=0$, i.e.,~the local measurements must anticommute on the support of the state.
One can then take this further and leverage these relations together with a well-chosen local isometry to deduce also the shared state.

The key component in this discussion is to first find the Tsirelson bound $\beta_Q$ and then find an SOS decomposition of the form of Eq.~\eqref{SOSdecomp}.
By considering a sufficiently high level of the NPA hierarchy, one can often recover $\beta_Q$.
To verify that the bound returned by the NPA hierarchy is optimal, one can for example match it with an explicit quantum strategy for the Bell test.
Then, an SOS decomposition can also be extracted.
As we have seen in Section~\ref{sec:NoncommutativePoly}, such SOS decompositions correspond to the dual of a noncommutative polynomial optimisation problem.
By considering both primals and duals \cite{Doherty2008b} of the NPA hierarchy, one can systematically approach the problem.
Indeed, this can be done analytically even for some Bell inequalities by identifying a suitable relaxation level \cite{Yang2013,Bamps2015}.
In particular, explicit SOS decompositions have been reported for the tilted CHSH inequalities \cite{Yang2013, Bamps2015}, a three party facet Bell inequality without a quantum violation \cite{Almeida2010}, multi-outcome generalisations of the CHSH inequality \cite{Kaniewski2019maximalnonlocality}, chained Bell inequalities \cite{Supic2016} and Bell inequalities tailored for graph states \cite{Baccari2020}.
Further methods have also been developed to generate Bell inequalities that will always admit SOS decompositions~\cite{Barizien2023}, resulting in a method to design Bell inequalities to self-test a given quantum state.
Notably, however, this general SDP approach is not guaranteed to lead to a self-testing statement for both states and measurements.

When the Bell inequality violation is non-maximal, self-testing is also possible, albeit with other techniques.
A useful approach employs SDP methods to place a lower bound on the fidelity of the state with the ideal state that would have been certified had the Tsirelson bound been reached.
This is achieved using the swap method \cite{Yang2014, Bancal2015}.
The main idea can be illustrated for the case of CHSH.
Since CHSH targets a two-qubit state in the registers $A$ and $B$, we can introduce qubit ancillas $A'$ and $B'$ into which the parties aim to swap their state.
To do this, they need to individually use the swap operator.
For Alice the swap operator can be written as $S_{AA'}=WVW$, where $W=\id\otimes\ketbra{0}{0}+\sigma_X\otimes \ketbra{1}{1}$ and $V=\ketbra{0}{0}\otimes \id+\ketbra{1}{1}\otimes \sigma_X$ are \textsc{cnot} gates.
Naturally, one cannot assume the specific operation on system $A$ because of the device-independent picture, but can instead try to emulate the swap operator by using an operation that on $A$ depends only on Alice's measurements.
While emulation is not unique and less immediate approaches can enhance the results, a concrete example is instructive.
Knowing that the local measurements are ideally qubits and anticommuting, it is reasonable to target a correspondence of $A_1$ and $B_1$ to $\sigma_Z$ and $A_2$ and $B_2$ to $\sigma_X$.
The optimal maximally entangled state is then a local rotation of the singlet $\ket{\psi^-_\text{target}}$.
Hence, the emulated swap operator corresponds to $W=\id\otimes\ketbra{0}{0}+A_2\otimes \ketbra{1}{1}$ and $V=\frac{\id+A_1}{2}\otimes \id+\frac{\id-A_1}{2}\otimes \sigma_X$ for Alice and analogously for Bob.
The swapped state is
\begin{align}
	&\rho'_{A'B'}=\tr_{AB}\left(S\psi_{AB}\otimes \ketbra{0}{0}_{A'}\otimes \ketbra{0}{0}_{B'}S^\dagger \right),
\end{align}
where $S=S_{AA'}\otimes S_{BB'}$.
Here, $\rho'_{A'B'}$ is a $4\times4$ matrix whose entries are linear combinations of moments (recall Eq.~\eqref{momentmatrixNPA}).
Its fidelity with the target state, $F=\bracket{\psi^-_\text{target}}{\rho'}{\psi^-_{\text{target}}}$, is therefore also a linear combination of moments.
Thus, if we relax the quantum set of correlations into a moment matrix problem \`a la NPA, we can view the fidelity as a linear objective and thus obtain a robust self-testing bound via SDP.
This applies also to other Bell inequalities and to other constructions of the swap operator.
The SDP relaxation for the fidelity, at the $k$-th level of the NPA hierarchy, becomes
\begin{equation}\label{swapmethod}
	\begin{aligned}
		\min_{\Gamma_k} \quad & F(\Gamma_k)\\
		\st \quad&\sum_{a,b,x,y} c_{abxy}\Gamma_k(A_{a|x},B_{b|y})=\beta, \\
		& \Gamma_k\succeq 0,
	\end{aligned}
\end{equation}
where $\beta \leq \beta_Q$ is the witnessed Bell parameter.
A practically useful relaxation typically requires selected monomials from different levels (i.e.,~an intermediate level, recall the remark in Section~\ref{rem:monomial-sets}) in order to ensure that all moments appearing in $F$ also appear in the moment matrix.
An important subtlety is that for some Bell scenarios and choices for emulating the swap operator, the operation may cease to be unitary in general.
This can be remedied \cite{Bancal2015} by the introduction of localising matrices in the SDP relaxation \eqref{swapmethod}.
In the literature, the swap method has been applied to noisy self-testing of partially entangled two-qubit states \cite{Bancal2015}, three-dimensional states \cite{Salavrakos2017}, the three-qubit $W$ state \cite{Wu2014}, four-qubit GHZ and cluster states \cite{Pal2014} and symmetric three-qubit states \cite{Li2020}.
The above relaxation naturally provides an upper bound on the minimal Bell inequality violation required to certify a non-trivial fidelity with $\ket{\psi^-_\text{target}}$.
Complementing this, SDPs were used in~\citet{Valcarce2020} to search for systems producing Bell inequality violations with trivial fidelities, thereby providing lower bounds on the minimal CHSH violation needed to obtain non-trivial robust self-testing statements.

\subsubsection{Entanglement dimension}\label{sec:entdim}
Hilbert space dimension roughly represents the number of controlled degrees of freedom in a physical system.
It is therefore unsurprising that it plays a significant role in quantum nonlocality: by creating entanglement in higher dimensions one can potentially increase the magnitude of a Bell inequality violation, and sometimes quantum correlations even necessitate infinite dimensions \cite{Coladangelo2018, Beigi2021separationof}.
We now discuss different approaches to characterising the set of quantum nonlocal correlations when states and measurements are limited to a fixed dimension $d$.

Dimensionally restricted quantum nonlocality can be linked to the separability problem of quantum states \cite{Navascues2014}.
To see the connection, let Alice and Bob share a two-qubit ($d=2$) state $\rho_{AB}$ on which they perform basis projections $\{\ketbra{a_x}{a_x},\id-\ketbra{a_x}{a_x}\}$ and $\{\ketbra{b_y}{b_y},\id-\ketbra{b_y}{b_y}\}$ respectively.
One can now shift the status of these operators from measurement projections to ancillary states.
To do that, we append the main registers $A$ and $B$ with additional $X$- and $Y$-qubit registers respectively, $A_1,\ldots,A_X$ and $B_1,\ldots,B_Y$, and define the $(2+X+Y)$-qubit state $\sigma=\rho_{AB}\bigotimes_{x=1}^X \ketbra{a_x}{a_x}\bigotimes_{y=1}^Y \ketbra{b_y}{b_y}$.
The quantum correlations can be recovered by Alice (Bob) applying operators $G_{a,x}$ ($H_{b,y}$), that act as the identity on all ancillary registers except $A_x$ ($B_y$), which instead are swapped with system $A$ ($B$) via the two-qubit swap operator $S$ for outcome $a\,{=}\,1$, and respectively its complement $\id\,{-}\,S$ for outcome $a\,{=}\,2$.
This yields $p(a,b|x,y)=\tr\left(\sigma G_{a,x}\otimes H_{b,y}\right)$.
The optimum of any Bell parameter for qubits is then converted into a type of entanglement witnessing problem where one must evaluate the optimum of $\tr\left(\sigma \mathcal{B}\right)$ for a known Bell operator, $\mathcal{B}=\sum_{a,b,x,y}c_{abxy}G_{a,x}\otimes H_{b,y}$, over a multiqubit state $\sigma$ that is separable with respect to the partition $AB|A_1|\ldots|A_X|B_1|\ldots|B_Y$ where all ancillary registers are separated from each other and from the joint register $AB$.
As discussed in Section \ref{sec:DPS}, this problem can be addressed systematically by the DPS hierarchy.
This method has also been extended to scenarios with more outcomes and parties.
However, it is useful mainly when the number of inputs is small owing to the increasing number of subsystems in $\sigma$.
It can be found to be more efficient when only some parties have a restricted dimension, since the other parties then can be treated as in the NPA hierarchy.

Another way to link fixed-dimensional quantum nonlocality problems to the separability problem is proposed in \citet{Jee2021}.
This method has the advantage of coming with good bounds on the convergence rate of the resulting SDP relaxations, and the disadvantage of poor performance in practice.
In the particular case of free games, i.e., nonlocal games where the probability distribution over the inputs is a product between a distribution for Alice and another for Bob, the complexity of computing an $\epsilon$-close approximation to the $d$-dimensional Tsirelson bound scales polynomially in the input size and quasi-polynomially in the output size.
In the general case the complexity is still quasi-polynomial in the output size, but becomes exponential in the input size.

The problem can also be approached without connecting it to a separability problem and instead adding dimension constraints directly to the NPA moment matrix.
This may consist of identifying operator equalities that hold only up to dimension $d$.
For example, the identity $[X_1,[X_2,X_3]^2]=0$ holds for all complex square matrices of dimension $d \le 2$.
However, this is difficult to do in practice since a complete set of operator equalities is not known for $d \ge 3$.
A handy alternative is instead to implicitly capture the constraints associated with a dimensional restriction, on the level of the NPA moment matrix, by employing numerical sampling in the $d$-dimensional space \cite{Navascues2015b}.
See Section~\ref{sec:NV} for more details.
This method is known to converge to the quantum set of correlations and is often useful for practical purposes when problems are not too large \cite{Navascues2015c}.

For the special case of restricting to bipartite maximally entangled states of dimension $d$, it is possible to to construct a sampling-free SDP relaxation hierarchy based on tracial moments of measurement operators that act only on a single Hilbert space \cite{Lang2014}.
This is a consequence of the identity $\tr\left(A \otimes B \phi^+_d\right)=\frac{1}{d}\tr\left(A B^T \right)$.
Systematic implementations based on projective measurements are reported in \citet{Lin2022naturallyrestricted}.

\subsubsection{Entanglement certification}
\label{sec:nonlocality:di:moroder}
Quantum nonlocality, understood as the absence of an explanation of correlations in terms of an LHV model \eqref{LHV}, implies entanglement and therefore allows for device-independent entanglement certification.
This is an inference of entanglement without any modeling of the experimental measurement apparatus.
Fundamentally, this black-box approach to entanglement comes at the cost of some entangled states not being detectable \cite{Werner1989, Augusiak2014}, although this can be at least partly remedied by considering more complicated nonlocality experiments, see e.g.~\citet{Sende2005, Cavalcanti2011, Bowles2018, Bowles2021}.
Nevertheless, a variety of interesting entangled states can still be certified, typically those that are not too noisy, and SDPs offer a powerful path for that purpose.

Local measurements performed on an $n$-partite fully separable quantum system always yield local correlations.
Hence, one can certify entanglement device-independently for a given correlation $p$ by evaluating the LP in Eq.~\eqref{LHVLP} that checks its membership to the local polytope.
However, this is demanding because the number of variables in the LP scales exponentially in the number of parties and inputs, and polynomially in the number of outputs.
Using simplex methods for linear programming, states of up to $n=7$ qubits have been certified in \citet{Gruca2010}.
This is increased to $n=11$ qubits in \citet{Gondzio2014} by adopting a matrix-free approach for interior-point solvers, which reduces memory requirements \cite{Gondzio2012}.

However, one can go further by considering SDP relaxations of the local polytope.
A key observation is that any local correlation between $n$ parties can be obtained from locally commuting measurements on a quantum state.
Take the fully separable state $\rho=\sum_\lambda p(\lambda)\ketbra{\lambda}{\lambda}^{\otimes n}$ and let the $x_l$-th POVM of the $l$-th party be $M_{x_l}^{a_l}=\sum_{\mu} D(a_l| x_l, \mu) \ketbra{\mu}{\mu}$, where $D(a_l|x_l,\mu)$ is a deterministic distribution.
All these POVMs commute.
Evaluating the Born rule, one finds the generic local model given in Eq.~\eqref{LHV}.
Thus, a sufficient condition for nonlocality is that $p$ fails some level of the NPA hierarchy under the extra constraint that all local measurements commute.
The latter appears in the form of additional equality constraints between the elements of the NPA moment matrix of Eq.~\eqref{momentmatrixNPA} that effectively turn it into a Lasserre hierarchy (recall Section~\ref{sec:lasserre}).
Using the second level relaxation\footnote{Note that local commutation is a trivial constraint at the first level of relaxation, i.e.,~the associated correlation set is still $\mathcal{Q}_1$.} of the commuting NPA hierarchy, nonlocality has been reported\footnote{For some states one requires small additions to the second level but these can be independent of the number of qubits.} for W states, GHZ states, and graph states, reaching up to $n=29$ qubits \cite{Baccari2017}.
A similar approach can also be used to certify entanglement in the steering scenario.
On Alice's side one imposes commutation in the NPA relaxation whereas on Bob's side one replaces the unknown measurements with known measurements and uses their algebraic relations to further constrain the moment matrix \cite{Kogias2015}.
A simple example of the latter is to assume that Bob's measurements are qubit and mutually unbiased, and impose it on the moment matrix by having the corresponding observables anticommute.

A natural next question is how one can device-independently quantify entanglement.
The main intuition is that a stronger violation of a Bell inequality ought to require stronger forms of entanglement.
This question can be addressed through a reinterpretation of the NPA hierarchy of \citet{Moroder2013}.
Consider that we apply local completely positive maps, $\Lambda_A:\mathcal{H}_A\rightarrow \mathcal{H}_{A'}$ and $\Lambda_B: \mathcal{H}_B\rightarrow \mathcal{H}_{B'}$, to a bipartite state $\rho_{AB}$.
Their action can be represented in terms of unnormalised Kraus operators, i.e.,~a set of operators $\{K_{i,A}\}_i$ and $\{K_{i,B}\}_i$.
Now, let us put Alice's and Bob's local POVM elements in lists $\mathbf{A}=\{\id, A_{1|1},\ldots, A_{N|X}\}$ and $\mathbf{B}=\{\id, B_{1|1},\ldots, B_{M|Y}\}$ respectively.
Then, we define Kraus operators of Alice as $K_{i,A}=\sum_{l_1,\ldots, l_k}\ketbra{l_1 \ldots l_k}{i}\mathbf{A}_{l_1}\ldots \mathbf{A}_{l_k}$ and analogously for Bob, where the index $k$ is the level of the hierarchy.
The moment matrix, $\Gamma=\Lambda_A\otimes \Lambda_B[\rho]$ becomes 
\begin{equation}
	\Gamma=\sum_{\bar{r},\bar{l}}\sum_{\bar{s},\bar{k}} \tr\left(\rho \mathcal{A}_{\bar{r},\bar{l}}\otimes \mathcal{B}_{\bar{s},\bar{k}}\right)\ketbra{\bar{l},\bar{k}}{\bar{r},\bar{s}},
\end{equation}
where $\bar{l}=(l_1,\ldots,l_k)$ and similarly for $\bar{k}$, $\bar{r}$ and $\bar{s}$, and where $\mathcal{A}_{\bar{r},\bar{l}}=(\mathbf{A}_{r_1}\ldots \mathbf{A}_{r_k})^\dagger \mathbf{A}_{l_1}\ldots \mathbf{A}_{l_k}$ and similarly for $\mathcal{B}_{\bar{s},\bar{k}}$.
This moment matrix would be the same as that of the NPA hierarchy if instead of local completely positive maps we had opted for global completely positive maps.
The advantage of this formulation is that it comes with an explicit bipartition on the level of the moment matrix.
For instance, this allows for imposing the PPT constraint, which is needed for device-independent entanglement quantification via the entanglement negativity measure.
The negativity, $N(\rho_{AB})$, is defined as the sum of the negative eigenvalues of $\rho_{AB}^{T_A}$, which can itself be cast as an SDP, see Eq.~\eqref{eq:negativity}.
On the level of the moment matrix, the SDP becomes: $N(\rho_{AB})=\min \tr\left(\chi_-\right)$ such that $\rho=\chi_+-\chi_-$ where $(\chi_\pm)^{T_A}\succeq 0$.
Thus, the negativity can be bounded by employing the above SDP relaxation for both the operators $\chi_\pm$, imposing their PPT property on $\Gamma$ and noting that the objective function is simply an element of the moment matrix.
The idea of building a moment matrix featuring a bipartition can also be extended to the steering scenario \cite{Pusey2013}.
This allows for one-side device-independent quantification of entanglement, which has also been considered for highly symmetric scenarios with multiple outcomes\cite{Huang2021}.

While the negativity is relevant as one possible quantifier of bipartite entanglement, other approaches are needed for multipartite entanglement.
If we have a multipartite quantum state, a natural question is to ask for the smallest cluster of subsystems that must be entangled in order to model the correlations $p$.
The smallest number of entangled subsystems required, $D$, is known as the entanglement depth \cite{Sorensen2001}.
Nonlocality alone gives only a device-independent certificate that some entanglement is present, i.e.,~that $D\geq 2$.
It was first noted in \citet{Bancal2011} that the NPA hierarchy with suitably imposed local measurement-commutation relations can be used to detect a maximal entanglement depth of $D=3$ in a three-partite system.
For systems of more particles, one can employ the hierarchy of Moroder \textit{et~al.} to relax separability across a given bisection of the subsystems to a PPT condition and use that to bound the entanglement depth \cite{Liang2015, Lin2019}.

Furthermore, by restricting to few-body correlators that are symmetric under permutations of parties, one can formulate a hierarchy of SDP relaxations of the corresponding party-permutation-invariant local polytope that benefits considerably from the imposed symmetry.
This is showcased in \citet{Fadel2017} where the local polytope is first relaxed to a semi-algebraic set, and then it is leveraged that such sets can be relaxed to SDPs \cite{Gouveia2010, GouveiaBook}.
The advantage of this approach is that the number of subsystems is featured as an explicit parameter in the SDP and does not impact the size of the moment matrix.
In this way, one can obtain party-permutation-invariant Bell inequalities for any number of parties via the dual SDP.
Permutation-invariant Bell expressions based on two-body correlators have been used to build device-independent witnesses of entanglement depth using relaxations to PPT conditions \cite{Aloy2019, Tura2019}.

\subsubsection{Joint measurability}\label{secJM}
It has been known since the early development of quantum theory that the values of some sets of measurements, such as position and momentum, cannot be simultaneously known~\cite{Heisenberg1925}.
This fundamental feature of quantum theory, known as measurement incompatibility, has been at the forefront of significant research and development within quantum theory~\cite{Heinosaari2016, Guhne2021}.

Formally, given a collection of POVMs $\{A_{a|x}\}_a$ indexed by some $x$, we say the collection is \emph{compatible} or \emph{jointly measurable} if there exist a parent POVM, $\{M_\lambda\}_\lambda$, and a conditional probability distribution, $p(a|x,\lambda)$, such that 
\begin{equation}\label{eq:compatibility_cond}
	A_{a|x} = \sum_{\lambda} p(a|x,\lambda) M_\lambda
\end{equation}
for all $a$ and $x$.
Operationally, the statistics of compatible measurements can be simulated by measuring the ``parent'' measurement $\{M_{\lambda}\}_\lambda$ and then  postprocessing the results.
If such a decomposition does not exist then we say that the measurements are incompatible.

It is well known that incompatible measurements are necessary for Bell nonlocality~\cite{Fine1982}, although not sufficient~\cite{Bene2018,Hirsch2018}.
However, when a party in a Bell test has access to more than two measurements it is possible that some subsets of their measurements are compatible whilst the entire set of measurements remains incompatible.
A collection of subsets for which the measurements are compatible is referred to as a compatibility structure.
In~\citet{Quintino2019} the authors investigate how different compatibility structures can be device-independently ruled out by large enough Bell inequality violations.
From the perspective of nonlocality, if the behaviour is restricted to the subsets of inputs for which the measurements are compatible then it necessarily becomes local.
Thus by combining linear programming constraints for compatible subsets~(see~Eq.~\eqref{LHVLP}) with the NPA hierarchy it is possible to explore relaxations of the sets of correlations with different compatibility structures and in turn find Bell-like inequalities that rule out different structures in a device-independent manner.

Once the presence of measurement incompatibility has been device-independently detected, a natural follow-up question is whether the degree of incompatibility can be quantified.
One such measure of incompatibility is the so-called incompatibility robustness~\cite{Haapasalo2015,Uola2015}.
For a collection of measurements $\{A_{a|x}\}_{a,x}$ this is defined as
\begin{equation}\label{incomprobustness}
	\begin{aligned}
		\min & \quad t \\
		\st & \quad \left\{\tfrac{1}{1+t}(A_{a|x} + t N_{a|x})\right\}_{a,x} \quad \text{are compatible}, \\
		& \quad \sum_{a} N_{a|x} = \id \qquad \forall\, x, \\
		& \quad N_{a|x} \succeq 0 \qquad \quad\hspace{0.9em} \forall\, a,x,\\
		& \quad t\geq 0,
	\end{aligned}
\end{equation}
which roughly captures the amount of noise, represented by a shift towards another set of POVMs $N_{a|x}$, that one needs to add to the measurements $A_{a|x}$ in order to make them compatible.
There are many other such measures of incompatibility and these measures can often themselves be expressed as SDPs~\cite{Guhne2021}.
This includes the incompatibility robustness which can be computed by the SDP
\begin{equation}
	\begin{aligned}
		\min & \quad \frac1d \tr(M_\lambda) \\
		\st & \quad \sum_\lambda D(a|x,\lambda) G_\lambda \succeq A_{a|x} \quad \forall\,a,\,x, \\
		& \quad \sum_\lambda G_\lambda = \id \frac1d \sum_\lambda \tr(G_\lambda), \\
		& \quad G_\lambda \succeq 0, 
	\end{aligned}
\end{equation}
where $D(a|x,\lambda)$ are deterministic distributions (recall Eq.~\eqref{lhsprimal}).

In~\citet{Chen2016} the authors relate the incompatibility problem to a steering problem to show that the incompatibility robustness can be lower bounded by a steering robustness quantity, which is an analogous quantity for quantifying steerability.
A hierarchy of SDP device-independent lower bounds on the latter quantity can then be derived to give a method to compute device-independent lower bounds on incompatibility.
Several additional lower bounds on the incompatibility robustness in terms of other robustness quantities were provided in~\citet{Cavalcanti2016b}, e.g., the consistent nonlocal robustness, which can similarly be turned into device-independent lower bounds on the incompatibility robustness using the NPA hierarchy.
This work, which surveys many robustness measures and their computability via SDPs, also introduces a new quantity called the consistent steering robustness, which provides tighter lower bounds on the incompatibility robustness than the standard robustness of steering.
By combining this with moment matrix techniques developed in~\citet{Chen2016}, even stronger device-independent bounds on the incompatibility robustness can be obtained which in some cases can even be shown to be tight~\cite{Chen2018}.
Later, a general method to obtain device-independent bounds on SDP representable incompatibility measures was presented in~\citet{Chen2021}.
By avoiding proxy quantities, this provides a much stronger characterisation of the incompatibility robustness and can even be shown to be tight for the correlations achieving the Tsirelson bounds of the tilted CHSH inequalities.


\section{Quantum communication }
\label{sec:communication}
In this section, we discuss quantum correlations in the prepare-and-measure scenario, introduced in Section~\ref{seccomscenario}, in which Alice prepares messages and Bob measures them.
Such correlations have been studied for several different types of communication and SDP relaxations play a central role in their characterisation and applications.

\subsection{Channel capacities}
\begin{figure}
	\centering
	\includegraphics[width=0.95\columnwidth]{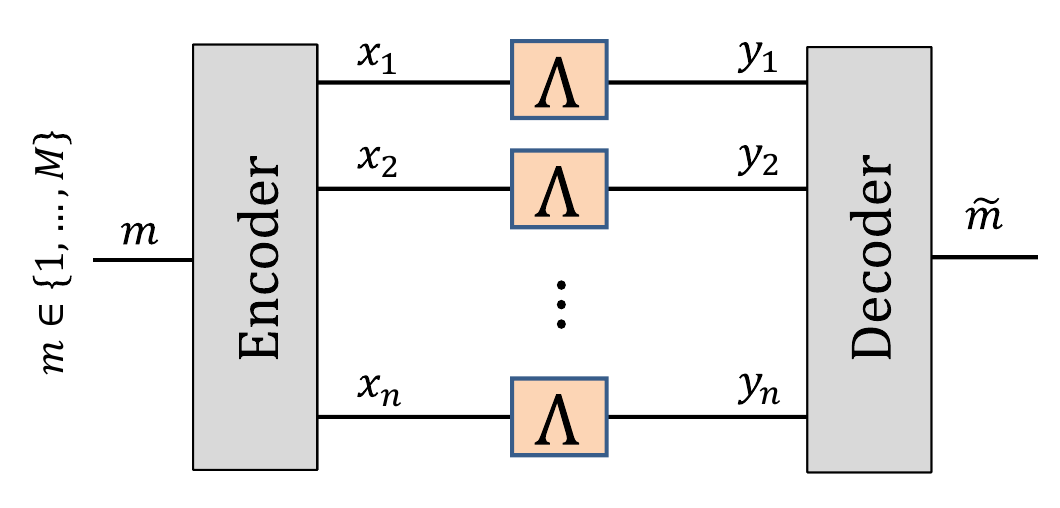}
	\caption{Unassisted classical channel coding scenario.
	A message, $m$, is selected from a given alphabet and encoded into a codeword consisting of $n$ letters.
	Each letter, $x_i$, is then sent through the channel, returning distorted letters, $y_i$, which are decoded into a guess, $\tilde{m}$, of the original message.}
	\label{FigChannelCoding}
\end{figure}

We provide a very brief overview of channel coding before proceeding to the role of SDPs in the topic.
In information theory, a paradigmatic task is to encode a message, send it over a channel, and then reliably decode it.
Specifically, the sender selects a message from an alphabet of size $M$ and encodes it into a codeword consisting of $n$ letters, where $n$ is called the block-length.
Each letter is then sent over the channel, $\Lambda$, which in the classical case can be represented as a conditional probability distribution $p_\Lambda(y|x)$ mapping the input $x$ to the output $y$.
Since the channel is noisy, it outputs a distorted codeword, which the receiver must decode into the original message, see Fig.~\ref{FigChannelCoding}.

The seminal work of Claude Shannon~\cite{Shannon1948} showed how to address the efficiency of the communication when the distributions of the letters in the codeword are independent and identical.
The key idea of Shannon was to allow a small error probability in the decoding, which then tends to zero in the limit of large $n$.
In this setting, when a memoryless noisy classical channel is used asymptotically many times, it is natural to consider the largest rate, $R=\frac{\log_2 M}{n}$, i.e.,~the ratio of the number of message bits and the number of channel uses, at which information can reliably be transmitted.
This rate, $\mathcal{C}_{cc}(\Lambda)$, is called the capacity of the channel and Shannon proved that it is given by the largest single-copy mutual information between the channel's input and output \cite{Shannon1948},
\begin{equation}\label{ShannonCapacity}
	\mathcal{C}_{cc}(\Lambda)=\max_{\{p_x\}} I(X;Y),
\end{equation} 
where $I(X;Y)=H(X)+H(Y)-H(X,Y)$, and $H$ denotes the Shannon entropy.

A natural endeavour is to extend this type of question to scenarios with quantum resources; see e.g.~ \citet{Holevo2012, Wilde2013, Gyongyosi2018} for a thorough discussion and \citet{Holevo2020} for a brief overview.
The Holevo-Schumacher-Westmoreland theorem \cite{Schumacher1997, Holevo1998} generalises Eq.~\eqref{ShannonCapacity} to the scenario where the channel instead is quantum, i.e.,~the message is encoded in a quantum state.
The classical capacity of a quantum channel is given by
\begin{equation}\label{HSW}
	\mathcal{C}_{cq}(\Lambda)=\lim_{n\rightarrow \infty} \frac{\chi(\Lambda^{\otimes n})}{n},
\end{equation}
where $\chi(\Lambda)=\max_{\{p_x\}, \{\rho_x\}} H\big(\sum_x p_x \Lambda(\rho_x)\big)-\sum_x p_x H\big(\Lambda(\rho_x)\big)$ is called the Holevo capacity of the channel and the maximisation is taken over all input ensembles to the channel.
In contrast to the classical case, the possibility of entanglement in the quantum codeword implies that the Holevo capacity is not additive\footnote{The failure of additivity for the Holevo capacity implies that several other entropic quantities are also not additive \cite{Shor2004}.} \cite{Hastings2009} and hence Eq.~\eqref{HSW} cannot in general be reduced to a single-letter formula, i.e.,~a closed expression based on a single use of $\Lambda$, but exceptions are known for convenient special cases; see e.g.~\citet{King2003, King2002, Bennett1997}.
Non-additivity makes the computation of $\mathcal{C}_{cq}(\Lambda)$ very difficult.

A contrasting situation is when the sender and receiver additionally are allowed to share entanglement.
Then, the resulting entanglement-assisted classical capacity of the quantum channel admits an elegant single-letter formula similar to Eq.~\eqref{ShannonCapacity}, but instead of maximising the classical mutual information one maximises the quantum mutual information of the bipartite state $(\openone\otimes \Lambda)[\rho_{AB}]$ over all entangled states $\rho_{AB}$ \cite{Bennett2002, Bennett1999}.

Another natural scenario is when the message itself is a quantum state.
One then speaks of quantum capacities.
In a general picture, it is possible to characterise the performance of a classical or quantum protocol in terms of the triplet $(R,n,\epsilon)$, where $\epsilon$ is the error tolerated in the decoding.
This error is favourably represented in terms of the fidelity between the maximally entangled state and the state obtained by sending half of it through the communication scheme.
The central question is then to characterise the set $(R,n,\epsilon)$ that is achievable for a given quantum channel.
In the asymptotic setting ($n\rightarrow \infty$) and independent channel uses, the quantum capacity of the quantum channel, $\mathcal{C}_{qq}(\Lambda)$, is the largest rate $R$ at which the error tends to zero.
It is given by the Lloyd-Shor-Devetak theorem \cite{Devetak2005b,Lloyd1997} in terms of the largest coherent information\footnote{The coherent information is defined as $I_\text{coh}(\rho_{AB})=H(\rho_B)-H(\rho_{AB})$.} when optimised over all bipartite input states after half of it is passed through the channel.
However, this quantity must be regularised, i.e.,~one must take a many-copy limit analogous to that in Eq.~\eqref{HSW}.
This renders the capacity non-additive and therefore difficult to compute.

\subsubsection{Classical capacities}
In the conventional setting, the decoding errors are tolerated as long as they vanish in the limit of large block-length.
A stricter approach, in which errors are exactly zero for any $n$, is called zero-error coding\footnote{This is not only of independent interest: the zero-error capacity is also relevant for how rapidly the error tends to zero for an increasing block-length in the standard capacity \cite{Shannon1967}.} \cite{Shannon1956}.
In zero-error capacity problems, one needs only to consider whether two distinct messages could be confused with each other after they are sent through the channel.
Therefore, if the channel is used only once, one can represent the zero-error problem as a graph where each vertex represents a message and each edge represents the possibility that two messages can be mapped onto the same output.
The one-shot zero-error capacity is given by the largest set of independent vertices in this graph, known in the literature as the confusability graph.
For larger block-length, one must consider the strong power of the graph.
However, computing the independence number is NP-hard \cite{Karp1972}.
In the celebrated work \citet{Lovasz1979}, it was shown that the independence number of a graph, $G$, can be upper bounded via the so-called Lov\'asz theta function, which admits an SDP formulation 
\begin{equation}\label{LovaszTheta}
	\begin{aligned}
		\vartheta(G)=\max \quad & \tr\left(X E\right) \\
		\st \quad & X_{ij}=0 \quad \text{if $i$ and $j$ are connected}, \\
		\quad & \tr\left(X\right)=1,\\
		\quad & X\succeq 0,
	\end{aligned}
\end{equation}
where $E_{ij}=1$.
The Lov\'asz theta function has the important property that it factors under strong products of graphs, which allows one to address the asymptotic limit for zero-error coding.
It has been proven that this SDP also bounds the entanglement-assisted zero-error classical capacity \cite{Beigi2010b, Duan2013}.
Bounds of this sort are important because it is known that the one-shot zero-error classical capacity can be increased by means of shared entanglement \cite{Cubitt2010}.
In fact, this is connected to proofs of the Kochen-Specker theorem, which in turn can be represented in the language of graph theory \cite{Cabello2014}.
Entanglement-assisted advantages are also possible for asymptotic zero-error coding.
Perhaps surprisingly, the zero-error capacity can sometimes even equal the classical capacity of a quantum channel \eqref{HSW} \cite{Leung2012}.
In contrast, the zero-error problem becomes simpler if the sender and receiver are permitted to share general bipartite no-signaling correlations.
The capacity can then be computed via an LP which corresponds to the fractional packing number of the confusability graph \cite{Cubitt2011}.
This can also be generalised to parties that share quantum no-signaling correlations\footnote{Quantum no-signaling correlations are completely positive and trace-preserving bipartite linear maps that forbid signaling of classical information in either direction.
}: in the one-shot setting, the capacity is given by an SDP and sometimes an SDP can also be formulated for the asympototic capacity.
If quantum no-signaling correlations are permitted, the Lov\'asz theta function corresponds to the smallest zero-error classical capacity of any channel associated with the same confusability graph \cite{Duan2016}.

A related problem considers a one-shot classical channel and a given number of messages and then addresses the largest average success probability with which they can be communicated through the channel.
The answer can be approximated in polynomial time only up to a factor $(1-\frac{1}{e})$, and obtaining a better approximation is NP-hard \cite{Barman2016}.
An upper bound is nevertheless known \cite{Polyanskiy2010}.
Interestingly, this bound admits a nice physical interpretation as it is equivalent to the optimal success probability obtained from assisting the classical channel with bipartite no-signaling correlations \cite{Matthews2012}.
As made intuitive from this connection, the solution can be bounded by means of relaxation to an LP.
Beyond point-to-point channels, the capacity regions of broadcast and multiple-access channels under no-signaling assistance can also be analyzed using linear programming~\cite{Fawzi2023, Fawzi2023b}.
This classical information problem can also be considered in the quantum case.
For a given block-length and a given tolerance for the error, upper bounds on the optimal transmission rate over a general quantum channel (also when assisted by entanglement) can be obtained by means of SDP by relating the problem to hypothesis testing relative entropies \cite{Matthews2014}.
These bounds were later made tighter in \citet{WangDuan2018}, applied also to the entanglement-assisted case, and used to give SDP upper bounds on the classical capacity of the qubit amplitude-damping channel.
For the large block-length limit, namely the asymptotic channel coding scenario, it was shown in \citet{Fawzi2022b} how to compute a sequence of upper bounds on $\mathcal{C}_{cq}$ via a hierarchy of SDP-tailored R\'enyi divergences \cite{Fawzi2021} but its convergence to $\mathcal{C}_{cq}$ is not presently known.
These SDPs are rendered considerably more efficient by invoking symmetry properties, in the spirit of the discussion in Section~\ref{sec:symmetries}.
SDP methods in this vein can sometimes also be used to obtain strong converse bounds\footnote{The strong converse property means that the error probability tends exponentially to $1$ if the rate exceeds the bound. This property is known for some quantum channels; see e.g.~\citet{Winter1999, Ogawa1999, Konig2009b, Wilde2014}.} on the classical capacity \cite{WangDuan2018, Ding2023}.
An alternative strong converse bound is obtained from the quantum reverse Shannon theorem \cite{Bennett2014, Berta2011}: if one sends messages at a rate above the entanglement-assisted classical capacity then the error tends to unit exponentially with $n$.

As highlighted in \citet{Fawzi2018}, this capacity can be approximated by SDP due to the possibility of establishing SDP bounds on the quantum relative entropy \cite{Fawzi2019b}.
As mentioned in Section~\ref{sec:qkd_dd_vn}, recent algorithmic developments have made it possible to optimize the quantum relative entropy directly, and thus compute the channel capacity, without relying on relaxations \cite{He2024}.

\subsubsection{Quantum capacities}

\begin{figure}
	\centering
	\includegraphics[width=0.95\columnwidth]{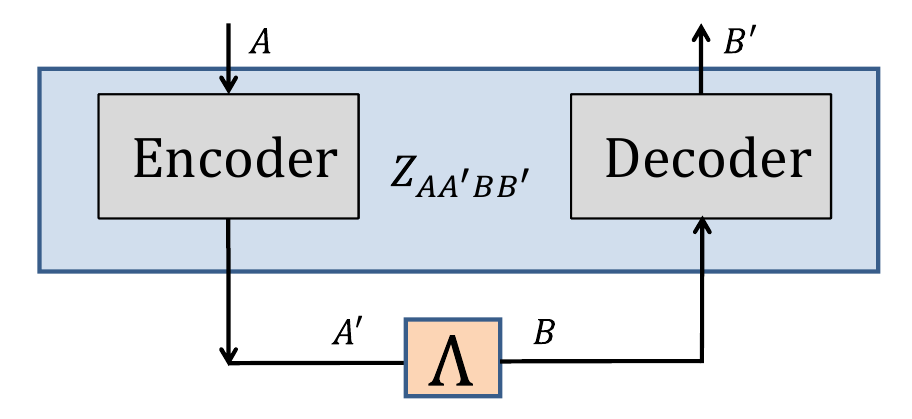}
	\caption{SDP relaxations of channel coding problems can be obtained by viewing the parties as single bipartite operations.}
	\label{FigSuperChannel}
\end{figure}

In general, it is difficult to determine capacities in a computable way.
A mathematically tractable framework is to view the encoding and decoding operations as a single global, bipartite, operation $Z_{AA'BB'}$ where $A$ and $B$ are input systems and $A'$ and $B'$ are output systems \cite{Leung2015}; see Fig.~\ref{FigSuperChannel}.
In practice, $B$ depends on $A'$, since system $A'$ is sent through the channel.
One can think of $Z$ as linearly transforming the channel $\Lambda$ into a new channel, and it is known as superchannel or supermap \cite{Chiribella2008}.
To it, we associate a Choi matrix $J_Z=d_{A}d_{B'}(\mathcal{I}\otimes Z)(\phi^+)$, where $\mathcal{I}$ is the identity channel.
In order for the operation $Z_{AA'BB'}$ to be completely positive and trace preserving, the Choi matrix must be PSD and satisfy $\tr_{A'B'}\left(J_Z\right)=\openone_{AB}$.
Moreover, it must not signal from receiver to sender, which corresponds to $\tr_{B'} \left(J_Z\right)=\tr_{BB'} \left(J_Z\right)\otimes \openone_B$; note that the first factor is the Choi matrix of the sender's transformation.
In addition, if the sender and receiver do not share post-classical resources, we need to ensure that they are not entangled with each other.
A handy relaxation of this constraint is to impose that $Z$ is PPT preserving, i.e.,~that every PPT state remains PPT after being sent through $Z$; which translates into $J_Z^{T_{AA'}}\succeq 0$.
The key observation here is that all of these constraints are either linear or semidefinite in the variable $J_Z$.

Now, to address the one-shot quantum capacity of $\Lambda$ we consider the fidelity between the maximally entangled state and the state obtained from sending half of it through the communication scheme.
The fidelity is particularly convenient because it too can be written in terms of the Choi matrix; $F=\tr\left(J_Z\left(J_\Lambda^T \otimes \phi^+\right)\right)$.
Putting it all together, \citet{Leung2015} obtains an SDP bound on the one-shot quantum capacity which applies both when the quantum channel is unassisted (relaxation to no-signaling and PPT-preserving) and when it is entanglement-assisted (relaxation to no-signaling only).
In \citet{Berta2022} this relaxation is extended, by means of connection to a separability problem and using symmetric extensions, into a hierarchy of SDPs which converges to the fidelity $F$.
The concept of bounding quantum capacities by means of relaxation to PPT-preserving and/or no-signaling codes was further explored in \citet{WangDuan2019}, where SDP bounds were given for the one-shot bounded-error capacity.
Complementarily, and in a similar spirit, it was shown in \citet{Tomamichel2016} how to determine an SDP upper bound the largest rate $R$ for a fixed number of channel uses $n$ and a fixed error $\epsilon$.

In the asymptotic setting, deciding whether a given number of quantum states can be sent over many channel uses with zero error is known to be QMA-complete \cite{Beigi2007}.
However, \citet{Duan2013} showed that, in the same way as the SDP-computable Lov\'asz theta function in Eq.~\eqref{LovaszTheta} can bound the classical problem, the quantum capacity can be bounded via an analogous SDP quantity.
Moreover, in the context of strong converse bounds on the asymptotic quantum capacity, one finds good use of SDPs.
It is known that taking the diamond norm\footnote{The diamond norm is defined by $\|\Lambda\|_\diamond=\max_{\rho_{AB}} \|\openone \otimes \Lambda(\rho_{AB})\|_1$.} of the channel after applying a transposition map, $T$, consitutes a strong converse bound.
Specifically if the rate exceeds $\log \|\Lambda \circ T\|_\diamond$ then the error tends to unit in the number of channel uses \cite{Hermes2016}.
Importantly, the diamond norm can be computed efficiently by SDP \cite{Watrous2009, Watrous2013, SDPbook}.
Another strong converse SDP bound was provided in \citet{WangDuan2016, WangDuan2019}.
This bound is stronger than the one based on the diamond norm but weaker than another known bound, based on the so-called Rains information of the channel, but not straightforward to compute \cite{Tomamichel2017}.
The SDP bound is given by
\begin{equation}
	\begin{aligned}
		\log \max_{\rho_{A},F_{AB}} \quad & \tr\left(J_\Lambda F_{AB}\right)\\
		\st \quad & -\rho_A\otimes \openone \preceq F_{AB}^{T_B}\preceq \rho_A\otimes\openone,\\
		\quad & \tr\left(\rho_A\right)=1,\\
		\quad & F_{AB},\rho_A\succeq 0, 
	\end{aligned}
\end{equation}
and it is additive under tensor products of channels.
In fact, this is closely related to the entanglement measure $E_W$ discussed around Eq.~\eqref{WangSDP} and it can be interpreted as the largest value of $E_W$ taken over all purifications of $\rho_A$ when half of the purification is sent through the channel.

In \citet{Fangb2021} it is shown how to systematically compute bounds on several different classical and quantum channel capacities via SDP.
These capacities are evaluated from a single use of the channel (no regularisation needed), they are computable for general channels and they admit a strong converse.
This method relies on a quantity known as the geometric R\'enyi divergence which has many convenient properties, for example additivity under tensor products and chain rule \cite{Matsumoto2018}.

Another type of quantum capacity where SDPs are useful concerns the ability of bipartite quantum channels (e.g.~a noisy control gate) to create entanglement \cite{BennettHarrow2003}.
The rate of distilling maximally entangled states over such an interaction can be bounded from above by an entropic quantity, which can in turn be evaluated by SDP \cite{Bauml2018, Das2020}.

\subsection{Dimension constraints}\label{sec:communication-dimension}
A natural quantifier of communication is the dimension of the alphabet of the message sent from Alice to Bob.
Classically, Alice's message is selected from a $d$-valued alphabet $\{1,\ldots,d\}$, whereas in the quantum case, the message is a state $\rho_x$ selected from a $d$-dimensional Hilbert space.
While Holevo's theorem \cite{Holevo1973} ensures that such quantum systems cannot be used to transmit a message more efficiently than the corresponding classical systems, it is well-known that there are many other communication tasks in which quantum messages provide an advantage over classical messages.
Examples are the random access codes previously mentioned in Section \ref{seccomscenario} \cite{Ambainis2002, Nayak1999} and distributed computation \cite{Galvao2001}.

A central goal is to characterise the set of quantum correlations, $\mathcal{Q}$, that can be generated between Alice and Bob for a fixed number of inputs ($X$ and $Y$, respectively) and a fixed number of outputs for Bob ($N$), when Alice sends a $d$-dimensional state to Bob (recall Section~\ref{seccomscenario}).
These correlations are given by the Born rule in Eq.~\eqref{BornCom}.
This problem is commonly investigated while granting Alice and Bob unbounded shared randomness.
This renders $\mathcal{Q}$ convex and the task particularly suitable to SDP methods.
It should be noted that the same problem without shared randomness deals with a non-convex correlation set, which leads to very different quantum predictions; see e.g.~\citet{Hayashi2006, Bowles2014, Vicente2017, Tavakoli2020indep}.
Outer and inner approximations to $\mathcal{Q}$ not only are important for determining the non-classicality enabled by quantum theory, but also serve as an important tool for quantum information applications.
SDP relaxations of $\mathcal{Q}$ are useful for this purpose, in particular since conventional analytical bounds on $\mathcal{Q}$ are possible only in rare special cases, see e.g.~\citet{Brunner2013q, Farkas2019, Frenkel2015}.

\subsubsection{Bounding the quantum set}\label{sec:NV}
SDP relaxation hierarchies can be constructed to bound the set of quantum correlations in the prepare-and-measure scenario for arbitrary input/output alphabets and arbitrary dimensions.
They are typically based on tracial moments, see e.g.~\citet{Burgdorf2012}, on which constraints specific to a $d$-dimensional Hilbert space must be imposed.
To this end, let $L=\{\id, \boldsymbol{\rho},\mathbf{M}\}$, where $\boldsymbol{\rho}=\left(\rho_1,\ldots,\rho_X\right)$ and $\mathbf{M}=\left(M_{1|1},\ldots,M_{N|Y}\right)$, be the list of all operators appearing in the quantum prepare-and-measure scenario.
We define $\mathcal{S}_k$ as a set of monomials over $L$ of length $k$.
Recall from the remark in Section~\ref{rem:monomial-sets} that it is interesting to consider subsets of all the monomials with a given length.
An $|\mathcal{S}_k|\times |\mathcal{S}_k|$ moment matrix can be constructed whose entries are given by,
\begin{equation}\label{qcmoment}
	\Gamma(u,v)= \tr\left(u^\dagger v\right),
\end{equation}
for $u,v\in \mathcal{S}_k$.
The moment matrix inherits many constraints from quantum theory.
First, normalisation implies that $\Gamma(\rho_x,\id)=1$.
Second, one can without loss of generality restrict to pure states, i.e.,~$\rho_x^2=\rho_x$.
Third, under the restriction of projective measurements\footnote{In general, one must also consider POVMs when fixing the dimension.}, it holds that $M_{b|y}M_{b'|y}=M_{b|y}\delta_{b,b'}$.
Fourth, the moment matrix contains elements that are equal to the probabilities observed between Alice and Bob, namely $\Gamma(\rho_x,M_{b|y})=p(b|x,y)$.
In addition, the cyclicity of the trace implies a number of additional equalities between the moments, e.g.~$\Gamma(\rho_x, M_{b|y}\rho_x)=\Gamma(\rho_x^2,M_{b|y})=\Gamma(\rho_x,M_{b|y})=p(b|x,y)$.
Last, by argument analogous to that in Eq.~\eqref{PSDcon}, determines that the moment matrix must be PSD.

The key issue is how to add constraints to $\Gamma$ that are specific to the dimension $d$.
One option is to identify polynomial operator identities or inequalities that pertain only to dimension $d$.
However, such relations are typically unknown and finding them is difficult.
Navascu\'es and V\'ertesi (NV) \cite{Navascues2015b} proposed a solution by employing numerical sampling to construct a basis of moment matrices.
The prescription is to randomly sample the states and measurements from the $d$-dimensional Hilbert space and then compute a moment matrix sample, $\Gamma^{(1)}$, which will automatically satisfy all the aforementioned constraints.
The sampling procedure is repeated until a moment matrix sample is found to be linearly dependent on all of the previous samples.
This can be quickly checked by vectorising the samples, arranging them in a matrix and computing its rank.
The process is then truncated and the collected samples $\{\Gamma^{(1)},\ldots,\Gamma^{(m)}\}$ are certain to span\footnote{The probability that the $m$ samples do not span the full space but nevertheless the next sample is found to be linearly dependent is essentially zero.} a relaxation of the subspace of moment matrices compatible with dimension $d$.
In order to preserve normalisation, namely $\tr\left(\id\right)=d$, the final moment matrix becomes an affine combination of the samples,
\begin{align}
	& \Gamma = \sum_{i=1}^m \gamma_i \Gamma^{(i)}, \quad \text{where} \qquad \sum_{i=1}^m \gamma_i=1.
\end{align}
The coefficients $\{\gamma_i\}$ serve as the SDP variables in the necessary condition for the existence of a $d$-dimensional quantum model for $p(b|x,y)$, namely that it is possible to find $\Gamma \succeq 0$ such that $\Gamma(\rho_x,M_{b|y})=p(b|x,y)$.
Note that by relaxing the latter condition, one can equally well employ the NV hierarchy to bound the extremal quantum value of an arbitrary linear objective function $\sum_{b,x,y} c_{bxy}p(b|x,y)$ characterised by some real coefficients $c_{bxy}$.
It is currently unknown whether this hierarchy converges to $\mathcal{Q}$ in its asymptotic limit.
Notably, one can also incorporate POVMs by explicitly performing a Neumark dilation in the measurements, albeit at the price of employing a larger dimension.
Alternatively, one can sample directly from the set of POVMs, but that requires the use of localising matrices to enforce the bounds $M_{b|y} \succeq 0$ and $\sum_{b=1}^{N-1} M_{b|y} \preceq \id$.

For scenarios with a reasonably small number of inputs and outputs or a fairly low dimension, the NV hierarchy is an effective tool; see examples in \citet{Navascues2015c,Bermejo2023}.
However, for middle-sized problems it becomes less handy.
A first reason is that the size of the moment matrix, for a fixed level of relaxation, scales polynomially in the size of $L$.
Second, that the number of SDP variables, $m$, increases rapidly with any one of the parameters $(X,Y,N,d)$\footnote{The memory required in a single iteration of a typical primal-dual solver scales quadratically in both $m$ and $|\mathcal{S}|$ while the CPU time scales cubically in both $m$ and $|\mathcal{S}|$.}.
Third, one typically obtains better bounds on $\mathcal{Q}$ by separately considering different rank combinations for the projective measurements and then selecting the best bound, but the number of combinations increases quickly with $(N,Y,d)$.

In \citet{Pauwels2022b}, an alternative SDP hierarchy is developed that applies to bounding correlations obtained from systems that can be represented nearly as $d$-dimensional.
This permits analysis of correlations of systems that approximate e.g.~qubits to any desirable extent, and in the special case in which the approximation is exact it provides bounds on $\mathcal{Q}$.
The main idea is to supplement the set $L$ with an additional operator $V$ which is meant to emulate the $d$-dimensional identity projection.
Therefore, it is given the properties $V^2=V$ and $\tr\left(V\right)=d$.
One can then proceed with building the moment matrix as described above after Eq.~\eqref{qcmoment}.
This method circumvents sampling and immediately takes POVMs into account, and has computational requirements that are constant in the dimension parameter.
The main drawback is that the hierarchy does not converge to the quantum set, because it inherently relaxes dimension-restricted communication to communication that on average has dimension $d$; such systems have been studied independently in the context of entanglement using SDP relaxations \cite{Gribling2017}.
In some concrete instances, this can lead to worse bounds on linear objective functions.
Another alternative method is based on transforming Bell scenarios into prepare-and-measure scenarios by considering the remote states prepared by Alice for Bob, ideally via a maximally entangled state \cite{Mironowicz2014}.
The dimension constraint is relaxed to a marginal constraint in the Bell scenario; one uses the NPA hierarchy under the extra constraint that one of Alice's outputs for each input has marginal probability $1/d$.
This, however, can be a crude relaxation and it does not converge in general \cite{Navascues2015c}.

\subsubsection{Applications}
Bounds on $\mathcal{Q}$, obtained by means of SDP relaxations, are broadly useful.
An evident application is determining upper bounds on the magnitude of quantum advantages in useful communication tasks.
An important class of examples is quantum random access codes, in which Bob aims to randomly access a piece of information in a larger database held by Alice \cite{Ambainis2002}.
Direct application of the NV hierarchy has given tight upper bounds in low dimensions \cite{Tavakoli2015} and lower bounds have been obtained via SDPs in seesaw heuristics when the channel is noisy \cite{Marques2022}.
The former can be made vastly more efficient by exploiting symmetries inherent to the problem in order to reduce the complexity of the SDP; see e.g.~\citet{Tavakoli2019, Pauwels2022b, Aguilar2018}.
Such symmetry methods are further discussed in Section~\ref{sec:symmetries}.

In other classes of communication tasks, inspired by the high-dimensional Bell inequalities of \citet{Collins2002}, SDP relaxations of $\mathcal{Q}$ can showcase dimensional thresholds, i.e.,~a critical dimension above which the optimal quantum strategy qualitatively changes \cite{Tavakoli2017, Martinez2018}.
Moreover, suitably chosen linear objective functions over $\mathcal{Q}$ can be linked to the long-standing problem of determining the number of mutually unbiased bases in dimension 6.
The hypothesis that no more than three such bases exist could, in principle and if true, be proven through a sufficiently precise SDP relaxation of $\mathcal{Q}$ based on a chaining of quantum random access codes \cite{Aguilar2018}.
There are also other SDP-based approaches to this problem; some that take the route of nonlocality \cite{Gribling2021b, Colomer2022} and others that consider the existence of so-called Gr\"obner bases \cite{Brierley2010}.
Nevertheless, owing to the computational complexity, the problem presently remains open.

A complementary consideration is to, in a given dimension, bound the optimal quantum violation of facet inequalities for the polytope of classical prepare-and-measure correlations in a given input/output scenario \cite{Navascues2015c, Mironowicz2014}.
From another perspective, such bounds can be seen as device-independent tests of quantum dimensions, which is a task where Alice and Bob aim to certify a lower bound on the dimension of their quantum channel without assuming any model for their preparation and measurement devices \cite{Gallego2010}.

Applications also pertain to semi-device-independent quantum information processing, i.e.,~practically motivated protocols performed solely under the assumption that the communicated quantum state is of a limited dimension.
Self-testing protocols based on the prepare-and-measure scenario have been developed in such settings.
Drawing inspiration from the swap method for noisy self-testing discussed in Section~\ref{sec:selftest}, one can for instance employ the NV hierarchy to bound the average fidelity of Alice's qubit preparations with the ensemble used in the paradigmatic BB84 quantum key distribution protocol \cite{Tavakoli2018}.
This type of self-testing has also been extended to quantum instruments in three-partite communication scenarios, featuring a sender, a transformer and a receiver, both with \cite{Miklin2020} and without \cite{Mohan2019} SDPs.
In \citet{Miklin2020} it is shown how the NV hierarchy can be extended to such prepare-transform-measure scenarios.

It is also possible to certify qualitative properties.
For instance, by restricting sampling to real-valued Hilbert spaces, the NV hierarchy has been used to test complex-valued quantum operations in a given dimension \cite{Navascues2015c}.
Furthermore, a particularly natural use for dimension-restricted systems is to certify non-projective quantum measurements, i.e.,~measurements that cannot be simulated with standard projective measurements and classical randomness \cite{Ariano2005}.
The reason is that such measurements cannot be certified when ancillary degrees of freedom are available due to the possibility of Neumark dilations \cite{nielsen2010}.
The NV hierarchy has been used to certify non-projective measurements of dimension 2 \cite{Tavakoli2020b, Mironowicz2019}, dimension 4 \cite{Martinez2022} and up to dimension 6 \cite{Tavakoli2019}.
All of these works rely on suitable witness constructions, but notably that is not essential for the SDP relaxation method to work.
By analogous means, SDP methods have enabled certification of non-projective measurements in Bell scenarios under the assumption of a limited entanglement dimension \cite{Smania2020, Gomez2016}.
The SDP seesaw methods for probing $\mathcal{Q}$ (recall Section~\ref{sec:introduction:entanglement}) can be used to reproduce the numerical estimates for the output rate of semi-device-independent quantum random number generators \cite{Li2011, Li2012}.
Proper randomness bounds can be obtained via the NV hierarchy \cite{Mironowicz2016} or other relaxation methods \cite{Mironowicz2014}.

However, many actual implementations of dimension-restricted quantum systems, such as spontaneous parametric down-conversion sources of single photons or weak coherent pulses, only nearly represent a proper dimension-restricted system.
This can be leveraged to hack a semi-device-independent protocol.
Using the SDP relaxation hierarchy of \citet{Pauwels2022b}, which was originally discussed in \ref{sec:NV} only for standard dimension constraints, the small inaccuracies of such ``almost qudit systems'' can be taken into account when constructing quantum information protocols.
The deviations from the dimension assumption can be quantified and incorporated as a linear inequality constraint on suitable elements of the corresponding moment matrix.

\subsubsection{Entanglement-assisted communication}\label{sec:EAPM}
\begin{figure}[t!]
	\centering
	\includegraphics[width=0.95\columnwidth]{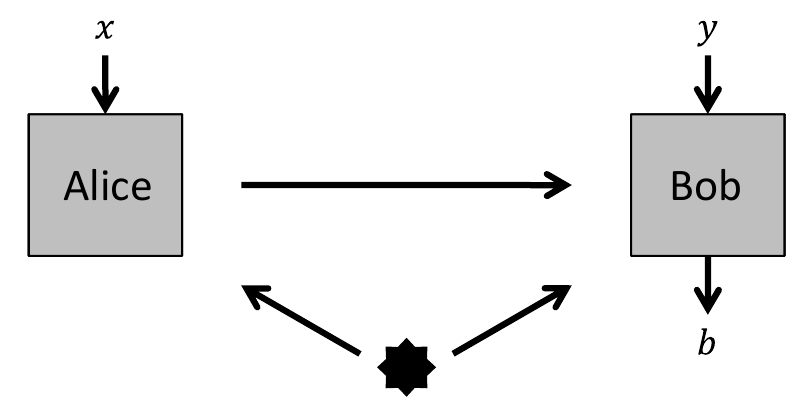}
	\caption{A bipartite entanglement-assisted prepare-and-measure scenario.
	A source distributes an entangled state between Alice and Bob.
	Conditioned on her input $x$, Alice maps her system onto a message that is sent over the channel to Bob.
	Conditioned on $y$, Bob measures both the systems and obtains the outcome $b$.}
	\label{FigEAPMScenario}
\end{figure}

In the previous section, the parties in the prepare-and-measure scenario communicated quantum states while sharing classical randomness.
In this section, we discuss the prepare-and-measure scenario when the parties additionally share an entangled state.
The entanglement-assisted prepare-and-measure scenario is illustrated in Fig.~\ref{FigEAPMScenario}.

Consider now a situation where the operations of Alice are not constrained to a binary choice.
Instead, they can be arbitrary but the message she sends over the channel to Bob is classical and of dimension $d$.
Such correlations are interesting because they can boost the performance of classical messages in various communication tasks \cite{Buhrman2010}.
The source emits an entangled state $\Phi$ of local dimension $D$.
Given $x$, Alice transforms her share into a $d$-valued classical message.
Hence, she applies a $d$-outcome POVM $\{N_{a|x}\}_{a=1}^d$ and upon receiving the outcome $a$ she sends the classical state $\ketbra{a}{a}$ to Bob.
Averaged over the probability $p(a|x)$, Bob's total state thus becomes $\sum_a \ketbra{a}{a}\otimes \tr_A\left(N_{a|x}\otimes \id \Phi^{AB}\right)$ where the second factor is the unnormalised state of Bob's quantum system conditioned on Alice's outcome.
In the most general situation, Bob can now first read the received message and then use the value $a$ to choose a POVM $\{M_{b|y,a}\}$ with which he measures his share of the entangled state.
The correlations become
\begin{equation}\label{BornEACC}
	p(b|x,y)=\sum_{a=1}^d \tr\left(N_{a|x} \otimes M_{b|y,a} \Phi^{AB}\right).
\end{equation}
This can be interpreted as marginalised Bell correlations with signaling from Alice to Bob and can also immediately be extended to non-identity classical channels connecting the parties.
In the most general case, the entanglement dimension $D$ is unrestricted and Bob can adapt his measurement to the incoming message.
Then, one can bound the set of quantum correlations from the exterior, when the alphabet size for the inputs and outputs is fixed, by a converging SDP relaxation hierarchy à la NPA \cite{Tavakoli2021c}, which was discussed in Section~\ref{sec:NPA}.
An alternative SDP relaxation hierarchy for this type of problems appears in \citet{Berta2016}At the first level, this hierarchy is more constraining than the NPA approach due to the possibility of requiring all elements in the moment matrix to be non-negative \cite{Sikora2017}.
Furthermore, when the entanglement dimension is known, one can instead employ SDP relaxations in the spirit of dimension-restricted NPA, as discussed in Section~\ref{sec:entdim}.
Another interesting situation arises when one requires Bob not to adapt his measurement to the message, i.e.,~that a Bell test is performed first and only afterwards does classical communication take place.
Correlations from such non-adaptive strategies can also be bounded by SDP relaxations \cite{Pauwels2022}, specifically by imposing the commutation relation $[M_{b|y,a},M_{b'|y,a'}]\,{=}\,0$ $\forall\,y,\,b,\,b',\,a,\,a'$ in the NPA-type matrix.

When the messages are $d$-dimensional quantum systems, it is well known from the dense coding protocol that stronger correlations are possible than are with classical $d$-dimensional messages \cite{Bennett1992}.
For quantum messages, Alice applies a quantum channel $\Lambda_x^{A\rightarrow C}$ which maps the incoming $D$-dimensional system to a $d$-dimensional message state that is sent to Bob.
The message space is denoted as $C$.
The total state held by Bob becomes $\tau_x^{CB}=\Lambda_x^{A\rightarrow C}\otimes \id_B[\Phi_{AB}]$ to which he applies a POVM $\{M^{CB}_{b|y}\}$.
The resulting correlations become
\begin{equation}
	p(b|x,y)=\tr\left(\tau_x^{CB} M^{CB}_{b|y} \right).
\end{equation}
The only non-trivial constraint on the total state is no-signaling, namely $\tau_x^B=\tau^B$.
In a given dimension, this allows for alternating convex search methods to be used for exploring the correlation set.
In particular, for a maximally entangled two-qubit state the entanglement-assisted correlation set is equivalent to the correlations achievable in an unassisted prepare-and-measure scenario when Alice sends real-valued four-dimensional systems \cite{Pauwels2022c}.
This permits SDP outer bounds via the NV hierarchy restricted to real Hilbert spaces.
More generally, when $D$ is unknown and when any quantum channel is used between Alice and Bob, outer bounds can be obtained by a convergent hierarchy of SDPs \cite{Tavakoli2021c}.
This hierarchy is based on an explicit Kraus operator parameterisation of the quantum message space.
It can be seen as a variation of the NPA hierarchy and therefore its convergence properties are inherited.
An important caveat is that this SDP hierarchy scales more rapidly than the adaption of the NPA hierarchy to the classical case: the number of operators used to build the SDP matrix scales quadratically in $d$, as compared to linearly in the classical case.
This quickly makes implementation demanding, although it may be possible to circumvent the issue via symmetrisation methods; see Section~\ref{sec:symmetries}.
The alternative, unrelated, SDP hierarchy of \cite{Tavakoli2021c} becomes relevant here, since it can be applied to efficiently obtain bounds.
This hierarchy is based on the concept of informationally restricted correlations \cite{Tavakoli2020a}, see Section~\ref{sec:communication-distinguish}, and relies on moment matrices with a size independent of $d$.
While this hierarchy does not converge to the quantum set in the entanglement-assisted scenario, it can give useful and even tight bounds in specific cases.
Notably, when Alice and Bob are connected by an identity channel and entanglement is a free resource, the correlation set for quantum messages is identical to the correlation set for classical messages of twice as many bits \cite{Vieira2022}.
Consequently, one can still use the SDP hierarchy for classical messages to address the scenario with quantum messages.

The SDP methods for the entanglement-assisted prepare-and-measure scenario have found several applications.
For instance, in order to fully capture the spirit of a device-independent test of classical or quantum dimensions, i.e.,~to place a lower bound on the dimension of a message without making assumptions about the internal working of the involved devices, one must allow for the possibility that the preparation and measurement devices share a potentially unrestricted amount of entanglement.
In \citet{Tavakoli2021c}, the SDP relaxations are used to make known dimension witnesses \cite{Gallego2010, Ahrens2014} robust to entangled devices.
Furthermore, bounds on the quantum correlations have enabled a number of quantum resource inequalities.
For instance, it has been shown that protocols in which Bob adapts his setting to a classical message are in general more powerful than the non-adaptive protcols, and that this distinction is crucial for using entanglement and one bit of classical communication to simulate correlations obtained from an unassisted qubit in the prepare-and-measure scenario depending on whether it is measured with projective measurements or POVMs \cite{Pauwels2022}.
For example, non-adaptive protocols based on Bell-inequality violations followed by classical communication are known, that improve the task of random access coding \cite{Pawlowski2010, TavakoliMarques2016}.
Using adaptive measurements and higher-dimensional entanglement can yield larger quantum advantages \cite{Pauwels2022c, Vaisakh2021}.
Moreover, while it is well known that entanglement cannot increase the capacity of a classical channel, the same is not true in general when the capacity is considered in the non-asymptotic setting \cite{Cubitt2010}.
For some noisy classical channels, the advantage of entanglement can even be linked to the CHSH inequality \cite{Prevedel2011} and SDP relaxations showcase that such strategies are in fact optimal \cite{Berta2016}.

\subsubsection{Teleportation}
Entanglement-assisted communication can also be considered when the inputs themselves are quantum states, rather than classical symbols.
The most famous instance of such a scenario is teleportation, where a quantum state is sent by means of a shared maximally entangled state and classical communication \cite{Bennett1993}.
However, if the state $\rho_{AB}$ is not maximally entangled, then the teleportation channel will not flawlessly simulate the quantum identity channel.
The traditional approach to quantifying the ability of an entangled state to perform teleportation is via the average fidelity of the target state and the teleported state.
It is known that teleportation fidelity is a function of the maximally entangled fraction \cite{Horodecki1999},
\begin{equation}
	F(\rho)=\max\,\, \bracket{\psi}{\rho}{\psi},
\end{equation}	
where $\ket{\psi}$ is any maximally entangled state.
However, if one has prior knowledge of $\rho$, one could consider applying a $\rho$-dependent LOCC protocol to potentially enhance the teleportation fidelity.
In \citet{Verstraete2003} it was shown that the optimal fidelity under such LOCC protocols, i.e.,~$\max_{\Lambda: \mathrm{LOCC}}\,\, F(\Lambda(\rho))$, can be determined exactly when $\rho$ is a two-qubit state, and moreover that a single round of one-way communication is sufficient.
To achieve this the authors first consider lower bounds by restricting to LOCC protocols of the form
\begin{equation}\label{eq:LOCC1}
	\begin{aligned}
		\Lambda(\rho) &= (A \otimes B) \rho (A^{\dagger} \otimes B^{\dagger}) \\ &+ \tr((\id-A^{\dagger}A) \otimes (\id-B^\dagger B) \rho ) \ketbra{v}{v} \otimes \ketbra{w}{w}
	\end{aligned}
\end{equation}
where $A$ and $B$ are Kraus operators with the corresponding outcome operators satisfying $0 \preceq A^\dagger A, B^\dagger B \preceq \id$, and $\ket{v}$ and $\ket{w}$ are any qubit states.
Through a careful choice of $\ket{v}$ and $\ket{w}$ and change of variables for $A$ and $B$ the authors show that $\max_{\Lambda}\,\, F(\Lambda(\rho))$, where the maximization is over LOCC channels of the form of Eq.~\eqref{eq:LOCC1}, is equivalent to the optimization
\begin{equation}\label{eq:teleportation-LB}
	\begin{aligned}
		\max_{C}& \quad \tfrac12 - \bra{\phi^+} (C \otimes \id) \rho^{T_B} (C^{\dagger} \otimes \id) \ket{\phi^+} \\
		\st& \quad C^{\dagger} C \preceq \id\,.
	\end{aligned}
\end{equation}
To derive upper bounds they relax the optimization of $F$ from LOCC channels to PPT channels, i.e.,~channels whose Choi matrix is PPT, leading to the SDP relaxation
\begin{equation}
	\begin{aligned}
		\max& \quad \tr((\rho_{AB}^T \otimes \ketbra{\phi^+}{\phi^+}) C_{ABA'B'}) \\
		\st& \quad \tr_{A'B'}(C_{ABA'B'}) = \id_{AB}, \\
		& \quad C_{ABA'B'}^{T_{BB'}} \succeq 0, \\
		& \quad C_{ABA'B'} \succeq 0.
	\end{aligned}
\end{equation}
Using the symmetries of $\ket{\phi^+}$ under unitary twirling, this problem can be reduced to the simpler SDP, $\max \,\, 1/2 - \tr{X \rho^{T_B}}$ with $X$ constrained to satisfy $0 \preceq X \preceq \id$ and $-\id \preceq 2 X^{T_B} \preceq \id$ which can be shown to be maximized by a rank-$1$ operator $X = (A^{\dagger} \otimes \id) \ketbra{\phi^+}{\phi^+} (A \otimes \id)$ for some matrix $A$ satisfying $A^{\dagger} A \preceq \id$.
However, for such rank-one operators, this optimization is exactly the same as the lower bound in~Eq.~\eqref{eq:teleportation-LB} and hence is achievable with a single-round LOCC protocol.
Notably, this gives an example where the relaxation to PPT channels is tight.

A more general approach to teleportation is proposed in \citet{Cavalcanti2017}, where a verifier supplies a given number of states $\psi_x$ to Alice and asks her to teleport them to Bob.
Alice applies a POVM $A_{a}^{VA}$ to $\psi_x$ and her share of the entangled state $\rho_{AB}$.
The resulting unnormalised states of Bob are $\sigma_{a|\psi_x}=\tr_{V}\left(A_a^{VB}(\psi_x\otimes \openone^B)\right)$ where $A_a^{VB}=\tr_A\left((A_a^{VA}\otimes \openone^B)(\openone^V\otimes \rho_{AB})\right)$.
However, if the state is separable, then this simplifies to separable operators, $A_a^{VB}=\sum_\lambda p_\lambda A_{a|\lambda}^V\otimes \varphi_\lambda^B$, where $A^V_{a|\lambda}=\tr_A\left(A_a^{VA}\left(\openone^V\otimes \rho^A_\lambda\right)\right)$.
Notice also that completeness of $\{A_a^{VA}\}_a$ implies that $\sum_a A_a^{VB}=\openone^V\otimes \rho^B$.
We can then quantify the amount of white noise that must be added to a given set $\{\sigma_{a|\psi_x}\}_{a,x}$ in order to model it classically,
\begin{equation}
	\begin{aligned}
		\min \quad t \\
		\st \quad & \frac{\sigma_{a|\psi_x}}{1+t}+\frac{t}{1+t}\frac{\openone^B}{dN}=\tr_V\left(A_{a}^{VB}\left(\psi_x\otimes \openone^B\right)\right),\\
		& \sum_a A_a^{VB}=\openone^V \otimes \left(\frac{\rho^B}{1+t}+\frac{t}{1+t}\frac{\openone^B}{d}\right),\\
		& A_{a}^{VB} \in \text{SEP} \quad \forall\,a,
	\end{aligned}
\end{equation}
where $N$ is the number of outcomes for Alice and $d$ is the dimension of system $B$.
If the solution has $t>0$, there is no classical teleportation model.
By relaxing the set of separable operators to a semidefinite constraint, for example PPT, the above becomes an SDP criterion for classicality of teleportation.
A resource theory for this type of teleportation, where SDPs again are relevant, was developed in \citet{Lipka2020}.
In \citet{Supic2019} it was shown how one can estimate entanglement measures by SDP analysis of the data generated in teleportation experiments.

Ideal teleportation can be seen as simulating a noiseless quantum channel using entanglement and classical communication.
In \citet{Holdsworth2023}, SDP methods are developed for bounding various forms of simulation errors for how well the teleportation channel approximates the noiseless quantum channel.
A key component for this analysis is to use SDP relaxations of the set of one-way LOCC channels, i.e.,~relaxations of procedures where Alice measures locally and sends her outcome to Bob who then performs a local channel.
To this end, as in Fig.~\ref{FigSuperChannel}, we view the actions of Alice and Bob as a single bipartite channel $\Lambda_{AB\rightarrow A'B'}$.
If this bipartite channel preserves PPT states, which is an SDP condition on the level of the Choi matrix associated with the channel \cite{Rains1999, Rains2001}, it is also a one-way LOCC channel.
Alternatively, it is possible to relax one-way LOCC channels by imposing that $\Lambda_{AB\rightarrow A'B'}$ is $k$-extendible\footnote{An alternative definition of $k$-extendible channels and their relevance to SDP appears in \citet{Berta2022}.} \cite{Kaur2019, Kaur2021}.
This concept is analogous to the constraints appearing in the DPS hierarchy.
It means that one can associate another channel, $\mathcal{N}_{AB_1\ldots B_k\rightarrow A'B'_1\ldots B'_k}$, which is invariant under permutations of Bob's inputs and outputs, and such that if all but one of Bob's systems are discarded $\Lambda_{AB\rightarrow A'B'}$ is recovered.
These two relaxations can be combined into a single SDP relaxation of one-way LOCC.
The former type of relaxation has also been used to address simulation errors when both Alice and Bob want to teleport states to each other \cite{Siddiqui2022}.

A related task is known as port-based teleportation \cite{Ishizaka2008}.
In it, Bob does not need to perform a correcting quantum channel upon receiving Alice's outcome.
To achieve this, Alice and Bob share $n$ copies of the maximally entangled state and Alice jointly measures her input state and all of her $n$ shares and sends the outcome to Bob.
The outcome tells Bob in which share he can find the teleported state.
Optimisation of protocols of this type has been cast as an SDP \cite{Studzinski2017, Mozrzymas2018}.

\subsection{Distinguishability problems}
\subsubsection{Distinguishability constraints for quantum communication}\label{sec:communication-distinguish}
It is often interesting to benchmark quantum communication not by its dimension, but instead by another property that is either physically or conceptually well-motivated.
This pertains partly to understanding the conditions under which quantum correlations go beyond classical limits and partly to building useful protocols for semi-device-independent quantum information processing, where deductions are made under weak and reasonable physical assumptions.
Many different frameworks have been proposed, all based on the general idea of limiting the distinguishability of the states sent from Alice to Bob.
What they have in common is that SDPs are typically crucial for their analysis.
Here we briefly survey the main idea of each of these frameworks from an SDP perspective.
Some of the cryptographic applications of these SDP methods are surveyed in Section~\ref{sec:SDI}.

A natural experimental setting is that Alice knows which state $\ket{\psi_x}$ she is trying to prepare for Bob.
However, since she does not have flawless control of her lab, she ends up preparing another state $\rho_x$ that is close but not identical to $\psi_x$.
The accuracy of her preparation can be quantified by the fidelity $F_x=\bracket{\psi_x}{\rho_x}{\psi_x}$.
Alice can either measure this quantity in her lab or estimate it from an error model and name $\epsilon_x$ the deviation in the fidelity from the ideal unit result.
The communication between Alice and Bob is then based only on the assumption that Alice can control her state preparation up to an accuracy $\epsilon_x$ \cite{Tavakoli2021b}.
The key observation for characterising such correlations, based on quantitative distrust, is that Uhlmann's theorem \cite{Uhlmann1976} allows one to substitute a mixed state $\rho$ for a purification $\ket{\phi}$ such that the fidelity is preserved.
Since $N$ pure states span at most an $N$-dimensional space, the correlations can be thought of as arising in a dimension-restricted space, to which the NV hierarchy applies as described in Section~\ref{sec:NV}.
One can therefore use the ideas of the NV hierarchy, but now extending the operator list to also include all of the target states $\{\psi_x\}$.
The fidelity constraints can then be imposed as additional linear inequality constraints, $F_x=\Gamma(\rho_x,\psi_x)\geq 1-\epsilon_x$, on the moment matrix.
This was used in \citet{Tavakoli2021b} to, for instance, certify collections of non-classical measurements as a function of $\epsilon_x$.

An alternative approach limits the distinguishability, not with respect to a target state but rather with respect to the collection of states prepared by Alice.
In \citet{Wang2019} it was considered that a set of $Z$ pure bipartite states, $\{\ket{\psi_z}\}$, are distributed between Alice and Bob.
The Gram matrix of the states is known, i.e.,~all pairs of overlaps $\lambda_{ij}=\braket{\psi_{i}}{\psi_j}$, are fixed.
The correlations then become $p(a,b|x,y,z)=\bracket{\psi_z}{A_{a|x}\otimes B_{b|y}}{\psi_z}$.
To bound the correlation set via an SDP hierarchy, consider a set of monomials $\mathcal{S}$ consisting of products of the global projective measurements $A_{a|x}\otimes \openone$ and $\openone \otimes B_{b|y}$.
Define the $|\mathcal{S}|Z\times |\mathcal{S}|Z$ moment matrix
\begin{equation}\label{eq:wang_grammethod}
	\Gamma(u,v)=\sum_{i,j=1}^{Z} G^{ij} \otimes \ketbra{i}{j},
\end{equation}
where for each pair $(i,j)$ we define the matrix $G^{ij}(u,v)=\bracket{\psi_i}{u^\dagger v}{\psi_j}$ for $u,v\in \mathcal{S}$.
The standard properties of projective quantum measurements and the known Gram matrix imply constraints on $\Gamma$.
In particular, one recovers the probabilities as $G^{zz}(A_{a|x},B_{b|y})=p(a,b|x,y,z)$ and the overlaps as $\sum_{a} G^{ij}(A_{a|x},\openone)=\lambda_{ij}$.
Combined with $\Gamma \succeq 0$, this gives an SDP relaxation that, in the limit of large relaxation level, converges to the quantum set of correlations.
In the special case of only two pure states, the Gram matrix trivialises to a single non-trivial entry and the correlation set can then be characterised completely with a single SDP \cite{Brask2017}.
This two-state case was for example used in \citet{Shi2019} to certify a genuine three-outcome measurement.
Furthermore, SDP relaxations based on the Gram matrix have been used to compute upper bounds on quantum state discrimination problems for optical modes with arbitrary commutation relations limited by a fixed average photon number \cite{Primaatmaja2021}.

A conceptually motivated framework for the prepare-and-measure scenario is to generalise dimension restrictions to information restrictions \cite{Tavakoli2020a}; see also \citet{Chaturvedi2020}.
The main idea is to consider the cost of creating correlations in terms of the amount of knowledge that Alice must make available about her input random variable $X$.
The classical information carried by Alice's ensemble, $\mathcal{E}=\{p_x,\rho_x\}$, is defined as the difference between the min-entropy before and after Bob has received the communication, $I(\mathcal{E})= H_\text{min}(X)-H_\text{min}(X|B)$.
Here, the first term is determined by the largest probability in Alice's prior, $H_\text{min}(X)=-\log \max_x p_x$.
The conditional min-entropy has an elegant operational interpretation in terms of minimal error quantum state discrimination \cite{Konig2009}: if $P_g$ is the largest average probability of state discrimination of the ensemble $\mathcal{E}$ then $H_\text{min}(X|B)=-\log P_g$.
The state discrimination task can be written as the following SDP:
\begin{equation}\label{MESD}
	\begin{aligned}
		P_g = \max_{\{M_x\}} \quad & \sum_{x} p_x \tr\left(\rho_x M_x\right)\\
		\st \quad & \sum_x M_x=\openone,\\
		& M_x\succeq 0.
	\end{aligned} 
\end{equation}

Notably, other natural communication tasks closely related to quantum state discrimination can also be expressed as SDPs.
Examples of this are the quantum guesswork, where one aims to minimise the number of guesses needed to learn $x$ \cite{Hanson2022}, discrimination of sets of labels \cite{Chaturvedi2021b}, maximum confidence state discrimination \cite{Lee2022} and quantum state exclusion, where one aims to rule out the possibility that Alice selected a subset of her input alphabet \cite{Bandyopadhyay2014, Russo2022}.
The central question then becomes to determine the relationship between the correlations $p(b|x,y)$ and the information $I(\mathcal{E})$ (or the guessing probability $P_g$).
In \citet{Tavakoli2022a}, convex programming methods are developed for analysing both informationally restricted classical and quantum correlations.
The former are fully characterised by an LP and the latter can be bounded by an SDP hierarchy.
The key step for constructing the hierarchy is to introduce an auxiliary operator, $\sigma$, when building the moment matrix.
This auxiliary operator comes from the SDP dual to Eq.~\eqref{MESD},
\begin{equation}
	\begin{aligned}
		P_g = \min_{\sigma} \quad & \tr\left(\sigma\right)\\
		\st \quad & \sigma\succeq p_x \rho_x \quad \forall\,x.
	\end{aligned}
\end{equation}
Note that strong duality holds.
Therefore, the properties that $\tr\left(\sigma\right)\leq P_g$ and that $\sigma \succeq p_x \rho_x$ are built into the moment matrix.
The former is simply the linear constraint $\Gamma(\sigma,\openone)\leq P_g$.
The latter are semidefinite constraints which can be imposed through localising matrices (recall Section~\ref{sec:NoncommutativePoly}).
Moreover, for informationally restricted correlations, one cannot restrict to pure states without loss of generality.
Taking this into account also requires localising matrices for imposing the condition that $\rho_x$ is a valid state, namely $\rho_x-\rho_x^2\succeq 0$.
Beyond its own domain, the SDP tools for informationally restricted quantum correlations can be applied to problems in quantum contextuality \cite{Tavakoli2021a} and entanglement-assisted correlations with classical or quantum messages \cite{Tavakoli2021c}.
For such ends, it has the convenient property that the complexity of the SDP is independent of the amount of information considered.
However, the convergence of the hierarchy to the quantum set is presently unknown.

A practical approach to quantum communication in the prepare-and-measure scenario is based on limiting the energy in the message from Alice to Bob.
In \citet{VanHimbeeck2017} it was proposed to limit the non-vacuum component of a weak coherent pulse through an upper bound of the form $\expect{\openone-\ketbra{0}{0}}_{\rho_x}\leq \omega_x$.
When Alice has two preparations, it was shown that the set of energy-restricted quantum correlations can be mapped onto a qubit problem, which in turn permits a complete characterisation in terms of a single SDP \cite{Himbeeck2019}.


\subsubsection{Discrimination tasks}
SDP techniques are useful for determining the relevance of quantum resources in various discrimination tasks. An important class of examples is quantum state estimation problems, which can with some generality be phrased as a source generating a pure state $\phi_x$ with probability $p(x)$.
The state is then encoded in another state $\varphi_x$ and given to a user who is tasked with performing a measurement $\{M_a\}$.
Upon observing the outcome, he outputs a quantum state $\psi_a$ as his estimate for $\phi_x$.
The average fidelity of the estimation becomes
\begin{equation}\label{fideq}
	F=\sum_{a,x} p(x) \tr(\varphi_x M_a) \tr(\phi_x \psi_a)=\tr(\rho_{AB} \Lambda_{AB}).
\end{equation}
In the second equality we have defined $\rho_{AB}=\sum_x p(x)\varphi_x\otimes \phi_x$ and $\Lambda_{AB}=\sum_a M_a\otimes \psi_a$, because this formulation permits us to connect the task of bounding $F$ with a separability problem \cite{Navascues2008bb}.
The key observation is that $\rho_{AB}$ is known whereas $\Lambda_{AB}$ is unknown but separable and subject to the constraint $\tr_B(\Lambda_{AB})=\openone$.
In fact, any separable operation with this property corresponds to a valid state estimation strategy.
Hence, one can apply PPT-symmetric extensions to system $A$ as given by the DPS hierarchy (recall section~\ref{sec:DPS}) and obtain a sequence of SDP bounds on $F$ which in the limit of large relaxation level converges to $F$.
A relevant variation of this problem is when $\varphi_x$ itself is a bipartite state and the measurement $\{M_a\}$ is restricted to admit implementation by LOCC.
Since it is difficult to mathematically characterise LOCC measurements, an often viable approach is to relax these to separable measurements.
As shown in \citet{Navascues2008bb}, this too can be connected to the DPS hierarchy.
To take the bipartite feature into account, Eq.~\eqref{fideq} is straightforwardly modified by replacing $\rho_{AB}$ and $\Lambda_{AB}$ with the three-partite operators $\rho_{ABC}=\sum_x p(x) \varphi_x^{AB}\otimes \phi_x^{C}$ and $\Lambda_{ABC}=\sum_a M_a^{AB}\otimes \psi_a^{C}$.
Since $\{M_a^{AB}\}$ itself is separable, one must impose that $\Lambda_{ABC}$ is fully (tripartite) separable and that $\tr_{C}(\Lambda_{ABC})=\openone$.
This makes the problem amenable to the multipartite variant of the DPS hierarchy of SDPs.

Discrimination of sets of entangled states using only measurements compatible with LOCC is a rich topic, and SDPs have played a role in its recent development.
The set of PPT measurements has frequently been used to bound the set of LOCC measurements in such tasks \cite{Cosentino2013}.
This relaxation can many times be useful.
For instance, it is known that any set of $k>d$ orthogonal maximally entangled states on $\mathbb{C}^d\otimes \mathbb{C}^d$ cannot be perfectly distinguished by LOCC measurements \cite{Ghosh2004} and the same is also true for PPT measurements \cite{Yu2012}.
Using duality theory for PPT measurements, it was shown that for any $d=2^t$ (with integers $t\geq 2$), the average success probability of discriminating $k$ orthogonal maximally entangled states is bounded by $\frac{7d}{8k}$ \cite{Cosentino2014}.
This proves that there exists cases with $k=d$ and even cases with $k<d$ for which LOCC discrimination cannot be exact; see also \citet{Yu2012}.
Using similar SDP techniques, the case of $k=d$ was then shown for any $d\geq 4$ \cite{LiMS2015}.
A variation of this discrimination problem is to allow for some auxiliary entanglement, to be consumed in the discrimination protocol.
Using duality theory, it was shown in \citet{Bandyopadhyay2015} that the optimal average success probability of discriminating four equiprobable Bell states by LOCC is $\frac{1}{2}\left(1+\sqrt{1-\epsilon^2}\right)$ where the auxiliary entangled state is $\sqrt{\frac{1+\epsilon}{2}}\ket{00}+\sqrt{\frac{1-\epsilon}{2}}\ket{11}$.
Note that a perfect discrimination is possible only with a maximally entangled state.
The latter is true also for discriminating three Bell states.
In contrast, using PPT relaxations, the entanglement cost for discriminating three Bell states corresponds to only $\epsilon=1/3$ \cite{Yu2014}.
Another variation of these problems is to consider having many copies of the entangled state.
It is well known that having many copies of orthogonal pure states permits perfect LOCC discrimination but that the same does not need to hold for orthogonal mixed states \cite{Bandyopadhyay2011}.
SDP techniques have shown that this behaviour for mixed states also persists under PPT measurements \cite{Yu2014, Li2017}.

Frequently, a quantum resource can be completely characterised in terms of its ability to yield an advantage in a certain discrimination task, which thereby lends it an operational interpretation.
We will now discuss how SDP techniques make these connections possible.
Consider for instance quantum steering, i.e.,~the inability of a state assemblage $\varrho_{a|x}=\tr_A\left(A_{a|x}\otimes \openone \rho_{AB}\right)$ to admit a local hidden state model, which was discussed in Section~\ref{sec:introduction:entanglement}; see Eq.~\eqref{eq:assemblage}.
It was shown in \citet{Piani2015} that a necessary and sufficient condition for $\rho_{AB}$ to be steerable is that it yields an advantage over all non-entanglement-based strategies in the task of sub-channel discrimination when measurements are limited to be implementable by one-way LOCC.
Specifically, an instrument is defined as $\mathcal{I}=\{\Lambda_x\}$ where each $\Lambda_x$ is a completely positive and trace non-increasing map (a sub-channel).
The instrument is normalised to the channel $\Lambda=\sum_x \Lambda_x$.
The discrimination task is to prepare a suitable state $\sigma$, apply the instrument $\mathcal{I}$ and then perform a measurement $\{M_{a}\}$ with the aim of determining the specific sub-channel, i.e.,~of outputting $a=x$.
The best protocol naturally involves an optimisation over both $\sigma$ and $\{M_a\}$, leading one to define the ``no entanglement'' performance as $p_{NE}=\max_{\sigma, M} \sum_{x}\tr(\Lambda_x[\sigma]M_x)$.
This can now be compared to the best protocol using the entangled state $\rho_{AB}$, where the instrument is applied on system $B$.
Let the measurements be adaptive product measurements, namely $M_a=\sum_b A_{a|b}\otimes B_{b}$ for some local POVMs $\{A_{a|b}\}$ and $\{B_{b}\}$.
One then finds that for every steerable state there exists $\mathcal{I}$ such that the success probability, $p_{E}=\max_{M}\sum_x \tr((\openone\otimes\Lambda_x)[\rho_{AB}] M_x)$, exceeds $p_{NE}$.
To arrive at this, one studies the quantity known as the steering robustness, $R(\rho_{AB})$.
It is defined as a supremum over a corresponding robustness quantity defined for assemblages, $R(\{\varrho_{a|x}\})$.
The latter quantity is the smallest $t\geq 0$ for which the assemblage can be decomposed as $\varrho_{a|x}=(1+t)\varrho_{a|x}^\text{US}-t\tau_{a|x}$ for some unsteerable assemblage $\{\varrho_{a|x}^\text{US}\}$ and some arbitrary assemblage $\{\tau_{a|x}\}$.
Thus, it is in a sense the smallest amount weight for a steerable assemblage necessary to recover $\{\varrho_{a|x}\}$.
It can be computed as the SDP
\begin{equation}\label{steeringrobustness}
	\begin{aligned}
		1+R(\{\varrho_{a|x}\})=\min& \quad \sum_{\lambda} \sigma_\lambda\\
		\st& \quad \sum_\lambda p(a|x,\lambda) \sigma_\lambda\succeq \varrho_{a|x}\quad \forall\,a,x, \\
		& \quad \sigma_\lambda \succeq 0 \qquad\qquad\qquad\quad\,\,\,\, \forall\,\lambda.
	\end{aligned}
\end{equation}
From this, one can show that $\frac{p_E}{p_{NE}} \leq 1+R(\{\varrho_{a|x}\})\leq 1+R(\rho_{AB})$, where the last inequality follows by definition.
In \citet{Piani2015}, it is shown that analysis of the dual SDP allows one to strengthen this to $\frac{p_E}{p_{NE}} = 1+R(\rho_{AB})$.
Hence, for any steerable state there exists local measurements that give rise to a steerable assemblage, which thus exhibits $R(\{\varrho_{a|x}\})>0$.
This implies that there exists some instrument for which the entanglement-assisted sub-channel discrimination (with one-way adaptive local measurements) exceeds the entanglement-unassisted limit, i.e.,~$p_E>p_{NE}$.
We note that while the steering robustness of an assemblage is useful for understanding sub-channel discrimination, there are many ways of quantifying steerability via SDPs, see, e.g., \citet{Skrzypczyk2014, Cavalcanti2016b} and the review paper \citet{CavalcantiReview2016}.

In general, quantum resources associated with convex sets (e.g.~entanglement, steering and joint measurability) can be systematically associated with resource quantifiers in the spirit of Eq.~\eqref{steeringrobustness}.
We already saw an example of this around Eq.~\eqref{entanglementprimal} for PPT states (which is a convex set).
When the quantum resource does not admit an SDP characterisation, the quantifiers typically also do not admit an SDP representation.
Still, they can often be formulated within the more general framework of conic programming.
While conic programs in general lack the important feature of being efficiently computable, they can still admit a duality theory that parallels the one discussed for SDPs.
For instance, a conic programming approach to separable measurements appears in \citet{Bandyopadhyay2015} and constructions of robustness-type measures for arbitrary convex measurement sets is developed in \citet{Uola2019, Takagi2019}.
Thus, regardless of the set admitting an SDP or a more general conic characterisation, a non-zero distance (in terms of a suitable robustness measure) from such a set can often be interpreted in terms of an operational advantage in a tailored discrimination task.

For instance, consider a quantum state discrimination task based on a pre-determined set of states $\rho_{a|x}$.
Alice selects $x$ from a prior $p(x)$, draws $a$ from some prior $p(a|x)$, and then sends both $x$ and $\rho_{a|x}$ to Bob.
Bob's task is to learn the value $a$.
There are two types of protocols permitted.
In either, Bob has a pre-defined set of incompatible measurmements, $\{M_{b|y}\}$.
Based on the knowledge of $x$, he stochastically selects which measurement to perform using some distribution $p(y|x,\mu)$ with prior $p(\mu)$.
The measurement outcome is then used, together with the knowledge of $(x,\mu)$, to select the final guess for $a$.
Contrasting this, an alternative protocol is for Alice to perform, in each round, a single measurement on $\rho_{a|x}$ and then use $y$ to  postprocess the outcome into the final guess for $a$.
It was shown in \citet{Carmeli2018} that $\{M_{b|y}\}$ must be incompatible in order for the first type of strategy to outperform the latter.
Using the knowledge of robustness-type resource quantifiers, one can show that measurement incompatibility is in fact both necessary and sufficient \cite{Carmeli2019, Skrzypczyk2019b, Uola2019}.
That is, for every set of incompatible measurements, there exists a tailored choice of a state discrimination task, of the above form, where they can outperform any protocol-based compatible measurements.
The relevant quantifier of the incompatibility resource is the incompatibility robustness discussed in Section~\ref{secJM}, which is computable via SDP.
Notably such ideas can also be extended to discrimination tasks for witnessing the incompatibility of channels \cite{Mori2020}.
A related but different example is when one considers measurements from a resource theory perspective.
A device that maps quantum states to classical outcomes is a measurement but not all such devices require one to actually interact with the quantum state.
A trivial example is a device that bins the state and then flips a coin and outputs the result.
In \citet{Skrzypczyk2019c}, such trivial measurements are identified as POVMs of the form $\{p(a)\openone\}_a$, for some probability distribution $\{p(a)\}$ and a robustness measure is formulated for quantifying the extent to which a given measurement deviates from such a trivial realisation.
This measure is computable via SDP and, by examining its dual, one can formulate a standard quantum state discrimination problem where the degree of advantage in the non-trivial measurements corresponds exactly to their robustness.

We also mention some examples of the previous ideas being applied to convex sets that are not fully characterised by SDPs, but instead by more general conic constraints.
A prime example is the set of bipartite separable states.
The previously discussed task of sub-channel discrimination assumed measurements that are compatible with one-way LOCC.
If one instead permits generic bipartite measurements, it can be shown that every entangled state (i.e.,~the unsteerable ones as well) yields an advantage in the task \cite{Piani2009}.
This can also be extended to high-dimensional entanglement, as quantified by the Schmidt number of $\rho_{AB}$.
It is shown in \citet{Bae2019} that every state with a Schmidt number larger than $k$ can outperform any state with a Schmidt number of at most $k$ in sub-channel discrimination.
Another example is witnessing the advantage of non-projective measurements with respect to protocols based on projective measurements.
The latter is a convex set but cannot be characterised as an SDP due to the projectivity condition.
Nevertheless, one can always find an advantage in a tailored state discrimination task \cite{Uola2019}.
The same also applies to sets of quantum states, as compared to the restricted set of quantum states that can be collectively diagonalised by a single unitary \cite{Designolle2021}.
These ideas have also been merged with the previous discussion of LOCC discrimination of entanglement in order to link advantages to local robustness measures \cite{Sen2024}.

\section{Randomness and quantum key distribution}
\label{sec:qkd}
Quantum cryptography offers a means to execute cryptographic tasks with information-theoretic security, with randomness generation and quantum key distribution (QKD) being the most well-studied primitives in this domain.
In this section we describe how SDP hierarchies can be used to quantify randomness and compute rates of QKD protocols, referring the reader to \citet{Gisin2002, Pirandola20020, Portmann2022, Scarani2009, Xu2020} for more in-depth reviews on the topic.

A QKD protocol consists of two parties, Alice and Bob, who want to establish a shared random string that is unknown to any potential adversary.
To do this they execute a procedure that generates a classical-classical-quantum system, $\rho_{ABE}$, where $A$ and $B$ are classical systems held by Alice and Bob, respectively, and $E$ is a quantum system held by a potential adversary.
From this system they can then try to postprocess the classical systems $A$ and $B$ to produce random strings $K_A = K_B$ that are not correlated with $E$.
In order to assess the security and performance of such a protocol one needs to compute (or at least lower bound) its asymptotic rate,\footnote{It is also possible to compute non-asymptotic rates from the asymptotic rates even against non-IID adversaries~\cite{Dupuis2020}.} i.e., the number of secret key bits generated per round of the protocol as the number of rounds tends to infinity.
For example, for QKD with one-way error correction, against an adversary who applies the same attack each round on the protocol independently of the other rounds (an independent and identically distributed, or IID, adversary) the asymptotic rate is given by the Devetak-Winter bound~\cite{Devetak2005},
\begin{equation}\label{eq:DW_bound}
	\min_{\rho_{ABE} \in \mathcal{S}} H(A|E) - H(A|B),
\end{equation}
where $H(X|Y):= H(XY) - H(Y)$ with $H(X) := -\tr(\rho_X \log_2 \rho_X)$ the von Neumann entropy, and the minimisation is over the set $\mathcal{S}$ of all classical-classical-quantum states $\rho_{ABE}$ that are compatible with the protocol.
Therefore the exact set $\mathcal{S}$ depends on the protocol used and the statistics observed.
The asymmetry in the Devetak-Winter bound comes from the restriction to protocols with one-way error correction, i.e., all the error correction is sent by Alice to Bob.
It is possible to interpret the second term $H(A|B)$ as approximately the rate of bits that Alice must send to Bob for him to successfully correct his raw key to be equal to hers.

Note that, as $A$ and $B$ are observed by Alice and Bob, one can estimate $H(A|B)$ directly from the statistics of the protocol.
Thus, the main task remaining is to bound from below 
\begin{equation}\label{eq:entropy_opt}
	\min_{\rho_{ABE} \in \mathcal{S}} H(A|E)\,.
\end{equation}
There are two main difficulties to overcome in order to compute bounds on Eq.~\eqref{eq:entropy_opt}.
First, the objective function is a nonlinear function of $\rho_{ABE}$, and second, one needs to characterise the set $\mathcal{S}$ of possible states $\rho_{ABE}$ output by the protocol when $E$ is a system unknown to Alice and Bob.
The latter depends significantly on the security model of the protocol, and so, in the following we will consider the different approaches suited to the different security models.
It is further possible to consider different adversaries.
For example, one could make a restriction to classical adversaries, that forces $E$ to be a classical system, and hence Eve cannot be entangled with the initial systems of Alice and Bob.

\subsection{Device-independent approach}
In the device-independent security model, pioneered by the ideas of \citet{Ekert1991} and \citet{Mayers1998}, Alice and Bob each have an untrusted device that they use to generate nonlocal correlations (see Fig.~\ref{FigBellScenario}).
It is assumed, without loss of generality, that the devices produce their outcomes given their inputs by measuring some projective measurements, $\{A_{a|x}\}_{a,x}$ and $\{B_{b|y}\}_{b,y}$, on a bipartite state $\rho_{Q_AQ_B}$ (see Eq.~\eqref{correlations}), where the subindex $Q$ indicates that the system is quantum in contrast to the classical systems $A$ and $B$ that hold the measurement outcomes of Alice and Bob.
In this security model the source $\rho_{Q_AQ_B}$ is not trusted, and hence there may exist an adversarial party holding a system $E$ that is potentially entangled with the systems $Q_A$ and $Q_B$.
In a device-independent protocol, Alice and Bob verify that the correlations $p(a,b|x,y)$ generated by their devices satisfy some linear constraints,
\begin{equation}\label{eq:di_constraints}
	\sum_{a,b,x,y}r_{abxy,i} \,p(a,b|x,y) \geq \omega_i \qquad \forall\,i,
\end{equation}
where $r_{abxy,i},\,\omega_i \in \mathbb{R}$ are specified by the protocol.
They can, for instance, verify that their devices achieve some sufficiently high average CHSH violation.
Thus, the set of states that are required to optimise over in Eq.~\eqref{eq:entropy_opt} are exactly the post-measurement states whose statistics are compatible with the corresponding version of Eq.~\eqref{eq:di_constraints} imposed by the protocol.
Using the NPA hierarchy it is in principle possible to relax the existence of such quantum systems satisfying Eq.~\eqref{eq:di_constraints} to a hierarchy of SDPs as was detailed in Section~\ref{sec:NPA}.
What then remains is to convert the objective function $H(A|E)$ into something that can be expressed within the framework of noncommutative polynomial optimisation.

\subsubsection{Bounding the min-entropy}\label{sec:qkd:di:minentropy}
For a classical-quantum state, $\rho_{XY} = \sum_x \ketbra{x}{x} \otimes \rho_Y(x)$, the conditional min-entropy of $X$ given $Y$ is defined as $H_{\min}(X|Y) := - \log_2 P_g(X|Y)$, where 
\begin{equation}\label{eq:guessing_prob}
	P_g(X|Y) := \max_{\{M_x\}_x} \sum_x \tr(M_x \rho_Y(x)),
\end{equation}
and the maximisation is over all POVMs $\{M_x\}_x$ on system $Y$.
Operationally, the quantity $P_g(X|Y)$ corresponds to the maximum probability with which someone who has access to system $Y$ can guess the value of system $X$ and hence $P_g$ is referred to as the guessing probability~\cite{Konig2009}.
This operational interpretation also implies that the min-entropy rates for a classical adversary coincide with those of a quantum adversary.
Eve effectively creates a classical system upon measuring, which implies the existence of a classical strategy achieving the same min-entropy bound.

Using the fact that $H \geq H_{\min}$ one can immediately get lower bounds on rates by lower bounding the min-entropy, or equivalently upper bounding the guessing probability.
In particular, when Alice inputs $X=x$ we have
\begin{equation}\label{eq:di-pguess1}
	P_g(A|E) = \max_{\{M_a\}} \sum_a \tr\left((A_{a|x} \otimes \id \otimes M_a) \rho_{Q_AQ_BE}\right),
\end{equation}
where $\{M_a\}$ is now some POVM given to the adversary (which can be assumed to be projective)~\cite{Masanes2011}.
By applying the tools of noncommutative polynomial optimisation the corresponding rate optimisation can be relaxed to a hierarchy of SDPs whose moment matrices are generated by the monomials $\{\id\}\cup \{A_{a|x}\}\cup \{B_{b|y}\}\cup\{M_a\}$ and the $k$-th level relaxation is given by 
\begin{equation}\label{eq:di_hmin_sdp1}
	\begin{aligned}
		\max& \quad \sum_a \Gamma^k(A_{a|x}, M_a) \\
		\st& \quad \sum_{a,b,x,y} r_{abxy,i}\,\Gamma^k(A_{a|x}, B_{b|y}) \geq \omega_i \qquad \forall\,i, \\
		& \quad \Gamma^k \succeq 0, 
	\end{aligned}
\end{equation}
where, as usual, we have not explicitly specified all the constraints present in Eq.~\eqref{eq:di_hmin_sdp1}, e.g., those coming from projectivity, commutativity, orthogonality and normalisation amongst others.
Taking $-\log_2$ of any solution to Eq.~\eqref{eq:di_hmin_sdp1} will therefore allow to lower bound the rates of device-independent QKD or device-independent randomness generation protocols.

By tracing out the $E$ system one can increase the efficiency of these relaxations in terms of the dimension of the SDP~\cite{Bancal2014,Nieto-Silleras2014} as Eve's operators are removed from the relaxation and subsequently the size of the moment matrix is reduced.
In particular, one can view Eve's measurement, upon obtaining the outcome $c$, as preparing the subnormalised state $\rho_{Q_AQ_B}(c) = \tr_{E}((\id_{Q_AQ_B}\otimes M_c)\rho_{Q_AQ_BE})$ for Alice and Bob, which satisfies a normalisation condition $\sum_c \rho_{Q_AQ_B}(c) = \rho_{Q_AQ_B}$.
Thus it is possible to create a relaxation for each of the states $\rho_{Q_AQ_B}(c)$, leading to several smaller moment matrix blocks instead of one large moment matrix for the state $\rho_{Q_AQ_B}$.
In particular, one can instead write the $k$-th level relaxation as
\begin{equation}
	\begin{aligned}
		\max& \quad \sum_a \Gamma^k_a(A_{a|x}, \id) \\
		\st& \quad \sum_{a,b,x,y,c} r_{abxy,i}\,\Gamma^k_c(A_{a|x}, B_{b|y}) \geq \omega_i \quad \forall\,i, \\
		& \quad \sum_{c} \Gamma^k_c(\id,\id) = 1, \\
		& \quad \Gamma^k_c \succeq 0 \qquad\qquad\qquad\qquad\qquad\qquad\hspace{0.6em} \forall\,c,
	\end{aligned}
\end{equation} 
where we have again omitted many implicit constraints.

The SDP bounds on $H_{\min}$ have been used extensively to analyse the device-independent randomness generated from different Bell inequalities in the presence of noise~\cite{Mironowicz2013, Bancal2014b, Law2014}, from non-inequality settings~\cite{LiHW2015}, in the presence of leakage~\cite{Silman2013,Tan2023}, from post-selected events~\cite{Thinh2016,Xu2022}, from PPT states~\cite{Vertesi2014} and from partially entangled states~\cite{Gomez2019}.
The efficiency of the SDPs allows them to be used to help optimise the experimental design to maximise randomness~\cite{Mattar2015} and through the dual it is possible to extract functions on which the experimental parameters can be optimised~\cite{Assad2016}.
The dual also provides a function on the space of correlations that lower bounds the certifiable device-independent randomness, which can then be used to create full security proofs of the corresponding protocols~\cite{Nieto-Silleras2018,Brown2019}.
Employing these SDP relaxations it is also possible to verify that a four-outcome POVM can be used to produce two bits of device-independent randomness using a maximally entangled qubit pair~\cite{Acin2016,Gomez2016} and similar advantages from non-projective measurements also appear in systems with higher dimensions \cite{Tavakoli2021MUB}.
The technique can also be applied to more exotic correlation scenarios like sequential measurements~\cite{Bowles2020} to show robust generation of more randomness than would be possible with just a single projective measurement or within the instrumental causal structure~\cite{Agresti2020}.
It was also used together with analytical investigations to demonstrate that a sequence of non-projective measurements can be used to generate unbounded amounts of randomness from a single maximally entangled qubit pair~\cite{Curchod2017}.

\subsubsection{Bounding the von Neumann entropy}\label{sec:qkd_di_vn}
The min-entropy approach provides a simple method to lower bound the rates of protocols.
However, secure asymptotic and non-asymptotic rates are often expressed in terms of the von Neumann entropy~\cite{Arnon2018}, which in general is larger than the min-entropy.
Thus, in order to find tighter lower bounds on the rate of a protocol it is necessary to find a way to lower bound the von Neumann entropy more accurately.

In \citet{Tan2021} the authors use duality relations of entropies to remove Eve from the problem, following a similar approach taken in \citet{Coles2016} to view the measurements of Alice and Bob through the lens of an isometry.
After rephrasing the problem in terms of only Alice and Bob, they provide a lower-bounding ansatz, which after applying a Golden-Thompson inequality~\cite{Sutter2017}, they express as a noncommutative polynomial optimisation problem that can in turn be relaxed to a hierarchy of SDPs.

In \citet{Brown2021a} the authors introduce a sequence of conditional Rényi entropies\footnote{R\'enyi entropies are usually single-parameter families of entropic quantities.
For an in-depth overview we refer the reader to \citet{Tomamichel2015}.}, that are all lower bounds on $H(A|E)$.
Each of these entropies is defined in terms of a solution to an SDP that emerges from the SDP representability of the matrix geometric mean \cite{Fawzi2017}.
Similar to the min-entropy, these R\'enyi entropies can be each used to give a hierarchy of SDPs that lower bound the rates.
In \citet{Gonzales-Ureta2021} the method was used to derive improved device-independent QKD rates in settings with more inputs and outputs.

While the above two methods improve over the min-entropy, it is not clear that either can compute tight lower bounds on the von Neumann entropy.
In \citet{Brown2021b} a sequence of variational forms that converge to the von Neumann entropy from below was introduced.
Let $m \in \mathbb{N}$ and let $t_1,\dots,t_m$ and $w_1,\dots,w_m$ be the nodes and weights of an $m$-point Gauss-Radau quadrature rule\footnote{A Gaussian quadrature rule approximates an integral $\int_a^b f(x) dx$ by a finite sum $\sum_{i} w_i f(t_i)$, where $w_i$ are referred to as weights and $t_i$ as nodes. We refer the reader to \citet{Davis1984} for further details.} on $[0,1]$ with $t_m=1$~\cite{Golub1973}.
For each $m \in \mathbb{N}$ the following noncommutative polynomial optimisation problem is a lower bound on the rate $\inf H(A|E)$, 
\begin{widetext}
	\begin{equation}\label{eq:diqkdsdp}
		\begin{aligned}
			\min & \quad c_{m} + \sum_{i=1}^{m-1}\frac{w_i}{t_i \log 2} \sum_a \tr\left\{ \rho_{Q_AQ_BE} \left[A_{a|x} (Z_{a,i} + Z_{a,i}^{\dagger} + (1-t_i) Z_{a,i}^{\dagger} Z_{a,i} ) + t_i Z_{a,i}Z_{a,i}^{\dagger}\right] \right\} \\
			\st & \quad \sum_{a,b,x,y} r_{abxy,i}\tr \left( \rho_{Q_AQ_BE} A_{a|x} B_{b|y}\right) \geq \omega_i, \\
			& \quad Z^{\dagger}_{a,i} Z_{a,i} \preceq \alpha_i^2 \id, \\
			& \quad [A_{a|x}, B_{b|y}] = [A_{a|x}, Z_{c,i}] = [B_{b|y}, Z_{c,i}] = 0,
		\end{aligned}
	\end{equation}
\end{widetext}
where $\alpha_i = \tfrac32 \max \{\tfrac{1}{t_i}, \tfrac{1}{1-t_i}\}$, $c_m = \sum_{i=1}^{m-1} \tfrac{w_i}{t_i \log 2}$, and the operators generating the noncommutative polynomial optimisation are $\{A_{a|x}\} \cup \{B_{b|y}\} \cup \{Z_{c,i}, Z_{c,i}^{\dagger}\}$.
This can then be relaxed to an SDP using the techniques detailed in Section~\ref{sec:NoncommutativePoly}.
In particular, the core of the relaxation consists of a moment matrix generated by the monomials $\{\id\}\cup \{A_{a|x}\}\cup \{B_{b|y}\}\cup\{Z_{c,i}, Z^\dagger_{c,i}\}$.
It is worth noting that the $Z_{c,i}$ operators are not Hermitian, and hence their adjoint must also be included in the generating set.

The construction leads to a double hierarchy, namely, a hierarchy of variational bounds indexed by $m$, and for each of these bounds, an SDP relaxation hierarchy.
For applications presented in \citet{Brown2021b}, a value of $m=8$ or $m=12$ was typically used alongside various tricks to speed up the computations.
This method has been shown to recover the known tight bounds on rate curves.
It has been used to compute randomness generation rates in mistrustful settings \cite{Metger2022}, i.e., when Alice does not trust Bob, as well as for assessing the optimal randomness certifiable in the binary input and binary output scenario \cite{Wooltorton2022}.

When Alice and Bob have only binary inputs and outputs, the analysis of the rate can be reduced to the analysis of qubit systems through the use of Jordan's lemma \cite{Masanes2006}.
For certain Bell inequalities it is then possible to solve the entropy optimisation analytically \cite{Pironio2009}.
In \citet{Masini2022} a hybrid analytical-numerical approach is introduced for binary settings.
It is shown that after reducing to qubit systems, it is possible to use further symmetries and properties of entropies to express the rate of the problem as an analytical function of some unknown correlators.
These correlators can then be bounded by a commutative polynomial optimisation of just a few variables.
This can then be relaxed to a hierarchy of SDPs using the Lasserre hierarchy (recall Section~\ref{sec:lasserre}).
The advantage over the techniques discussed previously is that the numerical optimisations are significantly smaller (and hence faster).
Furthermore, it can achieve tight bounds in certain cases.

A similar hybrid approach is taken in \citet{Schwonnek2021}, where the authors analyse key rates achievable in binary input and binary output settings when the key is extracted from different input settings.
After reducing to qubit systems, they reformulate the problem as a triplet of nested optimisations, with the innermost optimisation an SDP arising from the SDP formulation of the trace norm,
\begin{equation}
	\begin{aligned}
		\|K\|_1 = \min & \quad \frac12 \tr(X+Y) \\
		\st & \quad \begin{pmatrix}
			X & K \\
			K^\dagger & Y
		\end{pmatrix} \succeq 0,
	\end{aligned}
\end{equation}
where $K$ is any square matrix~\cite{Watrous2018}.

\subsubsection{Beyond entropy optimisations}
Thus far we have focused on randomness generation and QKD with one-way error correction, the rates of which both require solving a minimisation of some entropy.
When one moves beyond these protocols the relevant figure of merit will often change.
Nevertheless, SDP relaxation techniques remain readily applicable to these new settings.

Two-way error correction in QKD, i.e., when both Alice and Bob can communicate in the error correction step of the protocol, is known as advantage distillation.
\citet{Tan2020} showed that a sufficient condition for secret key generation in this binary-outcome protocol is
\begin{equation}
	\min F(\rho_{E|00}, \rho_{E|11}) > \sqrt{\frac{\epsilon}{1-\epsilon}},
\end{equation}
where $F(\rho,\sigma) := \|\sqrt{\rho}\sqrt{\sigma}\|_1$ is the fidelity, $\epsilon$ is the probability that Alice and Bob's outcomes disagree on the key-generating inputs $x$ and $y$, i.e., $\epsilon := \sum_{a\neq b} p(a,b|x,y)$, and $\rho_{E|ab} =\frac{1}{p(ab|xy)} \tr_{Q_AQ_B}[(M_{a|x} \otimes N_{b|y}\otimes \id) \rho_{Q_AQ_BE}]$ is the marginal state of Eve conditioned on Alice receiving outcome $a$ and Bob receiving outcome $b$.
Various approaches to compute the minimisation of the fidelity have been proposed.
In \citet{Tan2020} the fidelity was lower bounded by a guessing probability, resulting in an optimisation that could be tackled using techniques introduced in \citet{Thinh2016}.
In \citet{Hahn2021} it was shown that one can apply a fidelity-preserving measurement to the quantum states to reduce the objective function to just a function of probabilities and then, by using techniques similar to those introduced in \citet{Himbeeck2019}, arbitrarily tight bounds on the fidelity can be directly computed using noncommutative polynomial optimisation.
A stronger sufficient condition was introduced in \citet{Stasiuk2022}, based on the Chernoff divergence $Q(\rho,\sigma) := \min_{0 < s < 1} \tr(\rho^s \sigma^{1-s})$, and indirect lower bounds were analysed using a guessing probability lower bound and the resulting SDP relaxations, similar to \citet{Tan2020}.

In contrast to QKD, wherein Alice and Bob trust each other, mistrustful cryptography aims to execute a cryptographic task between two agents who do not trust each other.
The relevant figures of merit for these protocols are then the probability that each agent can cheat.
Bit commitment is such a protocol in which Alice commits a bit to Bob.
After commitment Alice should not be able to modify the bit, and Bob should be able to learn the bit only when Alice chooses to reveal it.
In \citet{Aharon2016} a device-independent bit-commitment protocol based on the CHSH game was introduced, and it was shown that the probability that Alice can cheat could be relaxed to a noncommutative polynomial optimisation problem similar to a guessing probability problem.
Similar SDP relaxations were also derived for the cheating probabilities of Alice and Bob in an \textsc{xor} oblivious transfer protocol based on the magic square game \cite{Kundu2020}.

In both randomness generation and QKD, regardless of the security model, various classical postprocessing of the data is performed.
In particular, a procedure known as randomness extraction is necessary to transform the measurement outputs of the devices into $\epsilon$-approximate secret uniform randomness.
In~\citet{Berta2015} it is shown that the condition for a procedure to be a valid randomness extractor can be recast as a quadratic program.
This quadratic program can then be relaxed to an SDP, giving a certificate for a procedure to be a secure randomness extractor.

\subsection{Device-dependent approach}
\label{sec:qkdB}
The opposite of the device-independent approach is having a full characterisation of the honest parties' devices involved in a protocol.
In this device-dependent scenario, Alice and Bob measure some fixed, known POVMs $\{A_{a|x}\}_{a,x}$ and $\{B_{b|y}\}_{b,y}$, on a bipartite state $\rho_{Q_AQ_B}$.
As before, the source is not trusted, and thus no assumptions on $\rho_{Q_AQ_B}$ are made, except that it must be compatible with the statistics measured by Alice and Bob.
In particular, an adversarial party holds a quantum system, $E$, that is potentially entangled with the systems $Q_A$ and $Q_B$, with the global state denoted by $\rho_{Q_AQ_B E}$.
The condition that is compatible with the measured statistics is then expressed as
\begin{equation}\label{eq:dd_constraints}
	\tr\big[(W_i \otimes \id_E \big)\rho_{Q_AQ_B E} \big] = \omega_i \qquad \forall\,i,
\end{equation}
where $W_i = \sum_{a,b,x,y} c_{abxy,i} A_{a|x} \otimes B_{b|y}$ is specified by the protocol and $\omega_i$ is given by the measured statistics.
For example, one can choose $\{A_{a|x}\}_{a,x}$ and $\{B_{b|y}\}_{b,y}$ to be mutually unbiased bases, and $W_i$ to compute the probability that Alice and Bob get equal results when measuring in the same bases \cite{Sheridan2010}.
Equation \eqref{eq:dd_constraints} is a simple linear constraint compatible with SDP, so in order to compute the key rate one simply needs to convert the objective function $H(A|E)$ from Eq.~\eqref{eq:DW_bound} into an expression that can be optimised via SDPs.

\subsubsection{Bounding the min-entropy}
As in the device-independent case, it is possible to lower bound the von Neumann entropy by the min-entropy and compute the latter from the guessing probability given by Eq.~\eqref{eq:di-pguess1}.
This equation can be linearised in the optimisation variables by absorbing the $\{M_c\}_c$ into $\rho_{Q_AQ_BE}$, this is, defining 
\begin{equation}
	\rho_{Q_AQ_B}(c) = \tr_E \left((\id \otimes \id \otimes M_c) \rho_{Q_AQ_BE}\right),
\end{equation}
so that $\rho_{Q_AQ_B} = \sum_c \rho_{Q_AQ_B}(c)$.
Computing the guessing probability is done by the following SDP:
\vspace*{-.3pt}
\begin{equation}
	\begin{aligned}
		\max_{\{\rho_{Q_AQ_B}(a)\}}&\quad \sum_a \tr\left((A_{a|x} \otimes \id) \rho_{Q_AQ_B}(a)\right) \\
		\st &\quad 	\sum_a \tr\left(W_i\rho_{Q_AQ_B}(a)\right) = \omega_i\quad \forall\,i, \\
		& \quad \sum_a \tr\big(\rho_{Q_AQ_B}(a)\big) = 1, \\
		& \quad \rho_{Q_AQ_B}(a) \succeq 0\qquad\qquad\qquad\, \forall\,a.
	\end{aligned}
\end{equation}
This was used in \citet{Doda2021} to demonstrate improved noise tolerances for QKD using high-dimensional systems.
The min-entropy, however, is a rather loose lower bound on the von Neumann entropy, so the key rates computed in this way are unnecessarily pessimistic.
The main advantage of this technique is simplicity.

\subsubsection{Bounding the von Neumann entropy}\label{sec:qkd_dd_vn}
The variational forms converging to the von Neumann entropy introduced by \citet{Brown2021b} can also be applied to compute the key rate in the device-dependent case \cite{Araujo2022}.
Having trusted measurement devices drastically simplifies the problem: for a given matrix element on Alice and Bob's side, one builds an NPA hierarchy for Eve together with the quantum state, resulting in a block matrix SDP as done in \citet{Navascues2014}.
This NPA-type hierarchy converges at the first level because there are no commutation relations to enforce, and thus for fixed $m$ one has a single SDP.

As shown in \citet{Coles2016}, the $H(A|E)$ term in the key rate can be rewritten as $D(\mathcal{Z}(\rho_{Q_AQ_B})\| \mathcal{Z}[\mathcal{G}(\rho_{Q_AQ_B})])$ where $D(\rho\|\sigma) := \tr( \rho\log_2(\rho)-\rho\log_2(\sigma))$ is the relative entropy and $\mathcal{Z}$ and $\mathcal{G}$ are quantum channels.
This rewriting achieves two things: it removes Eve from the problem, and it results in an objective function that is convex in the variables $\rho_{Q_AQ_B}$.
Thus, the key-rate calculation becomes a convex optimization problem.
In~\citet{Winick2018}, it is shown that a Frank-Wolfe-type algorithm~\cite{Frank1956} can be used to solve the problem.
The authors also provide a method to convert feasible points of the convex optimization problem into rigorous lower bounds on the key rate: by linearizing the objective at a given feasible point $\rho_{Q_AQ_B}$, the convex optimization problem becomes an SDP, and a guaranteed lower bound can be obtained using the weak duality of SDPs.
Dedicated interior-point algorithms to solve the optimization problem have also been developed~\cite{Hu2022} with improved stability and convergence rates over Frank-Wolfe-type methods.
The convex optimization can also be done using the semidefinite approximations of the matrix logarithm developed in~\citet{Fawzi2019b} giving an alternate method that uses SDPs~\cite{Bunandar2020,Hu2022}.

More recently, effective conic methods have been developed to optimize over non-symmetric cones \cite{skajaa2015,papp2017,coey2023}.
Together with the modeling of the relative entropy as a non-symmetric cone \cite{fawzi2022c}, this has made it possible to solve the key-rate problem directly, without relying on SDP relaxations or dedicated algorithms \cite{Lorente2024,He2024}.
Such methods offer a dramatic improvement of performance over previous techniques.

\subsection{Semi-device-independent approach}\label{sec:SDI}
Semi-device-independent (SDI) protocols offer a tradeoff between the high security of device-independence and the ease of implementation of device-dependent protocols.
Ideally, one looks to add one or more easily verifiable assumptions that enable the resulting protocol to be implemented in a significantly simpler manner.
Often these protocols reduce to a prepare-and-measure scenario (recall Fig.~\ref{FigCommunicationScenario}) similar to those discussed in Section~\ref{sec:communication}, wherein Alice randomly prepares one of the states $\{\rho_x\}_x$ and sends it to a measurement device (Bob) that performs one of several measurements to generate the data that becomes the source of randomness or the secret key.
Different assumptions can then be placed on the source device and the measurement device to generate different SDI protocols.
Interestingly, if one has an SDP relaxation of the set of correlations achievable within this setting, then one can often optimise entropies over these sets to bound the rates of randomness generation protocols.
For several assumptions these SDP relaxations of the correlation sets have already been discussed in Section~\ref{sec:communication}.

The first work to introduce SDI cryptography considers a QKD scheme in the prepare-and-measure setting that assumes that the quantum states prepared by Alice and sent to Bob are qubits \cite{Pawlowski2011}.
By restricting the dimension of the Hilbert space, the set of possible prepare-and-measure correlations $p(b|x,y)$ is also restricted.
Importantly, akin to the situation in Bell-nonlocality, there exist qubit correlations that cannot be realized by classical systems of dimension 2.
By witnessing these correlations under the qubit assumption, Alice and Bob can verify that their systems must be acting in a non-classical manner and crucially, non deterministically.
Using analytical results concerning dimension witnesses and random access coding, the authors are able to demonstrate that secret key can be extracted from certain correlations, although their analysis is limited to an ideal protocol.
Later, SDI protocols for randomness generation based on dimension bounds were introduced~\cite{Li2011, Li2012, Mironowicz2016}.
In \citet{Mironowicz2016} the authors use the dimension-restricted correlations hierarchy of \citet{Navascues2015b} (see also Section~\ref{sec:NV}) to compute a lower bound on the min-entropy that can be certified from devices that achieve some minimal success probability for a quantum random access code.
The issue with an upper bound on the dimension of a quantum system is that it is difficult to physically justify, hence works then looked for protocols that rely on other assumptions that are easier to verify.

\citet{Himbeeck2019} analyses SDI protocols for randomness generation under assumptions of bounded average and maximal energy of the quantum states sent from Alice to Bob, and a full security proof against classical adversaries is given.
The resulting correlation set has an SDP representation (see Section~\ref{sec:communication-distinguish}), which the authors use to analyse the randomness generated against a classical adversary.
To achieve this they have to deal with the nonlinearity of $\sum_x p(x) h(E_x)$, where $p(x)$ is the input distribution, $h(x) := - x \log_2 x - (1-x) \log_2(1-x)$ is the binary entropy, and $E_x$ is some observed quantity in the protocol.
As $h$ is concave, one can define a sequence of piecewise linear lower bounds on $h$ that approximates $h$ arbitrarily well in the limit.
This leads in \citet{Himbeeck2019} to a hierarchy of SDP bounds on the rates based on this approximation.

SDI randomness amplification protocols, i.e., SDI randomness generation with a partially trusted seed, can also be performed under the assumption of energy bounds~\cite{Senno2021}.
A similar approach was taken in \citet{Jones2022}, where it is shown that the formalism developed in \citet{Himbeeck2019} can also be applied to assumptions on the spacetime symmetries of the protocol, which further can be treated independently of quantum theory.

If the source is limited to the preparation of two states then the energy bounds give a physical justification for a minimal overlap $|\braket{\psi_x}{\psi_y}| > 0$ in the states prepared by the source.
A number of works have also analysed SDI randomness generation directly under the assumption of a minimal overlap between the states $\{\ket{\psi_x}\}_x$ sent by the source to the measurement device.
In \citet{Brask2017} it is noted that any attempt to unambiguously discriminate between $\ket{\psi_0}$ and $\ket{\psi_1}$ must contain some randomness in whether or not the discrimination is conclusive.
An SDP based on $H_{\min}$ is formulated to bound the randomness generated with respect to an adversary who may share classical correlations with the measurement device.
This SDP can also be generalised to more inputs and outputs to achieve higher randomness generation rates~\cite{Tebyanian2021}.
Employing time-bin encoding and single-photon detection, \citet{Carceller2024b} used SDP to certify more than one bit of randomenss per round in this framework.
In \citet{Roch2022} the minimal overlap assumption is used to show that quantum devices can generate more randomness than noncontextual devices that obey an analogous assumption, and an SDP to compute the randomness for noncontextual devices is also presented.
Note that, since these works assume the preparation of pure states, they are able to restrict the analysis to the finite-dimensional subspace spanned by those states.
As a result, the computation of the rates can be directly written as a single SDP.

An SDI randomness generation protocol under which all overlaps $\braket{\psi_x}{\psi_y}$ are known is presented in \citet{Ioannou2021}.
This work also analyses the randomness generated against a quantum adversary who has access to the quantum channel between the source and the measurement device and can use this to create entanglement between themselves and the measured states.
By adapting the SDP relaxation method of \citet{Wang2019} (see~Eq.~\eqref{eq:wang_grammethod}) one can then compute a hierarchy of min-entropy bounds against the adversary.
A similar SDI randomness generation protocol is also presented in \citet{Wang2023} in which the source is fully characterised but the quantum channel and Bob's measurement device is untrusted.
Using also the techniques of \citet{Wang2019} an SDP relaxation of min-entropy bounds is given and a full security analysis is provided.

An alternative bounded overlap condition is presented in \citet{Tavakoli2021b}, where it is assumed that the source prepares states that are close to some fixed set of target states (see Section~\ref{sec:communication-distinguish}).
The randomness generated by the measurement device with respect to an adversary who shares classical correlations with both the source and the measurement device is then analysed using the min-entropy and SDP relaxations of the underlying correlation set of the scenario.
An alternative to the various overlap assumptions is given in \citet{Tavakoli2022a}, wherein a bound on the information transmitted through the prepare-and-measure channel is assumed (see also Section~\ref{sec:communication-distinguish}).
The certifiable randomness is evaluated using the min-entropy for various bounds on the transmitted information, and comparisons to the dimension-bounded protocols are also given.

The setting of measurement-device-independent (MDI) QKD offers stronger security than device-dependent QKD as well as a longer distance~\cite{Lo2012}.
In~\citet{Winick2018,Hu2022} it is shown that the methods developed for device-dependent QKD can also be used to compute key rates of MDI-QKD protocols and variants like twin-field QKD~\cite{Lucamarini18}.
In \citet{Primaatmaja2019} an SDP method to bound the phase error rate of a variety of protocols is derived that can in turn be used to compute their rates via the Shor-Preskill formula~\cite{Shor2000}.
Other works in the setting of uncharacterised measurement devices have investigated randomness generated in steering scenarios~\cite{Passaro2015} and with trusted quantum inputs~\cite{Supic2017}.

\section{Correlations in networks}
\label{sec:networks}
A line of research in quantum correlations takes the study of entanglement-based correlations beyond the traditional Bell-type scenarios and into the domain of networks.
Networks are composed of a number of parties that are connected to each other through multiple independent sources that each emit physical systems.
A party can then perform measurements on the shares received from several different sources, see e.g.~Fig.~\ref{fig:triangle}.
Not only is the framework of networks the appropriate one to analyze long-distance entanglement-based and communication scenarios, but it has also provided new insights into the fundamentals of quantum theory \cite{Renou2021,Abiuso2022}.

Characterising classical, quantum or no-signaling correlations in networks is a major theory challenge.
The origin of the difficulty is that the presence of multiple independent sources renders the correlation sets non-convex and therefore one cannot rely on more standard tools from convex optimisation theory.
For this reason, relaxation hierarchies have become a useful way to approach correlations in networks.
In this section we provide brief introductions to these methods and their applications, referring to \citet{TavakoliPozas2022} for in-depth discussions of these and other aspects of correlations in networks.

\subsection{Inflation methods}
\label{sec:networks:inflation}
\begin{figure*}
	\centering
	\subfloat[\label{fig:triangle}]{
		\includegraphics[width=0.22\textwidth,clip,trim={0.5cm 0 0.5cm 0}]{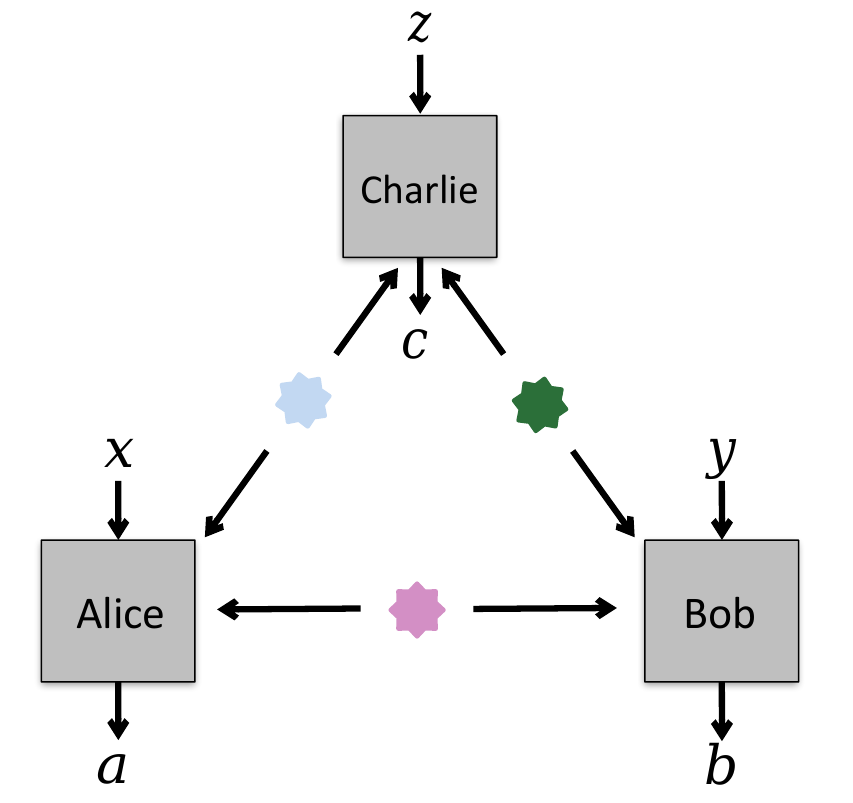}
	}
	\hspace{0.02\textwidth}
	\subfloat[\label{fig:cinf}]{
		\includegraphics[width=0.22\textwidth,clip,trim={0 0 0 -1cm}]{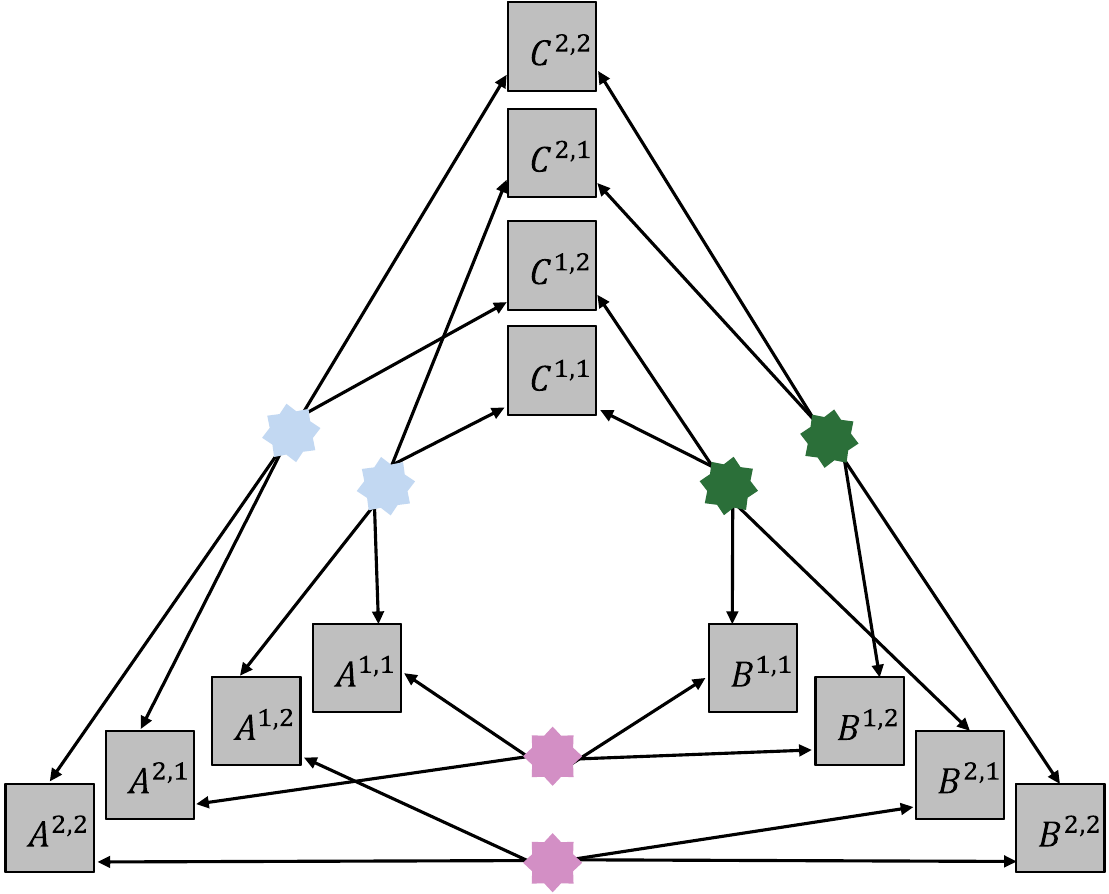}
	}
	\hspace{0.02\textwidth}
	\subfloat[\label{fig:qinf}]{
		\includegraphics[width=0.22\textwidth,clip,trim={0 0 0 -1.5cm}]{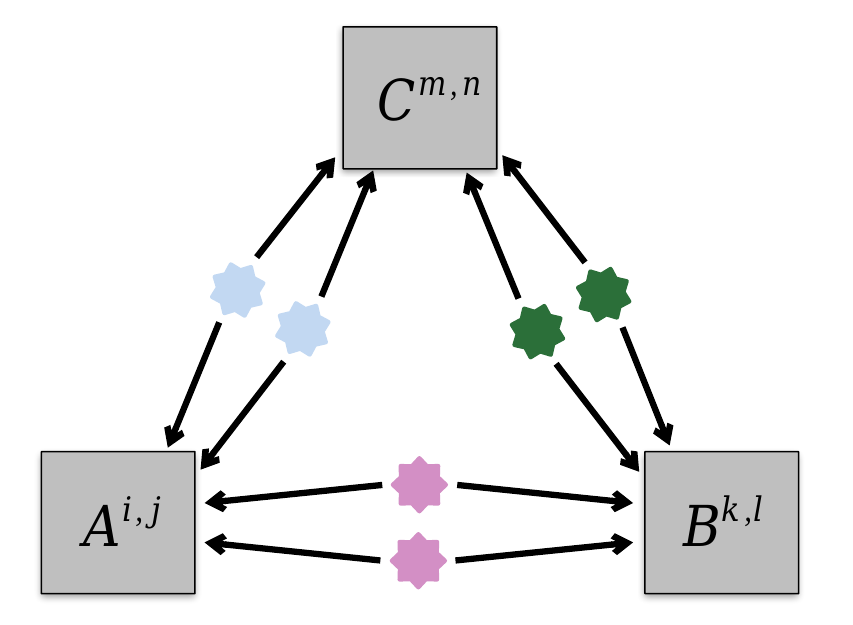}
	}
	\hspace{0.02\textwidth}
	\subfloat[\label{fig:ginf}]{
		\includegraphics[width=0.2\textwidth]{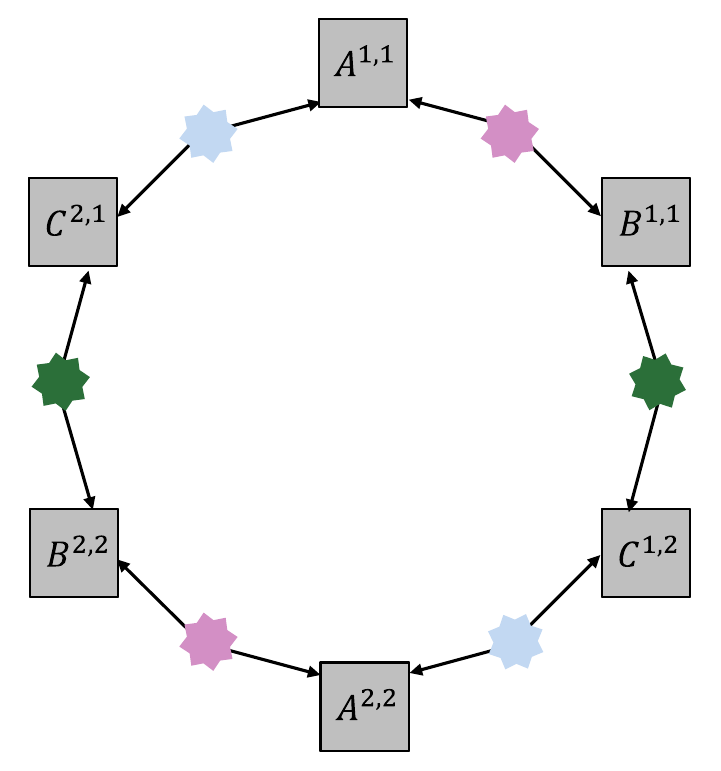}
	}
	\caption{\protect\subref{fig:triangle} The triangle network, and the second level of its \protect\subref{fig:cinf} classical, \protect\subref{fig:qinf} quantum, and \protect\subref{fig:ginf} NSI inflation hierarchies.
	Note that the inputs and outputs have been omitted in \protect\subref{fig:cinf}-\protect\subref{fig:ginf} for clarity.
	The superindices denote the copies of the sources that each party acts upon.}
	\label{fig:inflation}
\end{figure*}

Inflation is a general framework for characterising correlations in causal structures in general and networks in particular, via SDP or LP relaxations.
The main idea in inflation is to substitute the original network and its non-convex constraints for a larger network, created by copies of the sources and parties of the original network, in which the constraints are relaxed to linear symmetry constraints.
An illustration is shown in Fig.~\ref{fig:inflation}, where the problem of characterising the probability distributions $p(a,b,c|x,y,z)$ compatible with the triangle scenario in Fig.~\ref{fig:triangle} is relaxed by the characterisation of distributions $p_\text{inf}(\{a^{i,j}\},\{b^{k,l}\},\{c^{m,n}\}|\{x^{i,j}\},\{y^{k,l}\},\{z^{m,n}\})$ compatible with one of the inflations in Figs.~\ref{fig:inflation}\subref{fig:cinf}-\subref{fig:ginf}, where the superindices denote the particular copies of the sources that are used to produce a particular value.
Thus, one trades a simple but technically challenging problem for one that can be solved using standard methods in a more complicated network.
The construction of the inflated network strongly depends on the physical model underlying the network.
The three main models of interest are (i) classical models, corresponding to associating each source with an independent local variable, (ii) quantum models, in which each source is associated with an independent entangled quantum state, and (iii) models only constrained by no-signaling and independence (NSI) assumptions, where each source is independently associated to a general nonlocal resource required only to respect the no-signaling principle.
The key difference in the construction of the inflations stems from the fact that the case with local variables allows free copying of information, while the quantum and no-signaling cases do not.
We now discuss the basics of the three types of inflation.

\subsubsection{Classical inflation}
In classical networks, the sources can be described by random variables and the measurement devices can without loss of generality be seen as deterministic functions of the classical variables received by a particular party.
Since classical information can be copied, an inflated network may feature copies not only of the sources and measurement devices, but also of the concrete values of the local variables distributed by the sources \cite{Wolfe2019}.

An example of a classical inflation is presented in Fig.~\ref{fig:cinf} for the triangle-shaped network of Fig.~\ref{fig:triangle} where the parties' outputs are labeled $a$, $b$, and $c$ respectively.
Since the sources and measurement devices are copies of those in the original network, the correlations $p_\text{inf}$ seen in the inflation must satisfy the symmetries
\begin{equation}
	\begin{aligned}
		p_\text{inf}(&\{a^{i,j}\},\{b^{k,l}\},\{c^{m,n}\})\\
		&=p_\text{inf}(\{a^{\pi(i),\pi'(j)}\},\{b^{\pi'(k),\pi''(l)}\},\{c^{\pi''(m),\pi(n)}\}),
	\end{aligned}
	\label{eq:inflation:classical_sym}
\end{equation}
for independent permutations $\pi$, $\pi'$, and $\pi''$ of the different copies of the sources.
Moreover, marginals of $p_\text{inf}$ over parties that reproduce (parts of) the original network can be directly associated with the probability distribution $p_\text{orig}$ in the original network,
\begin{equation}
	p_\text{inf}(\Pi_i \{a^{i,i},b^{i,i},c^{i,i}\})=\Pi_i p_\text{orig}(a^{i,i},b^{i,i},c^{i,i}).
	\label{eq:inflation:classical_iden}
\end{equation}
For instance, one of such constraints in the inflation in Fig.~\ref{fig:cinf} is $p_\text{inf}(a^{1,1},a^{2,2},b^{1,1},b^{2,2},c^{1,1},c^{2,2})=p_\text{orig}(a^{1,1},b^{1,1},c^{1,1})p_\text{orig}(a^{2,2},b^{2,2},c^{2,2})$.
It is important to note that the constraints in Eq.~\eqref{eq:inflation:classical_iden} can only be imposed in feasibility problems, where $p_\text{orig}$ is given and thus the right-hand side is a number.
When optimising over the set of distributions compatible with a given inflation, $p_\text{orig}$ are also variables, and thus the linear constraints that can be imposed are just $p_\text{inf}(a^{i,i},b^{i,i},c^{i,i})= p_\text{orig}(a^{i,i},b^{i,i},c^{i,i})$ $\forall\,i$.

A third type of constraint is relevant for feasibility problems \cite{AlexThesis,Pozas2022b}.
These are constraints on marginals that factorise, and some of the factors can be associated with $p_\text{orig}$.
An example, also illustrated in the inflation of Fig.~\ref{fig:cinf}, is
\begin{equation}
	\begin{aligned}
		p_\text{inf}(&a^{1,1},a^{2,2},b^{1,1},b^{1,2},c^{1,1},c^{2,1})\\
		&=p_\text{orig}(a^{2,2})p_\text{inf}(a^{1,1},b^{1,1},b^{1,2},c^{1,1},c^{2,1}).
	\end{aligned}
	\label{eq:inflation:classical_lpi}
\end{equation}
Since all of the discussed constraints are linear for a given $p_\text{orig}$, the task of finding a probability distribution $p_\text{inf}$ compatible with the same constraints can be cast as an LP.
Interestingly, the inflation technique \cite{Wolfe2019} provides a complete solution to the characterisation of classical network correlations.
This was shown in \citet{Navascues2017} by identifying a sequence of inflation tests that, in the limit of large inflation, converges to the set of local correlations associated with the original network.
This sequence is constructed such that the $n$-th test features $n$ copies of each of the sources of the original network.
The inflation illustrated in Fig.~\ref{fig:cinf} corresponds to the second step in this sequence of converging tests for the case of the triangle network.
Although it guarantees convergence in the asymptotic limit, one must bear in mind that the complexity of the hierarchy of LPs (measured by the number of elements in the probability distribution $p_\text{inf}$) grows in $n$ as $N^{n^r}$, where $N$ is the number of outcomes for a party and $r$ is the amount of sources that send states to a party.

Inflation has become a standard tool in the analysis of network nonlocality and it has many times been employed in the depicted triangle network.
For a specific inflation, the set of compatible no-input binary-outcome correlations was completely characterised \cite{Wolfe2019}.
This characterisation was later found not to admit quantum violations, but other Bell-like inequalities for the triangle scenario, with four outcomes per party, were found that do admit noise-robust quantum violations \cite{Fraser2018}.
Classical inflation has also been used to show that a shared random bit cannot be realised in the triangle network with classical sources \cite{Wolfe2019}, and to find certificates of more genuine quantum nonlocality in the four-output triangle scenario by studying the dual of an inflation LP \cite{Pozas2022b}.
Inflation methods have also enabled examples of nonlocality that use a smaller number of outputs \cite{Boreiri2022,Pozas2023}.
Moreover, outside the domain of networks, it has been used to determine equivalences between causal structures that produce the same correlations \cite{Ansanelli2022}.

\subsubsection{Quantum inflation}
In quantum networks, the sources distribute entangled quantum systems and the parties perform quantum measurements on the subsystems at their disposal.
In contrast to the classical case, quantum theory must respect the no-cloning theorem \cite{Park1970,Wootters1982,Dieks1982}, which prevents valid inflations from copying the individual subsystems produced in the quantum sources and distributing them between additional parties.
This limits the inflations allowed for a network.
Taking again as example the triangle scenario of Fig.~\ref{fig:triangle}, Born's rule gives the quantum correlations
\begin{equation}
	\begin{aligned}
		p_Q^\triangle(a,b,c|&x,y,z)\\
		=&\tr(A_{a|x}\otimes B_{b|y}\otimes C_{c|z} \cdot \rho_{AB}\otimes\rho_{BC}\otimes\rho_{CA}),
	\end{aligned}
	\label{eq:qtriangle}
\end{equation}
which is analogous to Eq.~\eqref{correlations}.
Quantum inflation \cite{Wolfe2021} relaxes the fact that the global state in the scenario, $\rho_{AB}\otimes\rho_{BC}\otimes\rho_{CA}$, has a tensor-product form.
It does so via a \textit{gedankenexperiment} similar to that of the previous section: if states and operators satisfying Eq.~\eqref{eq:qtriangle} exist, then one can have multiple copies of them, and thus there exist (at least) one state $\rho$ and operators $A_{a|x}^{i_{CA},i_{AB}}$, $B_{b|y}^{i_{AB},i_{BC}}$, and $C_{c|z}^{i_{BC},i_{CA}}$ (where the new superindices indicate which copies of the corresponding sources are acted upon) that satisfy the analogs of Eqs.~\eqref{eq:inflation:classical_sym} and \eqref{eq:inflation:classical_iden} at the level of Born's rule, namely
\begin{equation}
	\begin{aligned}
		\tr&[\rho\cdot Q(\{A_{a|x}^{i,j}\},\{B_{b|y}^{k,l}\},\{C_{c|z}^{m,n}\})]\\
		& = \tr[\rho\cdot Q(\{A_{a|x}^{\pi(i),\pi'(j)}\},\{B_{b|y}^{\pi'(k),\pi''(l)}\},\{C_{c|z}^{\pi''(m),\pi(n)}\})],
	\end{aligned}
	\label{eq:inflation:quantum_sym}
\end{equation}
for any polynomial $Q$ of the operators, and
\begin{equation}
	\begin{aligned}
		\tr\bigg[\rho\cdot \bigotimes^n_{i=1}\Big(A_{a_i|x_i}^{i,i}&\otimes B_{b_i|y_i}^{i,i}\otimes C_{c_i|z_i}^{i,i}\Big)\bigg]\\
		& = \Pi^n_{i=1} p_Q^\triangle(a_i,b_i,c_i|x_i,y_i,z_i),
	\end{aligned}
	\label{eq:inflation:quantum_iden}
\end{equation}
for any $n$ and any independent permutations $\pi$, $\pi'$ and $\pi''$ of the copies of a same source.
Now, one can develop a hierarchy in $n$, which will denote the amount of copies of each source in the inflation network (Fig.~\ref{fig:qinf} contains the $n=2$ quantum inflation of the triangle scenario).
For each $n$ the characterisation of the states and operators that satisfy Eqs.~\eqref{eq:inflation:quantum_sym} and \eqref{eq:inflation:quantum_iden} has the same form of that discussed in Section~\ref{sec:nonlocality:npa} with some additional linear constraints at the level of expectation values.
Thus, this characterisation can be approximated by the NPA hierarchy.

Therefore, the implementation of quantum inflation comprises two different hierarchies.
The first hierarchy is that of inflations increasing the number of copies of the sources in the network.
Then, for each inflation, there is an NPA-like hierarchy to characterise the correlations compatible with such inflation.
The latter hierarchy is known to converge to the set of quantum correlations with a commutation structure (which, as explained in Section~\ref{sec:nonlocality:npa}, is a relaxation of the tensor-product structure that coincides with it for finite-dimensional Hilbert spaces).
For the former, it is not yet known whether the hierarchy where step $n$ represents the inflation with $n$ copies of each source (defined in \citet{Wolfe2021}) converges, except in the particular case of the bilocality network (see Fig.~\ref{fig:swapping} in Section~\ref{sec:networks:swapping}) \cite{Ligthart2022}.
There exists another, provably convergent, SDP sequence for quantum network nonlocality, namely that of \citet{Ligthart2021}.
However, this is not a hierarchy in the standard sense because it is not monotonic: failing a particular SDP test does not imply that the subsequent SDP tests will fail too.

\citet{Wolfe2021} outlines with examples various families of applications of quantum inflation.
These include certifying that distributions are impossible to generate in a concrete quantum network, optimizing over distributions that can be generated in a quantum network, extracting polynomial witnesses of incompatibility, and a concrete practical example bounding the information that an eavesdropper could obtain in cryptographic scenarios involving quantum repeaters.
Notably, additional commutation constraints can be added to the quantum inflation SDPs in order to constrain the resulting correlations to be classical.
These SDPs can be seen as semidefinite relaxations of the LPs of the classical inflation hierarchy of the previous section, where one can trade off computational power for accuracy.

\subsubsection{No-signaling and independence}
\label{sec:networks:inflation:nonfanout}
Correlations in networks can be characterised subject only to minimal physical constraints, namely only by the independence of the sources and by the no-signaling principle.
While noting that other tools also apply to this task (\citealp{Renou2019,Beigi2022}; see also Section~\ref{sec:networks:topology}), inflation methods present a systematic hierarchy approach for the purpose.
The principles for NSI inflation were already put forward in the original work \cite{Wolfe2019}: not only can physical systems not be cloned, but also the compatibility relations between the measurements that are performed on the physical states distributed are not characterised.
This means, in practice, that measurement devices receive only one copy of each relevant system.
These two requirements significantly constrain the set of allowed inflations (see, e.g., Fig.~\ref{fig:ginf}).
The characterisation of correlations compatible with NSI inflations can be formulated in terms of a single LP for each inflation.

NSI inflations (also known as \textit{non-fanout} inflations in the literature, see, e.g.,~\citet{Wolfe2019}) have been explicitly used in the context of extending the role of the no-signaling principle to networks, in situations where the parties do not have a choice of measurements to perform on their systems \cite{Gisin2020}, and in demonstrations of nonlocality in the simplest scenario in the triangle network, namely that in which all the parties do not have inputs and produce binary outcomes \cite{Pozas2023}.
This has led to the definition of the analogous of a Popescu-Rohrlich box \cite{Popescu1994} for network correlations, based on \citet{Bancal2021}.
Moreover, the agnosticity of the physical model has been used as a theoretical basis for proposing a definition of genuine $n$-partite nonlocality based on the idea that correlations cannot be simulated in any network using global classical randomness and nonlocal resources shared between $n-1$ parties \cite{Coiteux2021a,Coiteux2021b}.
However, in contrast to the classical and quantum versions of inflation, it is not true that the steps in the NSI inflation hierarchy describe sets of correlations that are contained in those corresponding to lower levels, although the behavior observed in practice is that of monotonically improving bounds.
Currently, how to define network correlations subject only to the existence of independent sources and no-signaling is a complicated matter \cite{Henson2014}, and in fact NSI inflations have been proposed as such a definition that is physically well motivated \cite{Beigi2022}.

It is interesting to note that the various inflation techniques can be combined in order to address correlations in networks where different sources distribute systems of different natures.
This is especially easy in the case of classical and NSI inflation, since both are naturally formulated in terms of LPs.
For example, such hybrid inflations are useful tools for tests of full network nonlocality \cite{Pozas2022,Wang2023exp,Gu2023,Luo2023}, where one aims to certify that every source in a network must uphold some degree of nonlocality in order to model observed correlations.
However, hybrid networks are still mostly \textit{terra incognita}.

\subsubsection{Entanglement in networks}
A natural question when studying quantum networks regards what sort of entangled states can be produced in a given network.
Indeed, this question has recently received considerable attention \cite{Kraft2021,Navascues2020,Luo2021,Kraft2021b}.
Taking again as illustration the triangle network of Fig.~\ref{fig:triangle}, if the sources distribute quantum states $\sigma\in\mathcal{H}_{A''B'}$, $\mu\in\mathcal{H}_{B''C'}$, and $\tau\in\mathcal{H}_{C''A'}$, and the parties perform local operations characterised as completely positive and trace-preserving maps $\Omega_P:\mathrm{B}(\mathcal{H}_{P'}\otimes\mathcal{H}_{P''})\rightarrow\mathrm{B}(\mathcal{H}_P)$, all states that can be produced in the triangle network take the form
\begin{equation}
	\rho^\triangle=[\Omega_A\otimes\Omega_B\otimes\Omega_C](\sigma\otimes\mu\otimes\tau).
	\label{eq:trianglestate}
\end{equation}
While the individual characterisation of any of the components of the expression above can be cast as an SDP (recall, e.g., the seesaw procedure in Eqs.~\eqref{seesaw1}, \eqref{seesaw2}), the complete characterisation of $\rho^\triangle$ cannot.

The problem of characterising the quantum states that can be produced in quantum networks is addressed in \citet{Navascues2020} via SDP relaxations based on inflation.
In a spirit similar to that described in Section~\ref{sec:networks:inflation}, these relaxations transform the conditions on the independence of the sources in the network into symmetry constraints in more complicated networks, created using copies of the original sources and operations.
The main difference is that while the symmetry constraints were enforced at the level of expectation values of operators when studying nonlocality in Section~\ref{sec:networks:inflation}, when studying network entanglement the constraints are enforced at the level of the quantum state in the inflation and its marginals.
The fact that the state must be a PSD operator allows the problem to be phrased as a single SDP for a fixed inflation.
In \citet{Navascues2020}, these relaxations are used to bound the maximum fidelities of known multipartite states with network realisations, which are later interpreted as witnesses of genuine network entanglement.
Similarly, \citet{Hansenne2022,Wang2022} provides no-go theorems regarding the preparation of cluster and graph states in networks.

\subsection{Other SDP methods in network correlations}
In addition to inflation, in some specific scenarios, there exist other methods for characterising network correlations
They range from analytic methods to alternative LP and SDP relaxations.
Below we review some of the latter.

\subsubsection{Relaxations of factorisation}
\label{sec:networks:swapping}
\begin{figure}
	\centering
	\includegraphics[width=0.95\columnwidth]{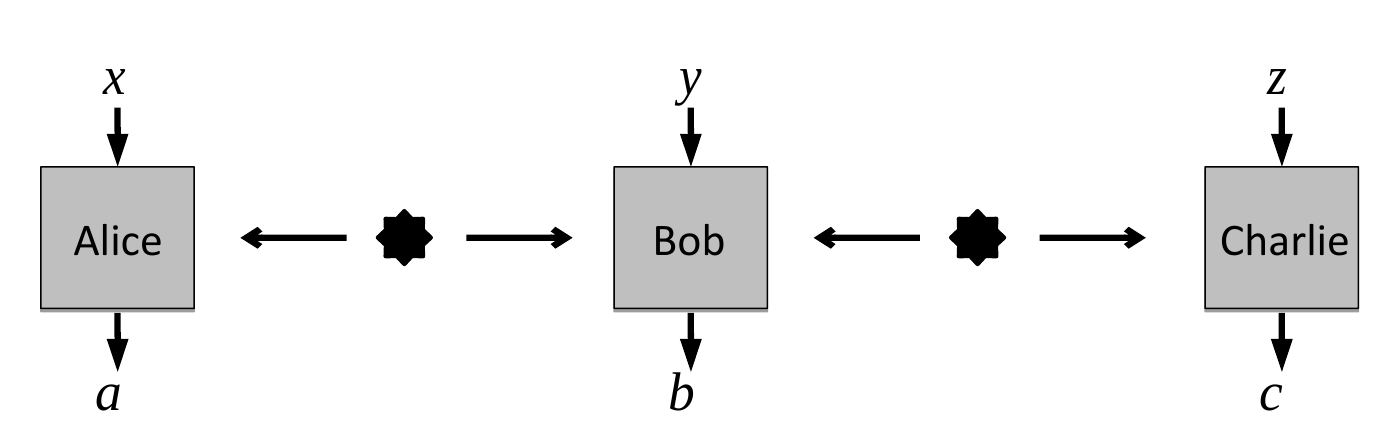}
	\caption{The bilocality network: a central party receives shares from two independent sources, each of which sends another share to a different party.
	This network underlies the simplest entanglement-swapping experiments and quantum repeaters.}
	\label{fig:swapping}
\end{figure}

Particles that have never interacted can become entangled via the seminal process of entanglement swapping \cite{Zukowski1993}.
The simplest entanglement-swapping scenario is that in which two parties, Alice and Charlie, share each a bipartite physical system with a central party, Bob, that performs entangled operations on the two systems received \cite{Pan1998, Jennewein2001} (see Fig.~\ref{fig:swapping}).
Recently this setting has been employed for showing that real Hilbert spaces have less predictive power than the complex Hilbert spaces postulated by quantum theory \cite{Renou2021}.
The associated entanglement-swapping scenario assumes that the two sources share no entanglement but may be classically correlated.
SDP relaxation methods based on the PPT constraints compatible with the interpretation of the NPA hierarchy discussed in Section~\ref{sec:nonlocality:di:moroder} are then employed to bound the predictive power of real quantum models.

However, it is also relevant to consider entanglement swapping when the sources are also classically uncorrelated.
Any correlation between Alice and Charlie must then be mediated by Bob, i.e.,~two-body expectation values factor as \mbox{$\expect{AC}=\expect{A}\expect{C}$}, thereby breaking the convexity of the sets of relevant correlations.
This was the first network scenario considered in the literature and for which specific network Bell inequalities were developed \cite{Branciard2010,Branciard2012}.

\citet{Pozas2019} presents SDP hierarchies that relax the non-convex sets of probability distributions generated in networks where some parties are not connected to others.
These hierarchies are modifications of the NPA hierarchy discussed in Section~\ref{sec:nonlocality:npa}.
The main idea of the modification consists in a \textit{scalar extension}; allowing the rows and columns of moment matrices to be labeled not only by operators, but also by sub-normalised operators that denote products of an actual normalised operator in the problem and a (possibly unknown) expectation value.
The elements in the moment matrices that are generated from these sets of operators will contain variables that represent products of expectation values, and can be associated via linear constraints to other variables upon which one wishes to impose factorisation.

Take, as an illustration, the entanglement-swapping scenario with Alice and Charlie performing two dichotomic measurements each, $\{A_0,A_1\}$ and $\{C_0,C_1\}$, respectively, and the moment matrix generated\footnote{See footnote \ref{fn:bordermatrix}.} by the set of operators $\{\id,A_0A_1,C_0C_1,\expect{A_0A_1}\id\}$:
\begin{equation}
	\Gamma = \kbordermatrix{
		& \id & A_0A_1 & C_0C_1 & \expect{A_0A_1}_\rho\id \\[0.3ex]
		\id & 1 & v_1 & v_2 & v_3 \\[1ex]
		(A_0A_1)^\dagger & & 1 & v_4 & v_5 \\[1ex]
		(C_0C_1)^\dagger & & & 1 & v_6 \\[1ex]
		\expect{A_0A_1}_\rho^*\id & & & & v_7
	}.
	\label{eq:scalarextension}
\end{equation}
Per the standard NPA prescription, we have that $v_1\,{=}\,v_3$ because they both evaluate to \mbox{$\expect{A_0A_1}_\rho$}, and that $v_5\,{=}\,v_7$ because both evaluate to \mbox{$\expect{\expect{A_0A_1}_\rho(A_0A_1)^\dagger}_\rho\,{=}\,|\expect{A_0A_1}_\rho|^2$}.
Moreover, and importantly, in the entanglement-swapping scenario it holds that, regardless of the particular operators and quantum state, $\expect{A_0A_1C_0C_1}_\rho\,{=}\,\expect{A_0A_1}_\rho\expect{C_0C_1}_\rho$.
This constraint is enforced in Eq.~\eqref{eq:scalarextension} by setting the linear constraint $v_4\,{=}\,v_6$.
Since all constraints between the variables in Eq.~\eqref{eq:scalarextension} are linear, the set of distributions admitting a PSD $\Gamma$ can be characterised via SDP.

A hierarchy can then be generated by taking the levels of the associated NPA hierarchy and, for each of them, adding as many new columns as are needed to impose at least one linear constraint per element of $\Gamma$ that should factorise.
For the entanglement-swapping scenario, the hierarchy constructed in such a way converges to the desired set of quantum distributions \cite{Renou2022}.
However, the proof technique used there is difficult to generalise to more complicated networks \cite{mukherjee2015,Tavakoli2014}, where the SDP method still applies.
The method has been used, for instance, to show that networks can activate the usefulness of measurement devices for detecting network nonlocality \cite{Pozas2019} and to provide quantum bounds on Bell inequalities tailored to the entanglement-swapping scenario \cite{tavakoli2021}.

\subsubsection{Tests for network topology}
\label{sec:networks:topology}
SDP relaxations can also be used to rule out a hypothesised causal structure, i.e.,~a network constellation, connecting a given number of measuring parties.
In some situations it is fairly easy to detect correlations that could not have been produced in a concrete network.
A simple illustration is the network in Fig.~\ref{fig:swapping}.
As discussed earlier, if one considers the marginal distribution of Alice and Charlie, the resulting correlations must factor, since there is no connection between the two parties.
Thus, any non-factoring correlations between Alice and Charlie are impossible to generate in the bilocal network.
While non-linear, these constraints can be linearised, for instance, by working with the entropies of the variables instead of the probabilities \cite{Weilenmann2017}.
However, in other networks like the triangle network of Fig.~\ref{fig:triangle}, such simple criteria do not exist because all parties are connected to all sources.

\citet{Aberg2020}, building on a similar characterisation for correlations admitting local models in networks \cite{Kela2020}, finds simple criteria that allow to discern whether correlations do not admit a realisation in a particular network, regardless of the physical model of the systems distributed by the sources.
The characterisation is based on the covariance matrix of the variables representing the outcomes of the measurements performed by the parties.
Covariance matrices are inherently PSD.
The important realisation is that the network structure imposes a decomposition of the covariance matrix in block matrices that are individually PSD as they determine the correlations established by each of the sources.
This is, for each source there is one PSD block matrix, that contains nonzero elements only in the rows and columns associated with the parties that receive systems from that particular source.
This leads to a simple and efficient way to characterise the correlations that can be generated in different networks via SDP, that has been found to be connected to the characterisation of block coherence of quantum states \cite{Kraft2021b}.
However, as discussed in \citet{Aberg2020}, this characterisation is not tight, in the sense that there exist alternative methods that better approximate the set of relevant correlations in some situations.


\section{Further topics and methods}
\label{sec:further}
In this section we collect additional topics where SDP relaxations are relevant and methods for reducing the computational load of SDP.

\subsection{Classical models for quantum correlations}\label{sec:grothendieck}
For some entangled states, the outcome statistics from performing arbitrary local measurements in Bell-type experiments can be simulated by models based on local variables.
Consequently, entanglement and nonlocality are two distinct phenomena \cite{Augusiak2014}.
The seminal example is Werner's model showing that the state $\rho_v=v\ketbra{\psi^-}{\psi^-}+\frac{1-v}{4}\id$, where $\psi^-$ is the singlet state, admits an LHV model for $v\leq \frac{1}{2}$ even though the state is entangled for any $v>\frac{1}{3}$ \cite{Werner1989}.
The critical $v$ up to which the $\rho_v$ admits an LHV model for all projective measurements turns out to be equal to the inverse of the Grothendieck constant of order 3 \cite{Acin2006}.

General methods are known for deciding whether a given entangled state admits an LHV model \cite{Hirsch2016, Cavalcanti2016}.
These methods are based on the idea of first choosing a finite collection of measurements and determining via LP how much white noise must be added to the state for there to be an LHV model of the resulting distribution.
Next, one can add sufficient white noise in the measurement space so that all the quantum measurements can be represented as classical  postprocessings of the measurements in the selected set.
In other words, the measurement space is shrunk until it is contained in the convex hull of the selected measurement set.
It then follows that for any such noisy measurement, the distribution must admit an LHV.
In a final step, one can pass the noise in the measurement space to the state space and obtain a generic LHV model for the final noisy state.
The bottleneck here is that one must choose a large measurement set to obtain a good approximation of the quantum measurement space.
This means solving an accordingly large LP or SDP.
To circumvent this, one can instead employ an oracle-based method known as Gilbert's algorithm \cite{Gilbert1966, Brierley2017} which allows one to approximate the distance between a point and a convex set in real space.
In \citet{Hirsch2017}, this algorithm is used together with a simple heuristic for the oracle to implement a polyhedral approximation of the Bloch sphere based on $625$ measurements and thereby obtain an LHV model for $\rho_v$ up to $v\approx 0.6829$.
Another option is to use instead of Gilbert's algorithm the more general Frank-Wolfe algorithm \cite{Frank1956, Bomze2021}.
This has further improved the LHV threshold to $v\approx 0.6875$ \cite{Designolle2023}.
Notably, this work also provided an improved upper bound at $v\approx 0.6961$ using 97 local measurements.
In fact, methods of this sort also work when building local hidden state models (recall Eq.~\eqref{eq:assemblage}) for entangled states.
The main difference is that one runs an SDP to check for the steerability of the assemblage, instead of the LP for membership to the local polytope.
However, when restricting to two-qubit entanglement, one can exploit geometric arguments and use increasingly large polytope circumscription and inscriptions of the Bloch sphere in order to determine bounds on the steerability of a state for infinitely many measurement settings, which can be evaluated via LP \cite{Nguyen2019}.

The idea of shrinking the quantum measurement space so that it can be inscribed in a polytope whose vertices comprise finitely many measurements can also be applied in other settings.
For example, in the case of steering, this was done in \citet{Bavaresco2017}.
A similar approach shows that there exists sets of incompatible measurements that can never be used to violate a Bell inequality \cite{Hirsch2018, Bene2018}.
This extends earlier SDP arguments that were restricted to sets of projective measurements for the uncharacterised party \cite{Quintino2016}.
Moreover, this type of approach can also be used in the prepare-and-measure scenario to determine whether the outcome statistics associated with performing arbitrary measurements on an ensemble of qubit states admits a classical model based on bits \cite{Gois2021}.
Using Gilbert's algorithm with up to $70$ measurements, non-trivial bounds have been established on the critical detection efficiency needed to violate classical constraints in the qubit prepare-and-measure scenario \cite{divianszky2022certification}.

\subsection{Generalised Bell scenarios}
A large portion of this review has focused on scenarios where all the parties share parts of the same quantum state, that are known as Bell scenarios.
These scenarios were generalised in Section~\ref{sec:networks} to account for multiple independent sources distributing systems between different collections of parties.
There exist further generalisations of the Bell scenario, that are relevant in randomness generation and in entanglement certification, and that can be characterised via SDP.

The first generalisation is that to sequential scenarios where, instead of performing only one measurement at every round, the parties perform sequences of measurements in the systems received \cite{Gallego2014,Silva2015}.
These scenarios are conceptually interesting, including in the context of randomness generation, since it is possible to extract more randomness from a given state when performing sequences of measurements \cite{Curchod2017}.
It is possible to modify the NPA hierarchy (recall Section~\ref{sec:nonlocality:npa}) in order to characterise the correlations that can be produced in these scenarios \cite{Bowles2020}.
This is achieved by considering operators that represent strings of outcomes, and requiring that these operators satisfy ``no-signaling to the past'' (i.e., that the measurement operators that define the first $k$ measurements do not depend on the $n-k$ remaining inputs, since these occur later in the sequence), which are linear constraints admitted in SDPs.
The result is a convergent hierarchy, that has been used to certify local randomness beyond two bits and for investigating monogamy properties of nonlocality.

The second generalisation that we discuss is that known as broadcast scenarios, where the systems sent to one or several of the parties are passed through channels that prepare new systems, and the outputs are distributed to multiple new parties that measure them \cite{Bowles2021}.
When the channel prepares quantum systems, the correlations in the scenario can be characterised with a slight modification of the NPA hierarchy.
This scenario has found particular interest in the activation of nonlocality.
With a quantum model, it allows to certify in a device-independent way the entanglement of Werner states in almost the entire range in which it is known to be entangled \cite{Boghiu2021}.

\subsection{Bounding ground-state energies}
A central problem in the study of many-body systems is computing or bounding the ground-state energy of the system, i.e.,~finding the minimal eigenvalue of its corresponding Hamiltonian.
This problem is known to be computationally hard, in particular it is in general QMA-complete~\cite{Kempe2006}.
Thus, computationally tractable relaxations have been sought and in particular several SDP approaches have been developed.

The structure of the problem naturally lends itself to a treatment in terms of noncommutative polynomial optimisation.
In particular, the problem takes the form $\min \tr(\rho H)$ where $H$ can be written as a polynomial of local operators.
Thus, lower bounds can be obtained from SDP relaxation techniques similar to those mentioned in Section~\ref{sec:sdp}~\cite{Baumgratz2012, Barthel2012}.

Interestingly, an apparent numerical paradox can be observed when these computations are preformed for bosonic systems~\cite{Navascues2013}.
Convergence of the semidefinite hierarchies for noncommutative polynomial optimisation problems is proven only when the operators are bounded~\cite{Pironio2010}.
Therefore, for problems involving the bosonic creation (annihilation) operators $a_i^\dagger$ $(a_i)$, the standard proof of convergence does not hold.
In fact, it can be shown that for the Hamiltonians in this setting the hierarchy collapses at level 1.
That is, higher levels give no improvement over level $1$ and the optimal value at level 1 is not equal to the optimal value of the original problem.
Nevertheless, when performing the computations numerically one can sometimes observe improving lower bounds that converge to the actual solution.
This apparent paradox is due to the finite precision of numerical computations implying that the solver is actually solving a slightly perturbed problem.
Mathematically, the set of SOS polynomials is dense in the set of positive polynomials generated by the ladder operators.
It is worth noting that a similar numerical paradox appeared in the setting of commutative polynomial optimisation~\cite{Henrion2005b} and it has a similar resolution~\cite{Lasserre2007}.

Another approach to obtaining SDP relaxations for the ground-state energy problem has been proposed in~\cite{Kull2022}.
There, the problem of computing the ground-state energy of a translation-invariant Hamiltonian with identical nearest-neighbour interactions on each pair of an infinite chain is considered.
Formally the problem can be stated as 
\begin{equation}
	\begin{aligned}
		\min & \quad \tr\left( H \rho^{(2)}\right) \\
		\st & \quad \tr(\rho^{(2)}) = 1, \\
		& \quad \rho^{(2)} \succeq 0, \\
		& \quad \rho^{(2)} \leftarrow \psi_{\mathrm{TI}},
	\end{aligned}
\end{equation}
where $\rho^{(m)}$ denotes the density matrix corresponding to an $m$-body marginal and $\rho^{(2)} \leftarrow \psi_{\mathrm{TI}}$ is the constraint that $\rho^{(2)}$ is a two-body marginal of some translation-invariant state for the entire chain.
This latter constraint is equivalent to a quantum marginal problem, asking whether there exists a global state consistent with the marginal states (see Section~\ref{sec:qmp}).
It can for instance be relaxed to the existence of all $m$-body marginals up to some finite $m_{\max}\in \mathbb{N}$, i.e., a collection of partial trace constraints $\rho^{(2)} \leftarrow \rho^{(3)} \leftarrow \dots \leftarrow \rho^{(m_{\max})}$.
This results in a hierarchy of SDP relaxations however the size of the SDPs grows exponentially in the number of sites considered.
The core idea of \citet{Kull2022} is to apply compression maps that retain the useful aspects of the marginal constraints whilst reducing the dimension of the variables significantly.

\subsection{Rank-constrained optimisation}
Several problems of interest in classical and quantum information theory can be formulated as an optimisation problem that includes a constraint in the rank of a matrix.
These include optimisation over pure quantum states, Max-Cut \cite{Goemans1995}, matrix completion \cite{Candes2010}, compressed sensing quantum state tomography \cite{Gross2010}, detection of unfaithful entanglement \cite{Weilenmann2020,Guhne2021b}, and many others \cite{Markovsky2012}.

The problem of optimising under rank constraints is in general NP-hard, and as such it is usually solved via heuristics or approximations \cite{Sun2017}.
It is possible, however, to formulate it as an SDP hierarchy similar to the DPS hierarchy discussed in Section \ref{sec:DPS} by reformulating it as a separability problem \cite{Yu2022}.
This allows one to obtain a sequence of global bounds to the problem that converge to the optimal value.

The idea is that a state $\rho$ of dimension $d$ has a rank of at most $k$ if and only if it is the partial trace of a pure state $\ket{\phi} \in \mathbb{C}^d \otimes \mathbb{C}^k$.
The set of such states is hard to characterise, but it can be handled by first noticing that one is interested only in its convex hull, and second by noticing that the convex hull is the partial trace over the last two subsystems of a state $\sigma \in \mathcal{D}(\mathbb{C}^d \otimes \mathbb{C}^k \otimes \mathbb{C}^d \otimes \mathbb{C}^k)$ that respects the constraints of being separable over the bipartition $(12|34)$, which invariant under a \textsc{swap} over the same bipartition, and recovering the initial $\rho$ through the appropriate partial trace.
Exploiting these, one can use the DPS hierarchy to characterise the separability constraint, thereby obtaining the SDP hierarchy for rank-constrained optimisation.

Note that although the idea we have explained here is in terms of quantum states, it also applies to bound ranks of general matrices.

\subsection{Quantum contextuality}
\label{sec:contextuality}
Quantum theory cannot be modeled with hidden variables that are both deterministic and non-contextual\footnote{This means that each projective measurement is assigned a definite value that is independent of other compatible measurements performed simultaneously.} \cite{Bell1966, Kochen1968}.
This is known as contextuality \cite{Budroni2022review}.
Contextuality scenarios can be cast in the language of graph theory, where each input/output tuple (event) is associated with a vertex, and an edge is drawn between two vertices if and only if the events can be distinguished by jointly measurable observables.
While many different contextuality tests can be associated with the same graph, both the non-contextual hidden variable and the quantum bounds associated with a given graph can be bounded in terms of the graph's independence number and the Lov\'asz theta function (recall Eq.~\eqref{LovaszTheta}), respectively \cite{Cabello2014}.
These quantities are computable via LP and SDP, respectively.
The connection with the Lov\'asz theta function has been leveraged to self-test quantum states in contextuality scenarios by examining the dual SDP \cite{Bharti2019}.
Generally, it is possible to adapt the NPA hierarchy to arbitrary tests of contextuality by leveraging the fact that compatible projective measurements commute, which adds constraints to the moment matrix \cite{Acin2015}.

A more operational notion of contextuality has also been proposed, that is not specific to quantum theory and not limited to projective measurements \cite{Spekkens2005}.
Two preparations (respectively, measurements) are considered indistinguishable if they cannot be distinguished through any measurement (respectively, preparation) allowed in the theory.
They are said to belong to the same context and are therefore assigned the same realist representation.
When operationally indistinguishable preparations give rise to statistics that do not admit such a realist model, the theory is said to be contextual.
This test can be cast as an LP, see e.g.~\citet{Selby2022}.
The set of preparation contextual quantum correlations can be bounded by hierarchies of SDPs.
Two different hierarchies have been proposed.
One leverages the idea and SDP methods of informationally restricted correlations \cite{Tavakoli2022a} reviewed in Section~\ref{sec:communication-distinguish}, by interpreting the indistinguishability of two preparations as the impossibility of accessing any information about which preparation was selected \cite{Tavakoli2021a}.
The other relies on using unitaries in the monomial representation, and connecting them to POVMs via the fact that every $0\preceq M \preceq \id$ can be written in terms of a unitary, $M=\frac{\id}{2}+\frac{U+U^\dagger}{4}$ \cite{Chaturvedi2021}.
Both methods require an extensive use of localising matrices to deal with mixed states and non-projective measurements.
Notably, these ideas also enable the addition of measurement contextuality.
The convergence of either hierarchy to the quantum set remains unknown.

Furthermore, in experimental tests of this type of contextuality, it is naturally not the case that the relevant lab preparations are precisely indistinguishable, including when the measurements used to probe their distinguishability are a small subset of the entire measurement space.
Upon accepting the latter limitation, the former issue can be resolved by means of LP by leveraging the linearity of the operational theory to postprocess the lab data into new data that corresponds to exactly indistinguishable preparations \cite{Mazurek2016}.
Simplified variants of this approach have also been used for qutrit-based contextuality tests \cite{Hameedi2017b}.

\subsection{Symmetrisation methods}\label{sec:symmetries}
Many of the most interesting problems in quantum information exhibit a degree of symmetry.
Exploiting them can lead to vast computational advantages: turning an intractable problem into a tractable one, or even making it simple enough to allow for an analytical solution.
Symmetries have been fruitfully applied to several problems: for example polynomial optimisation \cite{Gatermann2004}, nonlocal correlations \cite{Fadel2017, Moroder2013, Ioannou2021b}, quantum communication \cite{Tavakoli2019}, mutually unbiased bases \cite{Aguilar2018, Gribling2021b}, port-based teleportation \cite{Studzinski2017,Mozrzymas2018,Mozrzymas2021}, unitary inversion, transposition, and conjugation \cite{Yoshida2022,Ebler2023,Grinko2023}, rank-constrained optimisation \cite{Yu2022}, and measurement incompatibility \cite{Nguyen2020}.

The fundamental idea behind symmetrisation techniques is that if both the objective function and the feasible set of an SDP are invariant under transformation of the variable $X$ by some function $f$, one can exploit this symmetry to eliminate redundant variables and block diagonalise $X$.
Both of these reductions can drastically simplify the problem.

To be more precise, consider again an SDP in the primal form of Eq.~\eqref{primal}.
Assume that there exists a function $f$ such that $\langle C, f(X)\rangle = \langle C, X\rangle$, and furthermore that if $X$ is a feasible point, that is, $\langle A_i,X\rangle = b_i\,\forall\,i$ and $X \succeq 0$, then $f(X)$ is also a feasible point.
For all feasible $X$ and $\lambda \in [0,1]$, the point $g(X,\lambda) = \lambda X + (1-\lambda) f(X)$ will then be feasible and attain the same value of the objective, which follows from linearity and convexity.
Assume also that there exists $\lambda'$ such that $f(g(X,\lambda')) = g(X,\lambda')$, so that $g(X,\lambda')$ is a projection of $X$ onto a fixed point of $f$.
One can then add the constraint $f(X) = X$ to the SDP in Eq.~\eqref{primal} without loss of generality.
This is, it is possible to rewrite Eq.~\eqref{primal} as
\begin{equation}\label{symmetrized}
	\begin{aligned}
		\max_X \quad & \langle C, X\rangle \\
		\st \quad & \langle A_i, X\rangle = b_i \quad \forall\,i, \\
		& f(X) = X, \\
		& X \succeq 0.
	\end{aligned}
\end{equation}
To see why, consider a feasible (or optimal) point $X'$ for the SDP in Eq.~\eqref{primal}.
From the previous argument $g(X',\lambda')$ will also be feasible for the original SDP, with the same value of the objective.
Since by assumption it is a fixed point of $f$, it is also feasible for the SDP in Eq.~\eqref{symmetrized}.

The simplest example of symmetrisation is when $C$, $A_i$, and $b_i$ are all real.
Then a function $f$ that leaves the objective and feasible set of the SDP invariant is complex conjugation, and the projection onto its fixed point is taking the real part of $X$.
This symmetrisation often delivers significant performance improvements, as SDP solvers often have poor support for complex numbers.

In general, symmetrising an SDP boils down to identifying $f$, the projection onto the fixed point subspace, and using the constraint $f(X) = X$ to simplify the problem.
The theory for doing so is particularly simple and well-developed when $f$ is a group action \cite{Gatermann2004,Bachoc2011,Riener2013}, so we shall present it here, while noting that more general techniques exist \cite{deklerk2011,Permenter2019}.

Let then $G$ be a group, and let $\rho$ be a representation of the group, that is, a function $g \mapsto \rho_g$ such that for all $g,h \in G$ we have $\rho_{gh} = \rho_g \rho_h$.
Here we are going to consider only unitary representations, which are those in which $\rho_{g^{-1}} = \rho_g^\dagger$.
The group then acts on the SDP variable as $X \mapsto \rho_g X \rho_g^\dagger$.
We say that the SDP is invariant under this group action if for all $g$ we have that $\langle C, \rho_g X \rho_g^\dagger\rangle = \langle C, X\rangle$ and $\langle A_i, X\rangle = b_i$ imply $\langle A_i, \rho_g X \rho_g^\dagger\rangle = b_i$ for all $i$.
Note that we do not need to consider whether $\rho_g X \rho_g^\dagger \succeq 0$ for $X \succeq 0$, as this is always the case.

The projection onto the fixed point subspace is then given by the group average\footnote{In the case of an infinite but compact group, this is given by $\int_G \dint \mu(g) \rho_g X \rho_g^\dagger$, where $\mu$ is the Haar measure on $G$.},
\begin{equation}
	\overline{X} = \frac1{|G|}\sum_{g\in G} \rho_g X \rho_g^\dagger,
\end{equation}
which can be easily verified to satisfy $\overline X = \rho_g \overline X \rho_g^\dagger$ for all $g$, so we can add that as a constraint to the SDP.

This constraint allows one not only to eliminate redundant variables, but also to block diagonalise $\overline X$ using Schur's lemma.
The main idea is that a group representation can be decomposed as a direct sum of the irreducible representations with their multiplicities, so there exists a unitary matrix $V$ such that for all $g$
\begin{equation}
	V \rho_g V^\dagger = \bigoplus_i \id_{n_i} \otimes \rho_g^i,
\end{equation}
where $\rho_g^i$ is the $i$-th irreducible representation with dimension $d_i$ and multiplicity $n_i$.
Now the constraint $\overline X = \rho_g \overline X \rho_g^\dagger$ is equivalent to $[\overline X, \rho_g] = 0$, which implies that the same $V$ also block diagonalises $\overline X$,
\begin{equation}
	V \overline X V^\dagger = \bigoplus_i {\overline X}^i \otimes \id_{d_i},
\end{equation}
where $X^i$ is a Hermitian matrix of dimension $n_i$.

Computing $V$ can be challenging.
For small problems it can be computed analytically using computer algebra systems such as GAP \cite{GAP}.
In the particular cases where the representation in question is the tensor product of $n$ unitaries of dimension $d$, the permutations between $n$ tensor factors of dimension $d$, or a combination of them, the Schur-Weyl duality can be used to give an explicit construction of $V$ \cite{Bacon2007}.
In general, though, the unitary $V$ can only be computed numerically, using software such as RepLAB \cite{Rosset2021}.

To illustrate how symmetrisation works, let us consider a simple SDP:
\begin{equation}
	\begin{aligned}
		\min_{x_1,x_2} \quad & x_1 + x_2 \\
		\st \quad & X = \begin{pmatrix} 2 & x_1 & 1 \\
			x_1 & 2 & x_2 \\
			1 & x_2 & 2 
		\end{pmatrix} \succeq 0.
	\end{aligned}
\end{equation}
This SDP is invariant under permutation of the first and third rows and columns of $X$.
Since this permutation is its own inverse, the underlying group is the symmetric group over two elements, $G = \{e,p\}$, where $e$ is the identity and $p^2 = e$.
The group representation that we need is then $\rho_e = \id$ and
\begin{equation}
	\rho_p = \begin{pmatrix} 0 & 0 & 1 \\ 0 & 1 & 0 \\ 1 & 0 & 0 \end{pmatrix}.
\end{equation}

First we eliminate variables using the group average:
\begin{equation}
	\overline X = \frac12(\rho_e X \rho_e^\dagger + \rho_p X \rho_p^\dagger) = \begin{pmatrix} 2 & y & 1 \\
		y & 2 & y \\
		1 & y & 2 
	\end{pmatrix},
\end{equation}
where $y = (x_1+x_2)/2$ is now the sole variable of the SDP.

To perform the block diagonlisation, we note that the symmetric group over two elements has only two irreducible representations, $1$ and $-1$.
The representation that we are using consists of two copies of $1$ and one copy of $-1$, and the unitary that block diagonalises it is 
\begin{equation}
	V = \frac1{\sqrt 2}
	\begin{pmatrix}
		1 & 0 & -1 \\
		0 & \sqrt{2} & 0 \\
		1 & 0 & 1 
	\end{pmatrix},
\end{equation}
with which $V\rho_p V^\dagger = (1 \otimes -1) \oplus (\id_2 \otimes 1)$.
The same unitary block diagonalises $\overline X$ as $V \overline X V^\dagger = ({\overline X}^1 \otimes 1) \oplus ({\overline X}^2 \otimes 1)$, where ${\overline X}^1 = 1$ and ${\overline X}^2 = \begin{pmatrix} 2 & y \sqrt2 \\ y\sqrt2 & 3 \end{pmatrix}$.
All in all, the SDP was simplified to
\begin{equation}
	\begin{aligned}
		\min_{y} \quad & 2y \\
		\st \quad & X = \begin{pmatrix} 1 & 0 & 0 \\
			0 & 2 & y\sqrt2 \\
			0 & y\sqrt2 & 3 
		\end{pmatrix} \succeq 0.
	\end{aligned}
\end{equation}
This problem can now be solved as an eigenvalue problem of a $2\times 2$ matrix, with the optimal solution bein $-2\sqrt3$.


\section{Conclusions}\label{sec:conclusions}
Quantum theory promises many advantages in information-processing tasks.
However, in general, characterizing the correlations established by quantum systems is very demanding, both at the complexity theory level and in practice.
In this review we have shown that many questions related to quantum correlations can be written as, or relaxed to, semidefinite programming problems.
This has enabled researchers to obtain approximate or exact solutions to many problems regarding entanglement, nonlocality, quantum communication, quantum networks, and quantum cryptography, among others.
For this reason, semidefinite programming has become a central tool in the field.

\appendix

\section{Table of abbreviations}
Below we collect all the abbreviations that appear throughout the review.
Abbreviations that denote names (e.g., NPA for Navascués-Pironio-Acín) or complexity classes (e.g., QMA for Quantum Merlin Arthur) are not included.

\begin{table}[h!]
	\begin{tabular}{c|l}
		Abbreviation & \multicolumn{1}{c}{Meaning} \\ \hline\hline
		GME & Genuine multipartite entanglement \\
		LHV & Local hidden variable \\
		LP  & Linear program \\
		LOCC & Local operations and classical communication \\
		MDI & Measurement- device independent \\
		POVM & Positive operator-valued measure \\
		PPT & Positive partial transpose \\
		PSD & Positive-semidefinite \\
		QKD & Quantum key distribution \\
		QMP & Quantum marginal problem \\
		SDI & Semi- device independent \\
		SDP & Semidefinite program \\
		SOS & Sum of squares \\
	\end{tabular}
	\caption{List of abbreviations used in the review}
\end{table}

\section{Implementation guide}
\label{sec:implementations}
In this appendix we discuss publicly available computer code packages for SDP relaxation hierarchies addressing various physics problems.
We also discuss different SDP solvers and programming languages.

SDP solvers require the problem to be input in a standard form, that is roughly similar to Eqs.~\eqref{primal} and \eqref{dual}, but with details that vary with the specific solver.
This can be quite cumbersome for more complex problems.
To get around this, it is common to use modelers, which allow much more flexible forms of input, and automatically translate them to the format required by the solvers.

The available modelers are:
\begin{itemize}
\item YALMIP: open source, written in MATLAB/Octave \cite{yalmip}.
\item CVX: proprietary, written in MATLAB \cite{Grant2008}.
\item CVXPY: open source, written in Python \cite{Diamond2016}.
\item PICOS: open source, written in Python \cite{PICOS}.
\item JuMP: open source, written in Julia \cite{Lubin2023}.
\end{itemize}

There is a large number of solvers available.
Here we will mention only a few notable ones:
\begin{itemize}
\item SeDuMi: open source, bindings for MATLAB/Octave. Can handle complex numbers natively \cite{Sturm1999}.
\item SDPA: open source, bindings for C, C++, and MATLAB. The variants SDPA-GMP, SDPA-QD, and SDPA-DD can solve problems with high or arbitrary precision \cite{Nakata2001,Nakata2010}.
\item MOSEK: proprietary, bindings for C, C++, Java, Julia, MATLAB, .NET, Python, and R. Fast, parallelised implementation \cite{mosek}.
\item SCS: open source, bindings for C, C++, Julia, MATLAB, Python, R, and Ruby. Uses a first-order method in order to handle large-scale problems \cite{Odonoghue2016}.
\item Hypatia: open source, bindings for Julia. Can handle complex numbers natively and solve problems with arbitrary precision, and supports a wide variety of cones other than the SDP one \cite{Coey2021}.
\end{itemize}

There are also several software packages that implement some of the SDP relaxations discussed here.
Some notable ones are:
\begin{itemize}
\item QETLAB: open source, written in MATLAB. Works with CVX. Implements several of the algorithms discussed here, including the DPS and NPA hierarchies \cite{qetlab}.
\item Ket: open source, written in Julia. Works with JuMP. Implements several of the algorithms discussed here, including the DPS hierarchy \cite{ket}.
\item toqito: open source, written in Python. Works with CVXPY. Implements several of the algorithms discussed here, including the DPS and NPA hierarchies \cite{toqito}.
\item Moment: open source, written in C++ with bindings for MATLAB. Works with YALMIP and CVX. Performs noncommutative polynomial optimisation and classical inflation, with symmetrisation support \cite{moment}.
\item Ncpol2sdpa: open source, written in Python. Works with SDPA and MOSEK. Performs commutative and noncommutative polynomial optimisation, focusing on NPA-type problems \cite{Wittek2015}.
\item GloptiPoly: open source, written in MATLAB, works with YALMIP. Performs commutative polynomial optimisation \cite{Henrion2009}.
\item SOSTOOLS: open source, written in MATLAB. Performs commutative polynomial optimisation \cite{sostools}.
\item NCSOStools: open source, written in MATLAB. Performs noncommutative polynomial optimisation \cite{Cafuta2012}.
\item Inflation: open source, written in Python, works with MOSEK. It implements quantum inflation for quantum and classical correlations \cite{Boghiu2022}.
\item RepLAB: open source, written in MATLAB/Octave. Performs numerical representation theory for the purpose of symmetrisation \cite{Rosset2021}.
\end{itemize}

\section{Strict feasibility}\label{sec:strict}
In this appendix we prove that the unconstrained NPA hierarchy is strictly feasible, by explicitly constructing a positive-definite feasible point.
We thank Miguel Navascués for providing this proof.

Instead of the usual basis of projectors to represent Alice's and Bob's measurements, $\{A_{a|x}\}_{a,x}$ and $\{B_{b|y}\}_{b,y}$, we shall instead use the unitary basis of generalized observables, which is defined as
\begin{gather}
	A_x = \sum_{a=0}^{N-1} \omega_N^a A_{a|x}, \\
	B_y = \sum_{b=0}^{M-1} \omega_M^b B_{b|y},
\end{gather}
where $\omega_N=e^{-i\frac{2\pi}{N}}$ and $\omega_M=e^{-i\frac{2\pi}{M}}$.
The conditions that $\{A_{a|x}\}_{a,x}$ and $\{B_{b|y}\}_{b,y}$ are sets of projectors that sum to identity are mapped onto the condition that $A_x$ and $B_y$ are unitary and that their $N$-th and $M$-th powers evaluate to identity, this is, 
\begin{gather}
	\begin{gathered}
		\label{eq:unitary_relations}
		A_xA_x^\dagger = A_x^\dagger A_x = \id, \\
		B_yB_y^\dagger = B_y^\dagger B_y = \id,\\
		A_x^N = B_y^M = \id.
	\end{gathered}
\end{gather}

Note that this transformation from projectors to unitaries is reversible.
The inverse operation is given by 
\begin{gather}
	A_{a|x} = \frac1N \sum_{a'=0}^{N-1} \omega^{aa'}_N A_x^{a'}, \label{eq:Ainverse} \\
	B_{b|y} = \frac1N \sum_{b'=0}^{M-1} \omega^{bb'}_M B_y^{b'}. \label{eq:Binverse}
\end{gather}
Now, consider the set $\mathcal A$ of inequivalent sequences of products of elements in $\{A_x\}_x$, that is, monomials that are not equivalent under relations \eqref{eq:unitary_relations}.
Define then an infinite-dimensional, countable Hilbert space $\mathcal H_A$, with a canonical orthonormal basis whose elements are labeled by monomials of $A_1,\ldots,A_n$.
That is, 
\begin{equation}
	\mathcal H_A=\overline{\mbox{span}}\{\ket{a}:a\in \mathcal A\},
\end{equation}
with $\braket{a}{a'} = \delta_{a,a'}$.

Define now the operators $\{\pi_A(A_x)\}_x \in B(\mathcal H_A)$ through their action on this orthonormal basis as follows:
\begin{equation}
	\pi_A(A_x)\ket{a}=\ket{A_xa}\quad \forall\,a\in\mathcal A.
\end{equation}
For each $x$, $\pi_A(A_x)$ is a unitary operator satisfying $\pi(A_x)^N=\id$.
The representation $\pi_A$ is known in operator algebras as the \emph{left regular representation} of $\{A_x\}_x$, given the relations \eqref{eq:unitary_relations}.

Doing the analogous construction from Bob, we can now define the moment matrix
\begin{equation}
	\begin{aligned}
		\Gamma_{ab,a'b'}=\bra{\psi}\left[\pi_A(a)^\dagger\pi_A(a')\otimes\pi_B(b)^\dagger\pi_B(b')\right]\ket{\psi} \\ a,a'\in\mathcal A,\quad b,b'\in\mathcal B,
	\end{aligned}
\end{equation}
where $\ket{\psi}=\ket{1}_A\otimes\ket{1}_B$.

From the definition of the constructions we have
\begin{equation}
	\begin{aligned}
		\Gamma_{ab,a'b'} &= \bra{1}\pi_A(a)^\dagger\pi_A(a')\ket{1}\bra{1}\pi_B(b)^\dagger\pi_B(b')\ket{1} \\
		&= \braket{a}{a'}\braket{b}{b'} \\
		&=\delta_{a,a'}\delta_{b,b'},
	\end{aligned}
\end{equation}
so $\Gamma = \id$, which is positive definite, and the NPA hierarchy without constraints is strictly feasible, as we wanted to show.

Note that from this construction one can also obtain a positive-definite feasible moment matrix in the usual basis of projectors.
Using Eqs.~\eqref{eq:Ainverse} and \eqref{eq:Binverse} one constructs the linear transformation $C$ that takes monomials from the unitary basis to the projector basis.
The desired moment matrix is then $\tilde{\Gamma} = C\Gamma C^\dagger = C C^\dagger$.

\begin{acknowledgments}
We thank Carlos de Gois, Cyril Branciard, Denis Rosset, Felix Huber, Gaurav Saxena, Jef Pauwels, Kishor Bharti, Ludovico Lami, Marco Túlio Quintino, Mark Wilde, Miguel Navascués, Omar Fawzi, Otfried Gühne, Sander Gribling, Siddharta Das, Stefano Pironio, Ties Ohst, Yeong-Cherng Liang, Valerio Scarani and Vincent Russo for comments and feedback.

A.T. is supported by the Wenner-Gren Foundations and by the Knut and Alice Wallenberg Foundation through the Wallenberg Center for Quantum Technology (WACQT).

A.P.-K. is supported by the Spanish Ministry of Science and Innovation MCIN/AEI/10.13039/501100011033 (CEX2019-000904-S and PID2020-113523GB-I00), the Spanish Ministry of Economic Affairs and Digital Transformation (project QUANTUM ENIA, as part of the Recovery, Transformation and Resilience Plan, funded by EU program NextGenerationEU), Comunidad de Madrid (QUITEMAD-CM P2018/TCS-4342), Universidad Complutense de Madrid (FEI-EU-22-06), the CSIC Quantum Technologies Platform PTI-001, the NCCR SwissMAP (grant no. 205607), and the Swiss National Science Foundation (grant number 224561).

P.B. acknowledges funding from the European Union's Horizon Europe research and innovation programme under the project ``Quantum Secure Networks Partnership'' (QSNP, grant agreement No. 101114043).

M.A. acknowledges funding from the FWF stand-alone project P 35509-N and was also supported by the European Union--Next Generation UE/MICIU/Plan de Recuperación, Transformación y Resiliencia/Junta de Castilla y León.
\end{acknowledgments}

\addcontentsline{toc}{chapter}{References}

\bibliography{references}

\end{document}